\documentclass{aa}  
\usepackage{graphicx}
\usepackage{txfonts}
\usepackage[colorlinks=true]{hyperref}
\hypersetup{linkcolor=blue, citecolor=blue}
\usepackage[draft]{todonotes}
\usepackage{mathtools}
\usepackage{booktabs}
\usepackage{multirow}
\usepackage{xcolor}
\definecolor{kostas}{rgb}{1.00, 0.00, 0.0}

\begin{document} 
\newcommand{\Ma}{$M_{A}$}
\newcommand{\Ms}{$M_{S}$}
\newcommand{\NHI}{N$_{\rm{H_{I}}}$}
\newcommand{\HI}{H\,{\sc i}}
\newcommand{\CII}{[C\,{\sc ii}]}
\newcommand{\CI}{[C\,{\sc i}]}
\newcommand{\Hmolecular}{H$_{2}$}
\newcommand{\kms}{km~s$^{-1}$}
\newcommand{\mum}{$\mu$m}
\newcommand{\ColDens}{cm$^{-2}$}
\newcommand{\VolDens}{cm$^{-3}$}
\newcommand{\rate}{s$^{-1}$}
\newcommand{\CO}{$^{12}$CO}
\newcommand{\Tk}{T$_{K}$}
\newcommand{\Bpos}{B$_{\rm{POS}}$}
\newcommand{\Blos}{B$_{\rm{LOS}}$}
\newcommand{\DensUnits}{cm$^{-3}$}
\newcommand{\Uturb}{$\sigma(u_{\rm{turb}})$}
\newcommand{\Utherm}{$\sigma(u_{\rm{therm}})$}
\newcommand{\Pangles}{$\chi$}
\newcommand{\Av}{A$_{v}$}
\newcommand{\NH}{N$_{\rm{H}}$}
\newcommand{\NHmol}{N$_{\rm{H_{2}}}$}
\newcommand{\NHatom}{$\rm{N_{H_{I}}}$}
\newcommand{\NHtot}{N$_{\rm{H}}$}
\newcommand{\ICplus}{$\rm{I}_{[C_{\rm{II}}]}$}

   \title{\HI\ - \Hmolecular\ transition: exploring the role of the magnetic field}
   
   \subtitle{A case study towards the Ursa Major cirrus}

    \titlerunning{\HI\ - \Hmolecular\ transition: exploring the role of the magnetic field}

   \author{R. Skalidis \inst{1} \fnmsep \inst{2}\thanks{rskalidis@physics.uoc.gr}, K. Tassis \inst{1} \fnmsep \inst{2}, G. V. Panopoulou \inst{3}, J. L. Pineda \inst{4}, Y. Gong \inst{5, 6},  N. Mandarakas \inst{1} \fnmsep \inst{2}, D. Blinov \inst{1} \fnmsep \inst{2}, S. Kiehlmann\inst{1} \fnmsep \inst{2} \and J. A. Kypriotakis\inst{1} \fnmsep \inst{2}
          }
        
   \institute{
        Department of Physics \& ITCP, University of Crete, GR-70013, Heraklion, Greece
        \and
        Institute of Astrophysics, Foundation for Research and Technology-Hellas, Vasilika Vouton, GR-70013 Heraklion, Greece
        \and
        California Institute of Technology, MC249-17, 1200 East California Boulevard, Pasadena, CA 91125, USA
        \and Jet Propulsion Laboratory, California Institute of Technology, 4800 Oak Grove Drive, Pasadena, CA 91109-8099, USA
        \and Max-Planck-Institut f{\"u}r Radioastronomie, Auf dem H{\"u}gel 69, D-53121 Bonn, Germany
        \and Purple Mountain Observatory \& Key Laboratory for Radio Astronomy, Chinese Academy of Sciences, 10 Yuanhua Road, 210033 Nanjing, PR China
        }
        
    \authorrunning{Skalidis et al. 2021}
    
   \date{Received ; accepted }
   
  \abstract
   {Atomic gas in the diffuse interstellar medium (ISM) is organized in filamentary structures. These structures usually host cold and dense molecular clumps. The Galactic magnetic field is considered to play an important role in the formation of these clumps.}
   {Our goal is to explore the role of the magnetic field in the \HI-\Hmolecular\ transition process.}
   {We targeted a diffuse ISM filamentary cloud where gas transitions from atomic to molecular towards the Ursa Major cirrus. We probed the magnetic field properties of the cloud with optical polarization observations.
   We performed multi-wavelength spectroscopic observations of different species in order to probe the gas phase properties of the cloud. We observed the CO (J=1-0) and (J=2-1) lines in order to probe the molecular content of the cloud. We also obtained observations of the \CII\ 157.6 \mum\ emission line 
   in order to trace the CO-dark \Hmolecular\ gas and estimate the mean volume density of the cloud.}
   {We identified two distinct sub-regions within the cloud. One of the regions is mostly atomic, while the other is dominated by molecular gas although most of it is CO-dark. The estimated plane-of-the-sky magnetic field strength between the two regions remains constant within uncertainties and lies in the range $13 - 30~\mu$G.
   The total magnetic field strength does not scale with density. This implies that gas is compressed along the field lines. We also found that turbulence is trans - Alfvénic with $M_{A} \approx 1$. In the molecular region, we detected an asymmetric CO clump with its minor axis being closer, with a $24\degr$ deviation, to the mean magnetic field orientation than the angle of its major axis. The \HI\ velocity gradients are in general perpendicular to the mean magnetic field orientation, except for the region close to the CO clump where they tend to become parallel. The latter is likely related to gas undergoing gravitational infall. The magnetic field morphology of the target cloud is parallel to the \HI\ column density structure of the cloud in the atomic region, while it tends to become perpendicular to the \HI\ structure in the molecular region. On the other hand, the magnetic field morphology seems to form a smaller offset angle with the total column density (including both atomic and molecular gas) shape of this transition cloud.}
   {In the target cloud where the \HI-\Hmolecular\ transition takes place, turbulence is trans-Alfvénic, hence the magnetic field plays an important role in the cloud dynamics. Atomic gas probably accumulates preferentially along the magnetic field lines and creates overdensities where molecular gas can form. The magnetic field morphology is probed better by the total column density shape of the cloud, and not from its \HI\ column density shape.}

   \maketitle
     
\section{Introduction}

The diffuse interstellar medium (ISM) is the precursor of molecular clouds, where stars form. The structure of this medium is known to  harbor filamentary overdensities \citep[e.g.][]{heiles_troland1,kalberla_2016}. Density enhancements can shield the interior of the clouds from the interstellar radiation field, and hence reduce the photo-destruction rate of molecular gas \citep{goldsmith_2013}. As a result, molecular gas over-densities, hereafter molecular clumps, can form within the atomic (\HI) medium \citep{planck_2015_chameleon, kalberla_2020}. The details of how the transition from atomic to molecular gas occurs (\HI\--$\rm H_2$ transition) are still uncertain, but have important implications for star formation \citep[e.g.,][]{Sternberg2014} and the multiphase structure of the ISM \citep[e.g.,][]{Bialy2017}.  

One of the remaining challenges in the study of the formation of molecular gas concerns the role of the magnetic field. State-of-the-art numerical simulations show that the strength of the magnetic field can alter the column density statistics of the \HI\--$\rm H_2$ transition \citep{bellomi_2020}. However, observational constraints on the properties of the magnetic field are limited, especially in the column density regime relevant for the \HI\--$\rm H_2$ transition \citep[the column density threshold at high Galactic latitudes is $\sim 2 \times 10^{20} \rm cm^{-2}$,][]{Gillmon2006}. 

In recent years, a new observable has been introduced to quantify the properties of the magnetic field, specifically its topology with respect to density structures. This observable is the relative orientation between the magnetic field (projected on the plane of the sky) and column density filaments. In the diffuse, purely atomic medium, \HI\ filaments are found to be preferentially parallel to the magnetic field \citep{clark_2014, clark_2015, kalberla_2016}. However, at higher column density where the medium is molecular, the magnetic field lines are preferentially perpendicular to dense gas filaments \citep{planck_xxxv_2016}. The change in relative orientation occurs over a wide range of column densities \NH\ $\sim 10^{21.7} - 10^{24.1} \rm cm^{-2}$ for the molecular clouds studied in \citet{planck_xxxv_2016}. \cite{fissel_2019} studied various molecular transitions in the Vela C molecular cloud. They found that the change in relative orientation happens at a volume density of $\sim$1000 \VolDens; however this value is quite uncertain, by a factor of 10. \cite{alina_2019} studied a larger sample of clouds and found that the transition is more prominent where the column density contrast between structures and their surroundings is low.

A number of theoretical works have investigated the change in relative orientation with the use of magnetohydrodynamic (MHD) simulations. In general, the change in relative orientation is found in MHD simulations with dynamically important magnetic fields \citep{soler_2014, Seifried_2020}. The two configurations (parallel and perpendicular) are minimum energy states of the MHD equations \citep{soler_2017}.  However, the various theoretical studies propose different interpretations for the origin of the change in relative orientation. \cite{Chen_2016} use colliding flow simulations and find that the change in relative orientation happens when gas motions become super-Alfvénic. In contrast, \cite{soler_2017} investigate decaying turbulence in a periodic box and find the change can happen at much higher Alfvénic Mach numbers (\Ma), $M_{A} \gg 1$. Their simulations do not show a correlation between \Ma\ and the transition density, but instead point to the coupling between compressive motions and the magnetic field as the root cause of the change. The simulations of \citet{kroetgen_2020}, which do not include self-gravity, do show a correlation of \Ma\ with the transition density. \cite{Seifried_2020} propose that the transition happens where the cloud becomes magnetically super-critical.

The \HI-\Hmolecular\ transition could, in principle, be associated with qualitative changes in other ISM properties, such as the relative orientation between the magnetic field and the molecular clump shape \citep[as in the simulations of][]{girichidis_2021}. Simply, if the magnetic field is dynamically important, magnetic forces, which are always exerted perpendicular to the field lines, should suppress gas flows perpendicular to the field lines. On the other hand, gas can stream freely parallel to the field lines and condense. This process can form molecular clumps which are asymmetric and stretched perpendicular to the lines of force \citep[e.g.,][]{mouschovias_1978, heitsch_2009, chen_ostriker_2014, chen_ostriker_2015}. These changes are expected to leave imprints on the velocity field as well; for example, velocity gradients of the \HI\ line become parallel to the magnetic field at the onset of gravitational collapse \citep{hu_2020_vgt_gravity}.

In this work we provide new observational constraints on the role of the magnetic field in the \HI\--$\rm H_2$ transition, by performing the analysis of relative orientations between the magnetic field and the gas structure, and kinematics within a carefully chosen case-study region. Specifically, we aim to answer the following questions:  Is the \HI\--$\rm H_2$ transition connected to a change in relative orientation between the magnetic field and the gas structures within the same cloud \citep[as expected if the change in relative orientation is tied to the onset of gravitational collapse in a dynamically important magnetic field, found in][]{girichidis_2021}? Are the statistics of velocity gradients consistent with a magnetically-channeled gravitational contraction and associated with the \HI\--$\rm H_2$ transition \citep[as expected by e.g.,][]{hu_2020_vgt_gravity}? In addition, we aim to study the correlation of the magnetic field morphology with the various tracers, \HI\ or \Hmolecular.

We conducted our analysis towards the Ursa Major cirrus \citep{miville_2002, miville_2003}. This is a diffuse ISM cloud located at the North Celestial Pole Loop (NCPL) \citep{heiles_1989}. It is filamentary, and consists of one part where gas is dominated by \HI\ and another part where gas is mostly molecular \citep{miville_2002}. We performed polarization observations in the optical, in order to probe the magnetic field properties of the cloud. We obtained CO observations, tracing the J=1-0 and J=2-1 emission lines, in order to probe the molecular content of the cloud. Since molecular gas, especially in transition clouds such as this one, is not completely traced by CO, we obtained \CII\ observations at 157.6 \mum. With these data, we aim to probe the existence of \Hmolecular which is not traced by CO, referred to as CO-dark \Hmolecular \citep{Grenier_2005, pineda_2010, langer_2010}. 

The structure of the paper is the following: In Sect.~\ref{sec:data} we present our observations and data processing. In Sect.~\ref{sec:analysis}, we estimate the gas abundances and the magnetic field strength of the cloud. In the same section we present an analysis of the relative orientation of the magnetic field, with the molecular structure of the cloud and the \HI\ velocity gradients. In Sect.~\ref{subsec:rht_polarization}, we compare the relative orientation of the magnetic field morphology with the \HI\ structure of the target cloud. In Sect.~\ref{subsection:ncpl_B_molecule_formation} we provide evidence of the role of the magnetic field in the molecule formation of the NCPL. We discuss our results in Sect.~\ref{sec:discussion} and in Sect.~\ref{sec:conclusions} we present our main conclusions.

\section{Data}
\label{sec:data}

    \begin{figure}
        \includegraphics[width=\hsize]{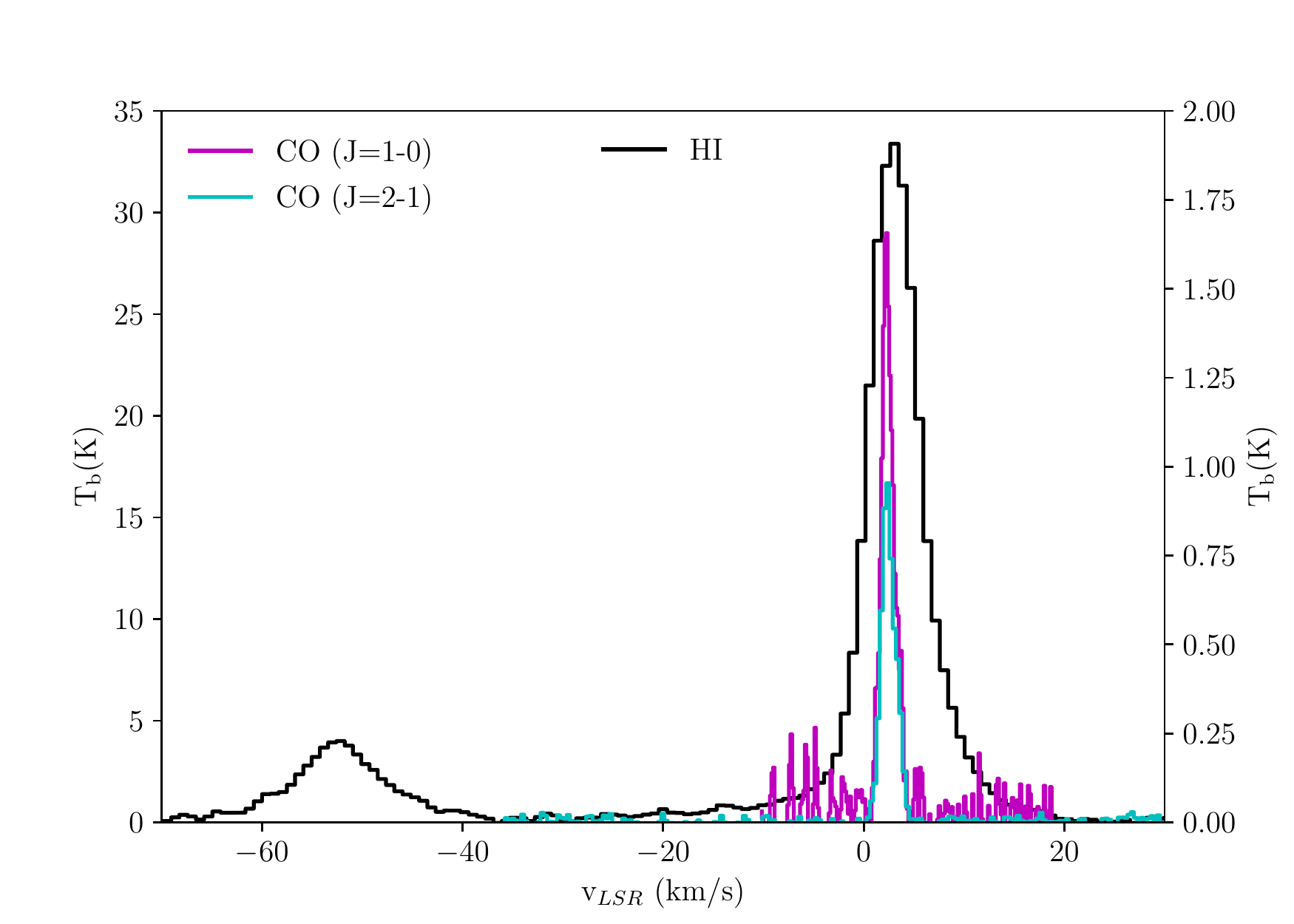}
        \caption{Brightness temperature versus gas velocity. The black histogram corresponds to the \HI\ emission spectrum, magenta to \CO\ (J = 1 - 0) and cyan to \CO\ (J = 2 - 1). The antenna temperature of \HI\ is shown in the left vertical axis, while the CO lines in the right vertical axis. The size of each bin is equal to the spectral resolution of each spectrum, i.e. 0.82, 0.16 and 0.32 \kms\ for \HI, CO (J=1-0) and CO (J=2-1) emission lines respectively. These spectra were extracted from RA, Dec = 09:33:00, +70:09:00.}
        \label{fig:spectra}
    \end{figure}
    
    \begin{figure*}
        \includegraphics[width=\textwidth]{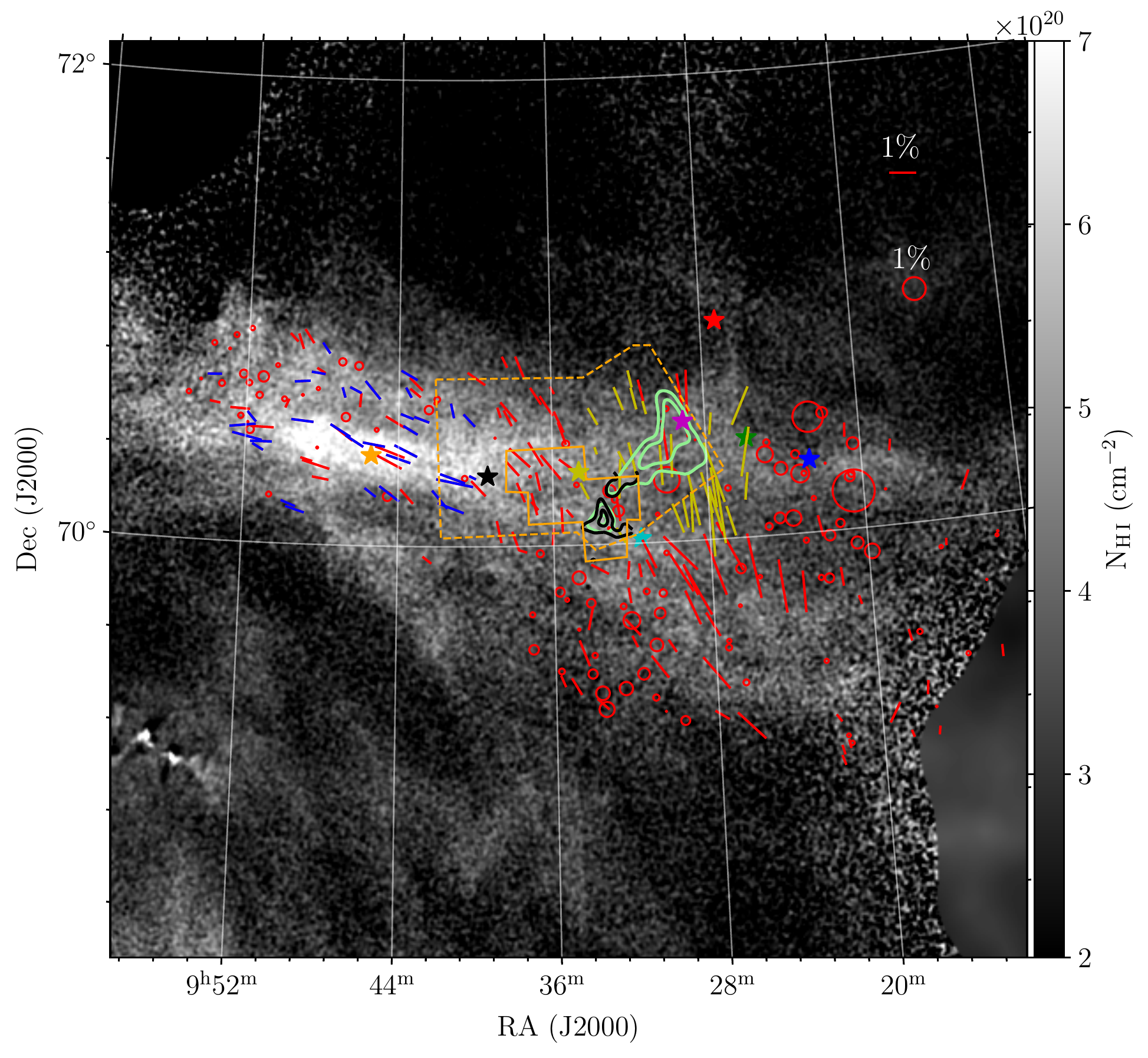}
        \caption{\HI\ column density map of the target cloud. Colored segments correspond to the RoboPol polarization data with S/N $\geq 2.5$ tracing the plane-of-the-sky magnetic field. Blue and yellow segments correspond to measurements included in the estimation of the magnetic field strength in the atomic and molecular region respectively. Red segments correspond to measurements with S/N $\geq 2.5$, but were not used in the estimation of \Bpos (Sect.~\ref{sec:bpos_strength_computation}). Measurements with S/N $<2.5$ are shown with red circles. The radius of each circle corresponds to the observed polarization fraction. A scale segment and circle are shown in the top right corner. Colored stars correspond to the C+ lines of sight from both ISO and SOFIA. The size of each star is much larger than the field of view of the instruments. Light green contours correspond to the CO (J=1-0) integrated intensity at the $3$, $5$ and $9$ K km/s levels respectively. The black contour shows the CO (J=2-1) integrated intensity at $2$ and $4$ K km/s levels respectively. The orange dashed and solid polygons show the CO J=1-0 and J=2-1 surveyed regions respectively.}
        \label{fig:hi_polarization}
    \end{figure*}

\subsection{\HI\ 21cm emission line}
\label{subsec:hi4pi_data}

We obtained the publicly available \HI\ data of the 21 cm emission line of the UM field from the DHIGLS survey \citep{DHIGLS}. The angular and spectral resolution of the data is $1\arcmin$ and $0.82$ km/s respectively. The data are in the form of a position-position-velocity (PPV) cube, $T_{b}(\alpha, \delta, {\rm v})$, where $T_{b}$ is the antenna temperature at given equatorial coordinates $\alpha$ and $\delta$, and ${\rm v}$ is the Doppler shifted gas velocity measured with respect to the local standard of rest (LSR). In Fig.~\ref{fig:spectra} we show with the black histogram a characteristic spectrum of the cloud, at a line of sight (LOS) where all of our spectraly resolved emission lines (\HI, CO (J=1-0), and CO(J=2-1)) have been detected. The \HI\ spectrum is characterized by a low velocity cloud (LVC) at $3$ km/s and an intermediate velocity cloud (IVC) at $-55$ km/s. \cite{tritsis_2019} used data from the \cite{green_2018} 3D extinction map and inferred the distance of the LVC to be $\sim 300$ pc and of the IVC to be $1$ kpc. Our polarimetric data also verify the LVC distance obtained by \cite{tritsis_2019} (Appendix~\ref{sec:cloud_distance}), and hence we adopt it as the reference cloud distance. In our analysis we explored the properties of the LVC only. We have computed the zeroth-moment map, $\rm I_0$, of the \HI\ emission in the [$-22$, $20$] km/s velocity range as
\begin{equation}
    \label{eqhi_integrated_intensity}
    \rm{I_{0}} (\alpha, \delta) =  \sum_{v=-22 \, km/s}^{20 \, km/s} T_{b}(\alpha, \delta, v) \, \Delta v,  
\end{equation}
where $\rm T_{b}$ is the brightness temperature, including the beam efficiency, measured in K and $\rm \Delta v = 0.82$ km/s is the spectral resolution. We computed the \HI\ column density, assuming optically thin emission following \cite{dickey_lockman}, \begin{equation}
    \label{eq:hi_column_density}
    \rm{N_{H_{\rm{I}}}} (\rm{cm}^{-2}) = 1.823 \times 10^{18} \times \rm{I_{0}}.
\end{equation}
In Fig.~\ref{fig:hi_polarization} we show the \HI\ column density map of the target cloud.

\subsection{Optical polarization}
\label{subsec:pol_data}

We performed polarimetric observations in the optical at the Skinakas Observatory in Crete, Greece\footnote{\url{http://skinakas.physics.uoc.gr}}. We used the RoboPol  four-channel imaging polarimeter \citep{king_2014, robopol_paper_2019}. The instrument measures the $q$ and $u$ relative Stokes parameters simultaneously and has a $13.6\arcmin \times 13.6\arcmin$ field of view. In the central part of the field there is a mask (Figs.~2 and 4 in \cite{king_2014}) which reduces the sky background. The instrumental systematic uncertainty in $q$ and $u$  within the mask is below $\sim 0.1 \%$ \citep{skalidis_2018, robopol_paper_2019}. All the stars were observed within the central mask of the instrument using a Johnson-Cousins R filter. Observations were carried out during four observing runs: 2019 October - November, 2019 May - June, 2019 September - October, 2020 May - June. The data processing was carried out by an automatic pipeline \citep{king_2014, panopoulou_2015}. We followed the same analysis as in \cite{skalidis_2018} for the calculation of the $q$ and $u$ Stokes parameters and their uncertainties. We computed the degree of polarization as
\begin{equation}
    p = \sqrt{q^{2} + u^{2}}.
\end{equation}
Uncertainties in $q$ and $u$ are propagated to $p$ assuming Gaussian error distributions for $q$ and $u$ respectively, Eq.~(5) in \cite{king_2014}. The degree of polarization is always a positive quantity, which is significantly biased towards larger values when the signal-to-noise ratio (S/N) in $p$ is low \citep{Vaillancourt_2006, Plaszczynski_2014}. We did not debias the $p$ values since we did not use them in our analysis. Thus, all the $p$ values in the accompanied data table of this paper are biased. The polarization angle was computed as in \cite{king_2014},
\begin{equation}
    \chi = 0.5~\rm{arctan \left( \frac{u}{q} \right )},
\end{equation}
with the origin of $\chi$ being the North Celestial Pole which follows the IAU convention and increases towards the East. Uncertainties in the polarization angles are computed using Eqs.~(1) and~(2) from \cite{blinov_2021}, which are based on \cite{Naghizadeh_clarke_1993}. 

We observed stars up to $\sim 15$ magnitude in the R-band of the USNO-B1.0 catalogue \citep{USNOB}. We extracted the geometric star distances from the \cite{bailer_jones_2021} catalogue, which uses stellar parallaxes from \textit{Gaia} DR3 \citep{gaia_dr3}. During each observing night we observed standard stars for instrumental calibration. We used the following standard stars:  BD + 28 4211, BD +32 3739, HD 212311 \citep{schmidt_1992}, BD +33 2642 \citep{skalidis_2018}, HD 154892 \citep{turnshek_1990}. The instrument calibration is performed as in \cite{skalidis_2018}.

In Fig.~\ref{fig:hi_polarization} we show our polarization measurements over plotted on the $\rm{N_{H_{I}}}$ map of the cloud. Measurements with S/N in fractional polarization higher than 2.5 are shown with colored segments, while measurements with S/N $<2.5$ are shown with red circles. The length of each segment and the radius of each circle is proportional to the degree of polarization. On the top right corner of the figure we show a segment and a circle for scale. 

    \begin{table}[h]
        \caption{Integrated \CII\ intensities}
        \begin{tabular}{lccccc}
        \hline \hline
         &RA$(\degr)$ & DEC$(\degr)$ & $\rm{I_{[C_{II}}]} \pm \sigma \rm{I_{[C_{II}}]}$ [K \kms] & Ref \\ 
        \hline
        $\textcolor{orange}{\ast}$ & 09:45:25 & +70:23:26 & $0.21  \pm 0.04$ & 1 \\
        $\textcolor{black}{\ast}$ & 09:39:22 & +70:18:08 & $0.39  \pm 0.07$ & 1 \\
        $\textcolor{yellow}{\ast}$ & 09:34:39 & +70:18:56 & $0.12  \pm 0.04$ & 1 \\
        $\textcolor{magenta}{\ast}$ & 09:29:09 & +70:31:00 & $0.33  \pm 0.07$ & 1 \\
        $\textcolor{green}{\ast}$ & 09:25:53 & +70:25:45 & $32.0  \pm 2.0 $ & 2 \\
        $\textcolor{blue}{\ast}$ & 09:22:41 & +70:18:55 & $52.0  \pm 5.0 $ & 2 \\
        $\textcolor{red}{\ast}$ & 09:27:13 & +70:56:28 & $43.0  \pm 4.0 $ & 2 \\
        $\textcolor{cyan}{\ast}$ & 09:31:32 & +70:01:15 & $57.9  \pm 2.0 $ & 2 \\
        \hline
        \end{tabular}
        \tablebib{(1)~\citet{inglalls_2002}; (2)~This work.}
        \tablefoot{Col.~1 corresponds to the colored-stars lines of sight shown in Fig.~\ref{fig:hi_polarization}.}
        \label{table:carbon_intensities}
    \end{table}

\subsection{\CII\ 157.6 \mum\ emission line}
\label{subsec:c+_data}

    \begin{figure*}
        \centering
        \includegraphics[width=\hsize]{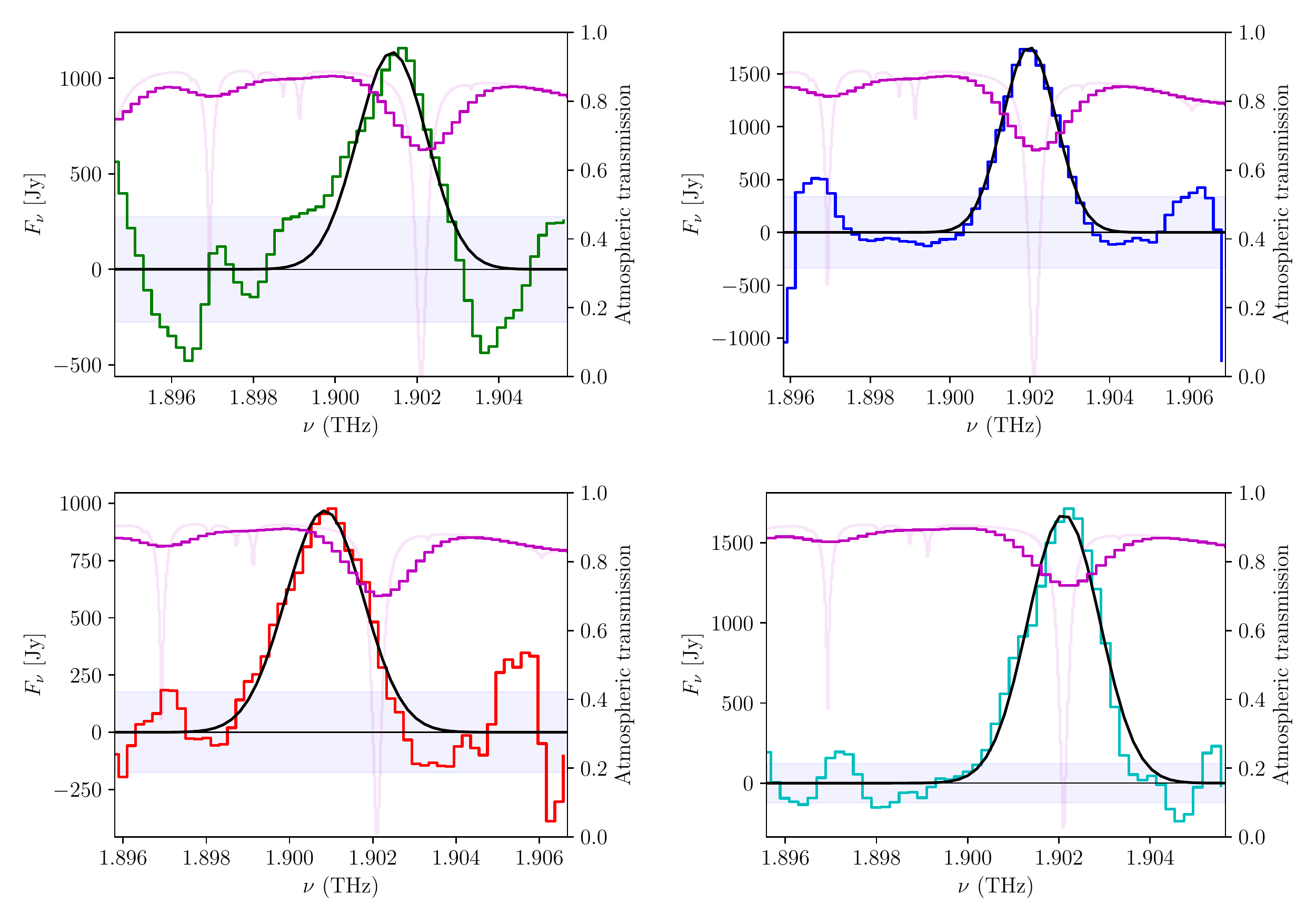}
        \caption{\CII\ emission line spectra of the four LOSs from the SOFIA FIFI-LS spectrometer. The continuum-subtracted fluxes are shown with the histograms. The color of each histogram corresponds to the LOS marked with the same colored star in Fig.~\ref{fig:hi_polarization}. Light magenta corresponds to the transmission atmospheric model and the dark magenta histogram to the transmission model convolved with the instrumental spectral resolution. The black solid line corresponds to the Gaussian fit. The blue shaded region shows the level of the continuum rms.}
        \label{fig:sofia_data}
    \end{figure*}

    \begin{figure*}
        \centering
        \includegraphics[width=\hsize]{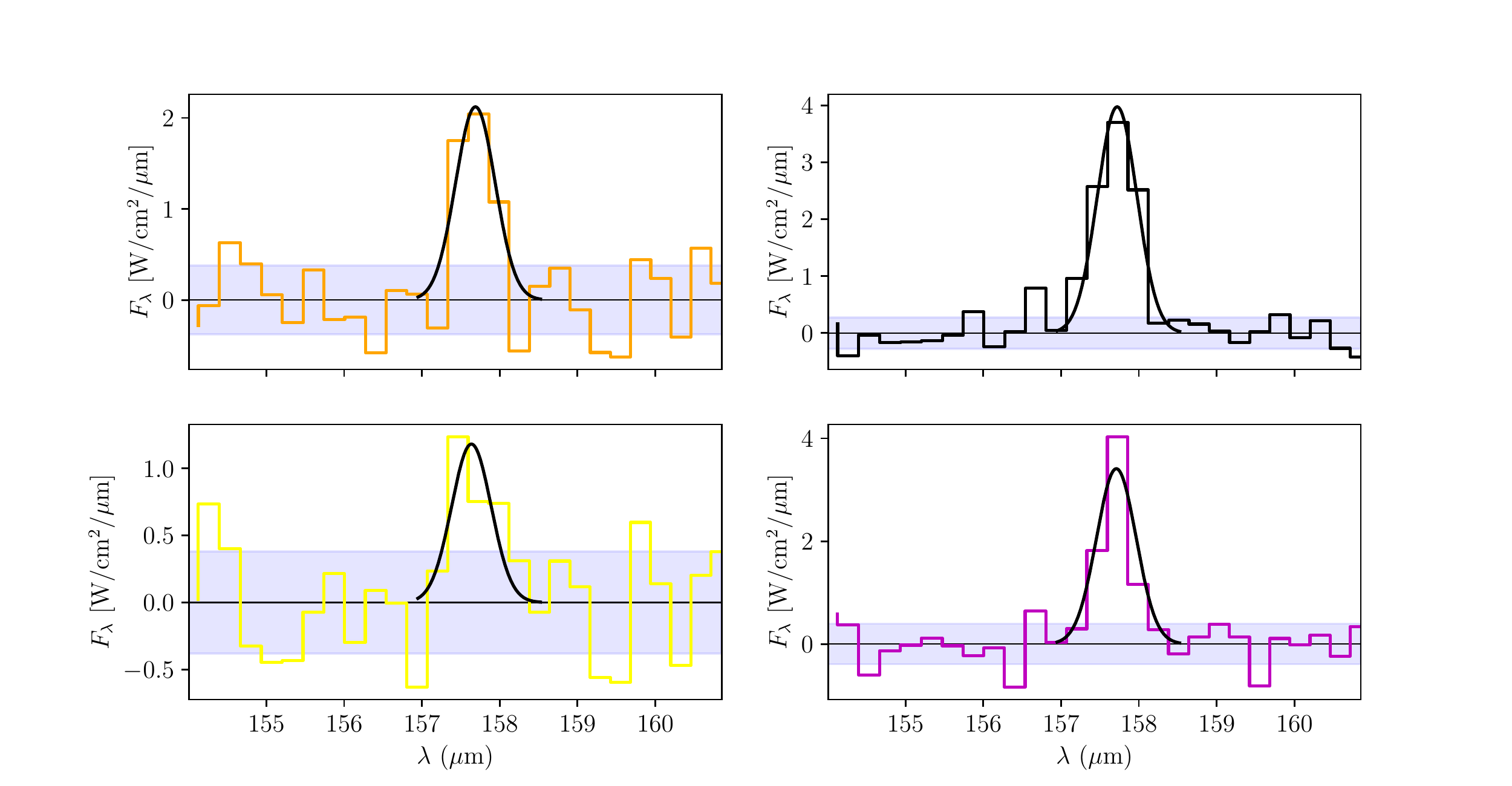}
        \caption{\CII\ emission line spectra, $I_{\lambda}$ vs $\lambda$, obtained with ISO \citep{inglalls_2002}. The amplitude of continuum fluctuations is shown with light blue. The black solid lines correspond to the Gaussian fits of the C+ emission line.}
        \label{fig:iso_carbon}
    \end{figure*}

We observed the $157.6$ \mum\ emission line of ionized carbon (C+) with the SOFIA FIFI-LS spectrometer \citep{colditz, klein, fisher_2018}. The data were obtained in the red channel of FIFI-LS, which covers the 115 - 203 $\mu$m wavelength range. The field of view consists of a $5 \times 5$ array of spectral pixels ("spaxels") with the size of each pixel equal to $12\arcsec \times 12\arcsec$. The spectral resolution at the targeted wavelengths is $1000$, which corresponds to a velocity resolution of $270$ \kms. The velocity resolution is much larger compared to typical linewidths of the C+ emission lines \citep[e.g.,][]{goldsmith_2018}, and hence the line is spectrally unresolved. 

The LOSs are shown in Fig.~\ref{fig:hi_polarization}, with the green, blue, red and cyan stars, and their corresponding coordinates are shown in Table~\ref{table:carbon_intensities}. Observations were carried out during the observing Cycle 08 as part of the proposal with plan ID 08\_0237 (PI: G. V. Panopoulou). The total observing time was 6.5 hours, out of which 2.5 hours corresponded to on-source time and the rest to off-source time used for calibration. We proposed for five different LOSs, but one of them was not used due to the presence of a negative baseline caused by bad weather conditions during the observations. In total we used four out of the five LOSs that were initially proposed. The observations were carried out in ``total power mode'', which does not include telescope chopping, but only nodding. We selected regions with low infrared flux (100~\mum) to extinction ($\rm A_{V}$) ratio as nod positions. The off-source (nod position) ratio was selected to be $\sim 100$ times lower than the on-source within a chopping angle limit of $45\arcmin$. Data reduction was performed using the FIFI-LS automatic pipeline. 
For the SOFIA data processing we have used the open source software SOSPEX\footnote{\url{https://github.com/darioflute/sospex}} \citep{sospex}. We used the highest level pipeline product (level 4) which includes the chop subtraction, wavelength and flux  calibration. We trimmed the frequency axis in order to eliminate frequency bins that were sampled with an exposure time less than half of the maximum exposure. In practice this affects only the bins at the edges of the frequency range. We computed the total flux for each LOS within a circular area of $33\arcsec$ radius, in order to increase the S/N. The observed frequency of the \CII\ emission line, ~1.9 THz, is highly affected by deep telluric absorption lines. We corrected the fluxes for the telluric transmission using the atmospheric models created by ATRAN \citep{Lord_atran}. Different options exist to parameterize this correction and we followed the most conservative approach, which corrects the fluxes with the transmission value at the reference frequency. Other options resulted in fluxes larger by a factor of $\sim 2$. We fitted a third order polynomial to the continuum baseline and subtracted it from the observed fluxes. The continuum-subtracted fluxes are shown in Fig.\ref{fig:sofia_data}. The color of each spectrum corresponds to the LOS shown with a star of the same color in Fig.~\ref{fig:hi_polarization}. The light magenta curve in each panel corresponds to the normalized atmospheric transmission model, while the dark magenta to the atmospheric model convolved with the FIFI-LS spectral resolution. The secondary vertical axis at the right edge of each subplot shows the scale of the atmospheric transmission. The blue shaded regions show the root mean square (rms), $\sigma_{rms}$, of the continuum. We fitted Gaussians to the observed spectra, shown with the black lines. We integrated the Gaussians and divided these values by the beam area (in steradians). The output integral was converted to erg/s/cm$^{2}$/sr and then divided by $6.99 \times 10^{-6}$ \citep{goldsmith_2012} in order to convert to K km/s. We computed the error on the integrated flux ($\sigma {\rm{I_{C_{II}}}}$) as follows,
\begin{equation}
    \label{eq:c+_noise}
    \sigma {\rm{I_{C_{II}}}} = \sqrt{N} \sigma_{rms} \, \delta \nu,
\end{equation}
where $\delta \nu$ is the frequency bin size and $N$ the number of bins within the integrated interval. The integrated \CII\ intensities can be found in Table~\ref{table:carbon_intensities}.

We have extracted \CII\ ancillary data from \cite{inglalls_2002} towards the four LOSs shown with the orange, black, yellow and magenta stars in Fig.~\ref{fig:hi_polarization}. These data have been obtained with the LWS spectrometer of the Infrared Space Observatory (ISO) and processed by an automatic pipeline. The LWS beam size is $71 \arcsec$, which corresponds to a beam solid angle $\Omega = 9.3\times10^{-8}$sr. For this data fitting a first order polynomial was sufficient for the background subtraction. In Fig.~\ref{fig:iso_carbon} we show with the colored histograms the baseline subtracted spectra of the four LOSs. The color of each spectrum corresponds to the LOS shown with a star of the same color in Fig.~\ref{fig:hi_polarization}. The blue shaded region corresponds to the continuum rms. We computed the error in the integrated flux of each Gaussian line, using Eq.~(\ref{eq:c+_noise}). In Table~\ref{table:carbon_intensities} we show the integrated fluxes with their corresponding errorbars and the LOS coordinates.

    \begin{figure}
        \centering
        \includegraphics[width=\hsize]{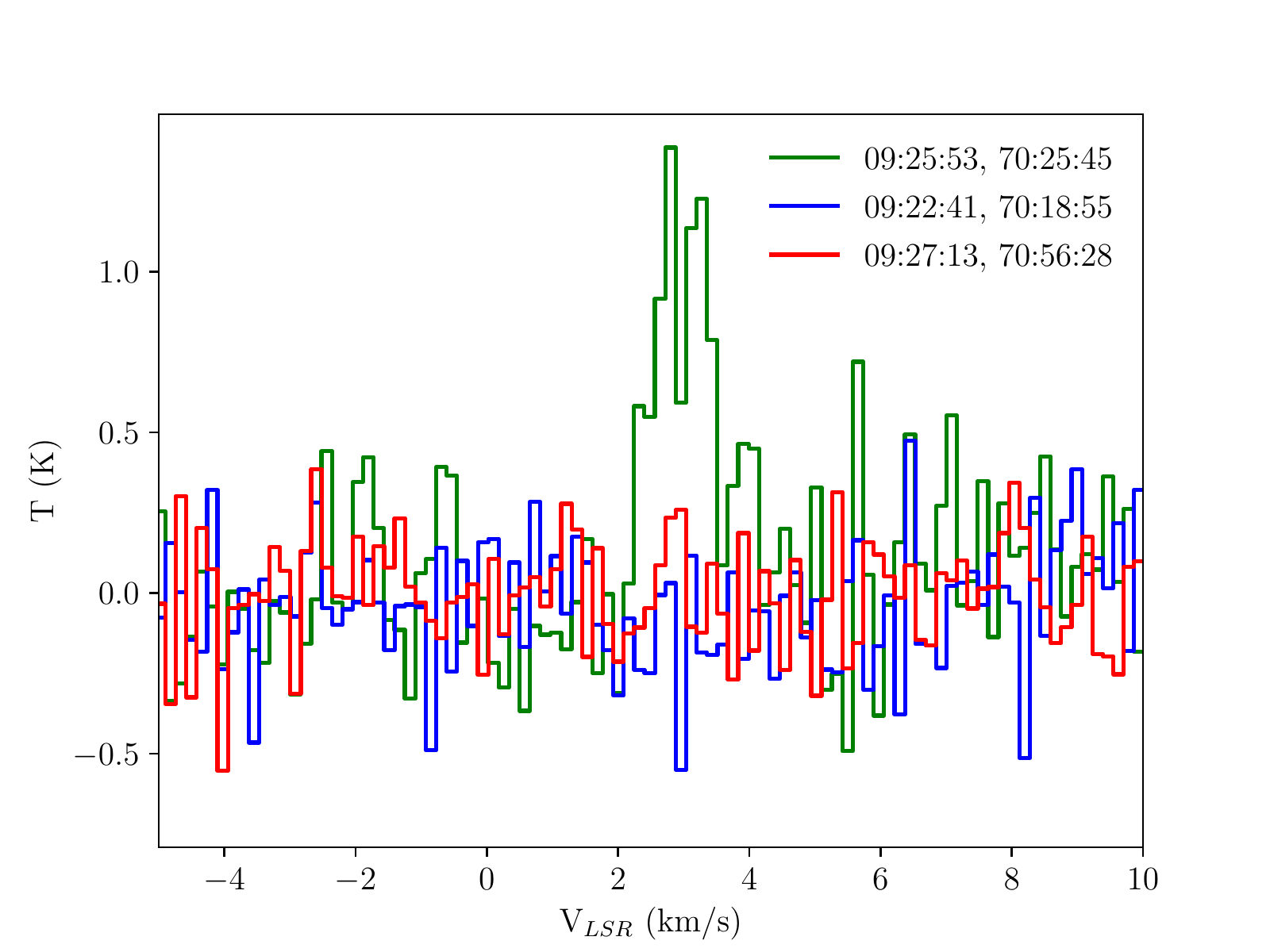}
        \caption{CO (J=1-0) spectra towards the three lines of sight overlapping with our SOFIA data. The horizontal axis is the velocity in the local standard of rest. The labels show the spectrum coordinates. The color of each spectrum corresponds to the LOS shown with colored stars in Fig.~\ref{fig:hi_polarization}.}
        \label{fig:co_sofia}
    \end{figure}
    
\subsection{CO(J=1-0) emission line}
\label{subsec:co_purple_mountain}

We carried out CO (1--0) observations towards the target cloud with the Purple Mountain Observatory 13.7 m telescope (PMO-13.7 m) from June 3 to 15, 2019 (project code: 19C002). The 3$\times$3 beam sideband separation Superconducting Spectroscopic Array Receiver \citep{2012ITTST...2..593S} was used as front-end, while a set of 18 Fast Fourier Transform Spectrometers (FFTSs) were used as backend to record signals from both sidebands. Each of the 18 FFTS modules provides 16384 channels, covering an instantaneous bandwidth of 1 GHz. This results in a channel spacing of 61 kHz, that is, 0.16~\kms\,at 115 GHz. The region was mapped in the on-the-fly mode (OTF) \citep{2018AcASn..59....3S} at a scanning rate of 50\arcsec\,per second and a dump time of 0.3 seconds. These observations encompass a total of $\sim$30 observing hours.

The standard chopper-wheel method was used to calibrate the antenna temperature \citep{1976ApJS...30..247U}. The antenna temperature, $T_{\rm A}^{*}$, is converted into the main beam brightness temperature scale, $T_{\rm mb}$, with the relation, $T_{\rm mb}= T_{\rm A}^{*}/\eta_{\rm eff}$, where $\eta_{\rm eff}$ is the main beam efficiency. The main beam efficiency is 52\% at 115 GHz according to the telescope's status report\footnote{\url{http://www.radioast.nsdc.cn/ztbg/ztbg2015-2016engV2.pdf}}. The flux calibration uncertainty is estimated to be roughly 10\%. Typical system temperatures were 332 - 441 K on a $T_{\rm A}^{*}$ scale during our observations. The half-power beam width (HPBW) is about 48\arcsec for CO (1--0). The pointing was found to be accurate to within 5\arcsec. Data reduction was performed with the GILDAS/CLASS\footnote{\url{https://www.iram.fr/IRAMFR/GILDAS/}} software \citep{2005sf2a.conf..721P}. A typical \CO\ spectrum is shown in Fig.~\ref{fig:spectra} with the blue histogram. 

In Fig.~\ref{fig:hi_polarization} we show with green contours the integrated CO (J=1-0) flux intensity and the total surveyed area with the large orange polygon. A CO clump is prominent in the right edge of the surveyed area indicating the existence of molecular gas. In addition, we targeted the three LOSs shown with the green, blue and red stars in Fig.~\ref{fig:hi_polarization} in order to overlap with the SOFIA \CII\ data. In Fig.~\ref{fig:co_sofia} we show the spectra of these three LOSs. The color of each spectrum matches the LOS shown with a star of the same color in Fig.~\ref{fig:hi_polarization}. CO was detected only towards the green-star LOS.

\subsection{CO(J=2-1) emission line}
\label{subsec:co_data_smt}

We used the Heinrich Hertz Submillimeter Telescope (SMT)\footnote{The telescope is operated by the Arizona Radio Observatories, a division of Steward Observatory at the University of Arizona.} on Mt. Graham, Arizona, to measure the J=2-1 transition of CO towards the target cloud. The data were collected on the dates of March 1, 3, and 4 in 2019. Following \cite{Bieging_2014}, we used the ALMA Band 6 sideband separating mixers dual polarization receiver to observe the CO (J=2-1) (230.5 GHz) line. The spectrometers were filter banks with 0.25 MHz bandwidth and 256 filters, providing a spectral resolution of 0.33 km/s. 

The observed region was centered around RA = 9h~34m~27.63s, Dec = $+70\degr~11\arcmin~12.2\arcsec$ (J2000) and was divided into six square subregions of $10\arcmin \times 10\arcmin$ size. Mapping was conducted in OTF mode, by scanning the telescope along lines of constant declination at a scanning rate of $15\arcsec$/s. Spectra were sampled every 0.1 s and later smoothed by 0.4 s, corresponding to a telescope shift of 6" per spectrum. Lines spaced by 10" in declination were observed in this way. 
The full-width-at-half maximum (FWHM) of the beam is 32" for the CO (J=2-1) observing frequency and is well-sampled with the aforementioned observing strategy. Reference spectra were observed after every other row in good weather and after every row if atmospheric conditions caused high noise levels. Each $10\arcmin \times 10\arcmin$ subregion required 2 hours of total observing time. 
Telescope pointing was checked and corrected for every 2 hours, resulting in pointing errors less than $5\arcsec$ \citep[see][]{Bieging_2014}. The intensity was acquired on the $T^*_A$ scale \citep{kutner_1981} and scaled to main beam brightness temperature by dividing with the measured beam efficiency of 0.776. Measurements of the efficiency were conducted in 2018 by J. Bieging (priv. comm.) following the procedure of \cite{Bieging_2010} and \cite{Bieging_2011}.

We used standard routines within the CLASS software to remove a linear baseline from each spectrum and map the irregularly sampled data onto a regular grid. During gridding the data are convolved with a Gaussian kernel of FWHM 1/3 of the beam size and are resampled on a regular grid in RA and DEC with $16\arcsec$ grid spacing. The data from all $10\arcmin \times 10\arcmin$ subregions were combined in one map in fits format within CLASS. The median rms noise of the map is 0.36 K. In Fig.~\ref{fig:hi_polarization} the small solid polygon corresponds to the CO (J=2-1) surveyed region and the black contours to the CO (J=2-1) integrated intensity.

\section{Connection of the \HI-\Hmolecular\ transition to the magnetic field, gas structure and kinematics}
\label{sec:analysis}

\subsection{Characterization of the gas phase properties}
\label{sec:gas_properties}

    \begin{figure}
        \centering
        \includegraphics[width=\hsize]{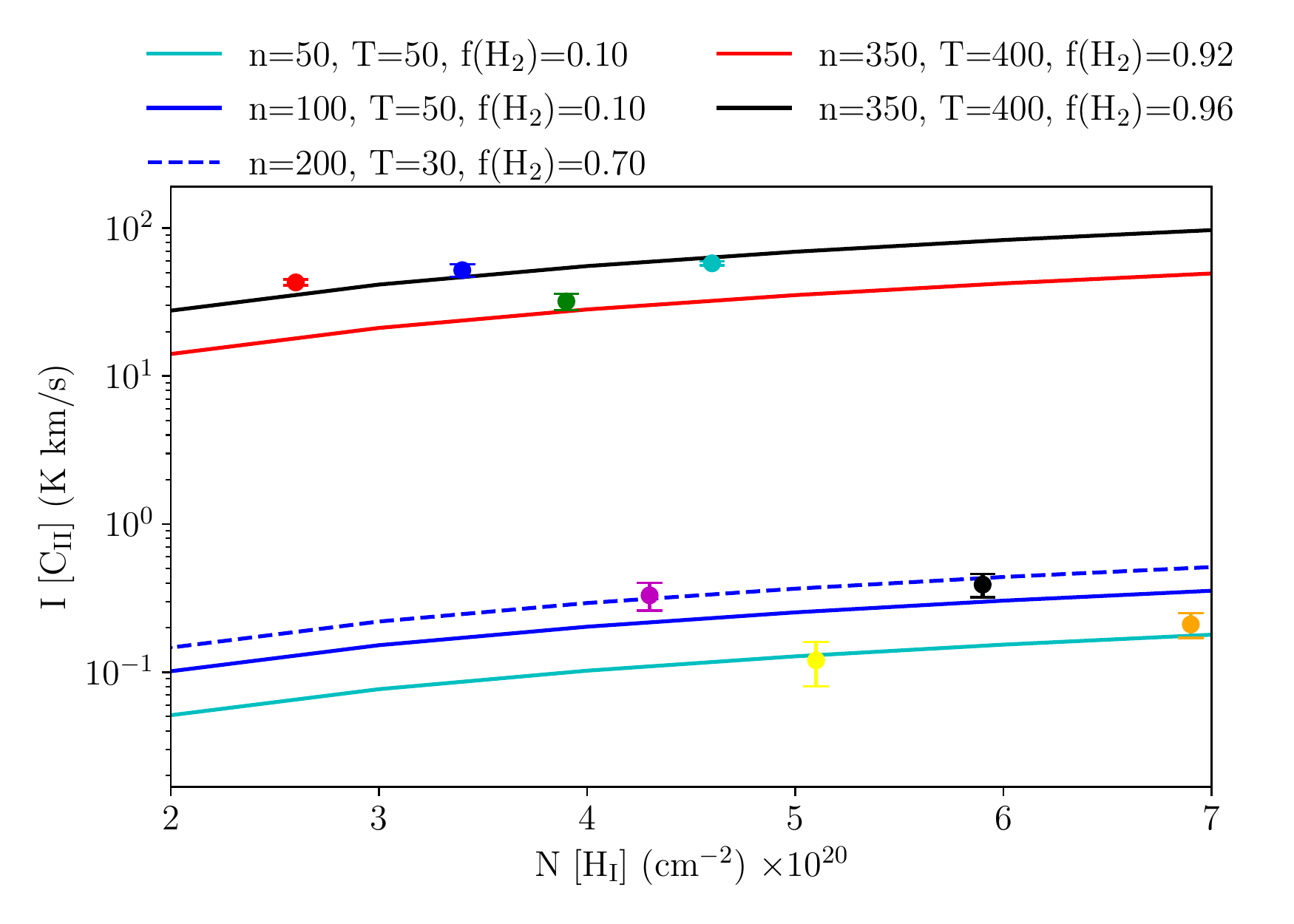}
        \caption{\CII\ integrated intensity versus \HI\ column density. Colored lines show the  expected \CII\ intensity assuming that both \HI\ and \Hmolecular\ contribute to the ionization of carbon, Eq.~(\ref{eq:carbon_model_htotal}), with varying temperature, gas density, and molecular fractional abundance. Colored points correspond to the colored-star LOSs shown with the same color in Fig.~\ref{fig:hi_polarization}.}
        \label{fig:carbon_models}
    \end{figure}
    
     \begin{figure}
        \centering
        \includegraphics[width=\hsize]{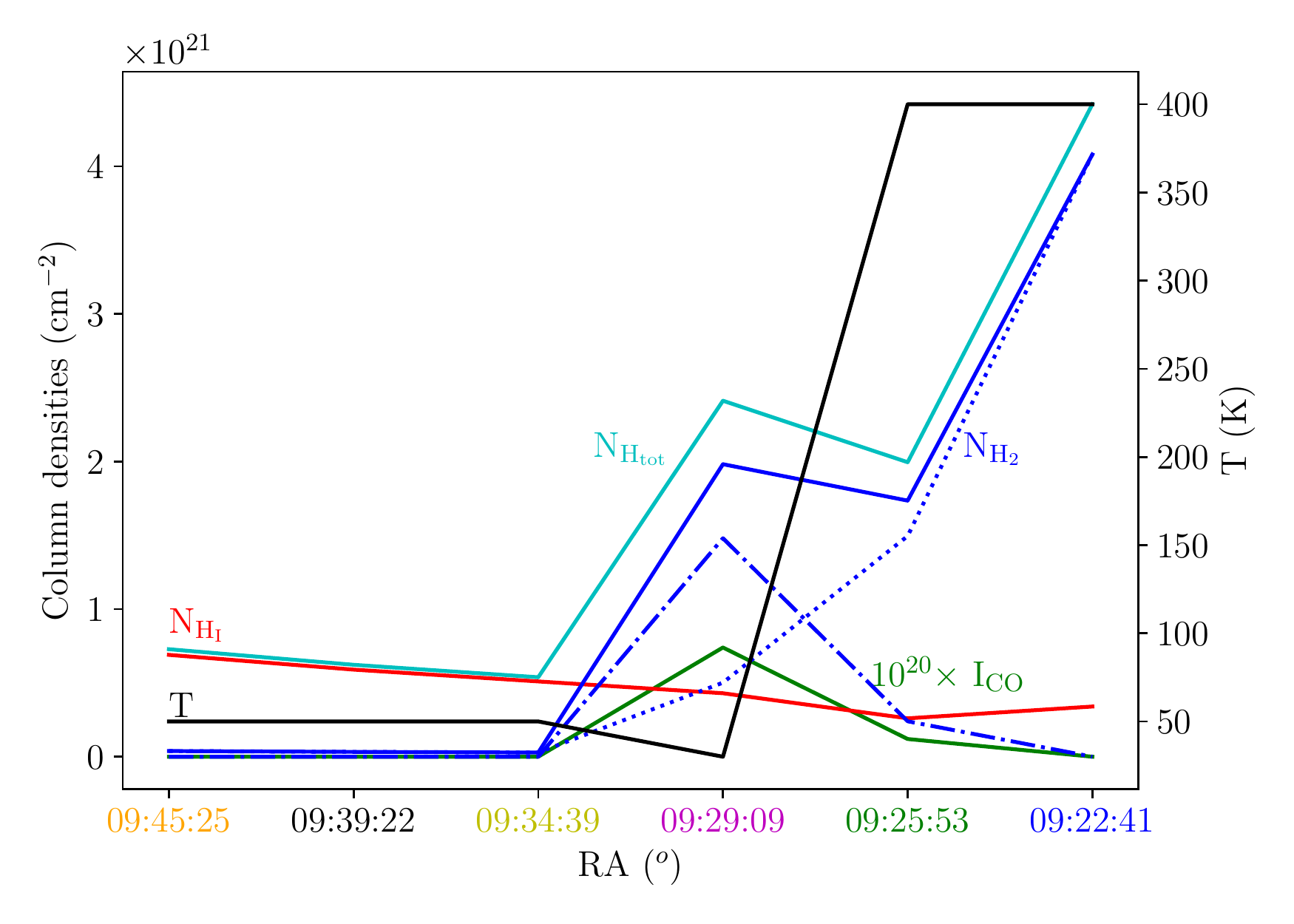}
        \caption{Column densities of different species across the main ridge of the cloud. The color of the horizontal tick labels matches with the colored-star LOSs shown in Fig.~\ref{fig:hi_polarization}. Red line corresponds to \NHI, blue solid line to \NHmol\ and cyan to the total hydrogen column density \NHtot. The blue solid line corresponds to the total column density of \Hmolecular, the blue dotted to CO-dark \Hmolecular, and the blue dash-dotted to CO-bright \Hmolecular. The green solid line shows $\rm{I_{CO}}$ multiplied by $10^{4}$ for visualization purposes. The black solid curve shows the gas temperature T(K).}
        \label{fig:pdr_model}
    \end{figure}
    
The excitation of the \CII\ emission line is due to collisions between C$^{+}$ and the following collisional partners: hydrogen (atomic and/or molecular) and electrons. The \CII\ emission originates from different layers within a cloud and enables the characterization of its average gas phase properties; emission may originate from a diffuse $\rm{H_{I}/C^{+}}$ layer or/and from a cold and dense $\rm{H_{2}/C^{+}}$ layer surrounded by the $\rm{H_{I}/C^{+}}$ envelope \citep{velusamy_2010}. Our observations do not coincide with star forming regions, and thus we do not expect significant contribution, to the observed \CII\ emission, from dense ionized gas. Ionized gas can still be present, but it is expected to contribute negligibly in this case, $\sim 4\%$ \citep{pineda_2013}. Assuming optical thin conditions, the total \CII\ line intensity can be written as \citep{goldsmith_2018},
\begin{equation}
    \label{eq:carbon_model_htotal}
    \rm{I}_{[C_{\rm{II}}]}  \text{ (K km/s)} = N({\rm C}^+) \frac{6.86 \times 10^{-6} n \times e^{-91.21/T}}{2.4 \times 10^{-6} R_{ul:mix}^{-1} + 4n \times e^{-91.21/T}},
\end{equation}
where $N({\rm C}^+)$ is the ionized carbon column density and $n$ the total hydrogen number density; $R_{ul;mix}$ is the de-excitation coefficient rate of gas mixed with \HI\ and \Hmolecular\ and it is equal to $R_{ul:mix} = f(H_{\rm{I}})R_{ul;H_{\rm{I}}} + f(H_{\rm{2}})R_{ul:H_{\rm{2}}}$, where $f(H_{\rm{I}})$ and $f(H_{\rm{2}})$ are the atomic and molecular hydrogen fractional abundances respectively. The molecular fractional abundance is defined as,
\begin{equation}
    \label{eq:molecular_fractional_abundance}
    f(H_{2}) = \frac{2N_{H_{2}}}{N_{H_{\rm{I}}} + 2N_{H_{2}}}.
\end{equation}
The atomic fractional abundance is $f(H_{I})=1-f(H_{2})$. The de-excitation rate of \HI\ is $R_{ul;H_{\rm{I}}}=7.6\times 0^{-10}(T/100)^{0.14}$ \citep{barinovs_2005, goldsmith_2012} and of \Hmolecular\ is $[4.9 + 0.22(T/100)]\times10^{-10} n(H_{\rm{2}})$\ \rate, valid when $20 \leq T \leq 400$ (K) with ortho-to-para ratio = 1 \citep{wiesenfeld_2014}.

Following \cite{langer_2010}, we computed the expected \ICplus\ as a function of $\rm{N_{H_{I}}}$ (Eq.~\ref{eq:carbon_model_htotal}) for different densities ($n$), and gas temperatures ($\rm T$) and compared it against the observed values, shown in Table~\ref{table:carbon_intensities}. Even though the physical quantities may vary along the LOS,  Eq.~(\ref{eq:carbon_model_htotal}) allows us to consider the mean gas properties ($n$, $f(H_{2})$, $\rm T$) averaged along a LOS. We assumed a \CII/H ratio equal to $\rm{X_{C_{II}}} = 1.4 \times 10^{-4}$  \citep{sofia_2001} and computed the expected $N({\rm C}^+)$ from the $\rm{H_{I}/C^{+}}$ layer as, $\rm{N}_{C_{\rm{II}}} = \rm{X_{C_{II}}} \rm{N_{H_{I}}}$. We also assumed different $f(H_{2})$ values in order to estimate $\rm{N_{H_{2}}}$, and then we computed the corresponding $\rm{N}_{C_{\rm{II}}}$ of the $\rm{C^{+}/H_{2}}$ layer as $\rm{N}_{C_{\rm{II}}} = 2 \rm{X_{C_{II}}} \rm{N_{H_{2}}}$. Overall, the expected \ICplus\ versus \NHI\ are determined by the following free parameters: $n$, $\rm T$ and $f(H_{2})$. 

The colored curves in Fig.~\ref{fig:carbon_models} show the estimated \ICplus - \NHatom\ relation for different $n$, $\rm T$ and $f(H_{2})$, as indicated in the legend. Colored points correspond to the observed \ICplus - \NHatom\ values; the color of each point corresponds to the LOS shown with a star of the same color in Fig.~\ref{fig:hi_polarization}. From these points, we detected CO only towards the green and magenta points; this indicates the existence of molecular gas there. We do not have  CO  observation coverage towards the orange and cyan points. 

The orange, black and yellow points have a relatively low \ICplus. There is no CO at the black and yellow points, which means that gas is either atomic or there is significant amount of \Hmolecular\ but with no corresponding CO emission (CO-dark \Hmolecular). We can use dust extinction, $\rm A_V$, to trace the total column density towards these LOSs. The total column density is related to the molecular fractional abundance as measured by UV spectroscopy towards background AGN \citep{Gillmon2006,shull_2021}. For these LOSs, we obtained $\rm{A_{V}}$ from \cite{planck_2016_av}, and then converted to reddening, E(B-V), via $\rm{R_{V} = A_{V}/ E(B-V)}$, where $\rm{R_{V}} = 3.1$. For the orange, black, and yellow stars we obtained E(B-V) = 0.25, 0.30, and 0.20 mag, respectively. We then used the relation $\rm N_H/ E(B-V) = 6.07 \times 10^{21} cm^{-2}/mag $ \citep{shull_2021} to obtain total column densities for the aforementioned sightlines. 
The total column density is $\rm N_H \sim 1.2 - 1.8 \times 10^{21} cm^{-2}$. Sightlines with this range of column densities have at most $f(H_2) = 0.4 $ \citep[Fig. 6 of][]{Gillmon2006}. Therefore, in the following we use 0.4 as an upper limit on the fractional molecular abundance for these sightlines. 

The relatively low estimated $f(H_{2})$ values for the aforementioned sightlines indicate that \HI\ gas is more abundant than \Hmolecular, there. This means that in the \CII\ emission there has a maximum contribution by a $\rm{H_{I}/C^{+}}$ layer. If we assume that $T=50$ K\footnote{This is the median of the kinetic temperature distribution over a large sample of diffuse clouds in our Galaxy \citep{heiles_troland1}.}, then the yellow and orange points are well represented by the cyan line, which corresponds to $T=50$ K, $n=50$ \VolDens\ and $f(H_{2})=0.1$. For the black point, however, a larger $n$ is necessary in order to match with the observed \ICplus. This point is consistent with $T=50$ K, $n=100$ \VolDens\ and $f(H_{2})=0.1$. To sum up, gas seems to be mostly atomic towards these LOSs where $T=50$ K, $f(H_{2})=0.1$ and $n=50 - 100$ \VolDens.

We detected significant \CO\ both J=2-1 and J=1-0 towards the magenta point. This indicates that there is sufficient amount of molecular gas there; CO emission originates from \Hmolecular-CO cores formed within $\rm{H_{2}/C^{+}}$ envelopes \citep{langer_2010}. Typical \Hmolecular\ temperatures in the Milky Way are $0.7$ times lower than their surrounding \HI\ envelopes \citep{goldsmith_2016}. In order to be consistent with the $T=50$ K that we assumed for the atomic part of the cloud, we adopted that $T = 30$ K for the \Hmolecular\ layer there. We obtain that the observed \ICplus-\NHI\ relation at the magenta point can be well represented by $n = 200$ \VolDens, and $f(H_{2})=0.7$, shown with the blue dotted line in Fig.~\ref{fig:carbon_models}.

At the green, red, cyan and blue points \ICplus\ is extremely large; $\sim 2$ orders of magnitude larger than the other points. These large values can only be obtained if there is a hot ($T >100$ K) and dense ($n > 100$ \VolDens) $\rm{H_{2}/C^{+}}$ layer. Towards the green point we have detected significant CO (J=1-0), which means that there is \Hmolecular\ traced by both \CII\ and CO. There, the observed \ICplus-\NHI\ values can be well represented by the red solid line which corresponds to the following gas properties: $T=400$ K, $f(H_{2}) = 0.92$ and $n = 350$ \VolDens.

At the blue, cyan and red points, we did not detect any CO, despite their large \CII\ intensities. This could be related to the fact that these points are close to the edges of the cloud (Fig.~\ref{fig:hi_polarization}) and self-shielding is not sufficient to allow for the \CII/\CI/CO transition. The observed large \CII\ intensities can only originate from a hot and dense CO-dark \Hmolecular\ gas layer. This is consistent with previous works, which suggest that CO-dark gas is located at the edges of diffuse clouds \citep[e.g.][]{langer_2010, pineda_2013}. There, the observed values are consistent with the black solid line which corresponds to the following conditions: $T=400$ K, $f(H_{2}) = 0.96$ and $n = 350$ \VolDens. We provide estimates of the uncertainty in the derived quantities in Section \ref{subsubsec:gas_phase_uncertainties}.

\subsubsection{Uncertainties in the estimated gas properties}
\label{subsubsec:gas_phase_uncertainties}

    \begin{figure}
        \centering
        \includegraphics[width=\hsize]{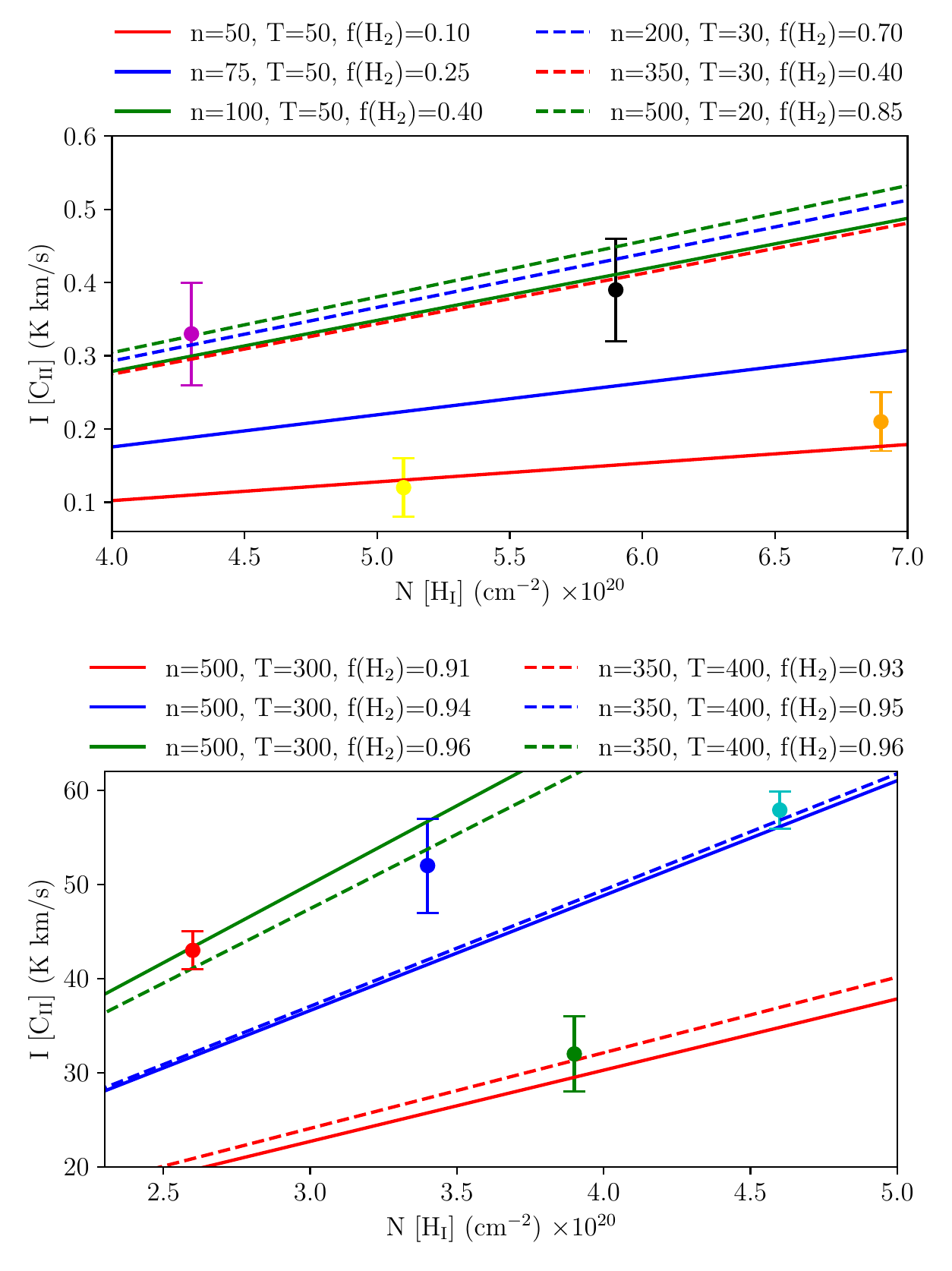}
        \caption{Same as in Fig.~\ref{fig:carbon_models}. Points correspond to the colored-star LOSs shown with the same color in Fig.~\ref{fig:hi_polarization}.}
        \label{fig:carbon_models_err}
    \end{figure}
    
We used Eq.~(\ref{eq:carbon_model_htotal})  to constrain the gas properties ($n$, $f(H_{2})$, $\rm T$) based on our \CII\ and \NHI\ data. However, it is essential to add uncertainties on our estimated values since we use them in the estimation of the magnetic field strength (Sect.~\ref{sec:bpos_strength_computation}). The large \CII\ intensities of the green, red, blue, and cyan points can be only represented with high $n$, $\rm T$, and $f(H_{2})$ values. In general the large \CII\ intensities can be fitted only if $\rm T > 200$K, otherwise the exponential in Eq.~(\ref{eq:carbon_model_htotal}) significantly reduces the expected \CII\ intensity. If we assume $\rm T = 300$ K, which is typical of CO-dark \Hmolecular\ dominated regions \citep{langer_2010}, then we find that $f(H_{2})\geq 0.9$. In the bottom panel of Fig.~\ref{fig:carbon_models_err}, the colored solid lines correspond to models with $\rm T = 300$ K, while the dashed lines to models with $\rm T = 400$ K. Overall, for these points we find that $n = 350 - 500$ \VolDens, and $f(H_{2}) = 0.91 - 0.96$. Temperatures larger than 400 K are not typical of cold neutral medium clouds, and hence were not considered in our analysis; also the molecular deexcitation rate used in Eq.~(\ref{eq:carbon_model_htotal}) has not been constrained for $\rm T > 400$ K \citep{wiesenfeld_2014}.

In the upper panel of Fig.~(\ref{fig:carbon_models_err}), we show the four points with low \ICplus\ intensities. The orange, black, and yellow points are dominated by \HI\ gas with $f(H_{2}) \leq 0.4$ (Sect.~\ref{sec:gas_properties}). If we take this into account the low \CII\ intensities can be only fitted with the red and green solid lines, which correspond to $n = 50 - 100$ and $\rm T \approx 50$ K. The \CII\ intensity of the magenta point is consistent with the green solid line. However, the significant amount of CO that we detected there indicates that \Hmolecular\ should be more abundant. For the magenta point we found that E(B-V)$\approx 0.55$ mag which is $\sim 2$ times greater than the extinction at the black point, hence $f(H_{2})$ should be larger at the magenta point. According to \cite{shull_2021}, such extinctions are characterized by $f(H_{2}) \geq 0.4$. By setting $f(H_{2})=0.4$, and considering that temperature should be lower there ($\rm T < 50$ K, due to the presence of CO, we find that $n\approx350$ \VolDens\ (red dashed line). Given the aforementioned constraints, we found that $ n \approx200$ \VolDens\ should be the minimum density towards that point (blue dashed line), and $ n \approx500$ \VolDens\ the maximum density with $\rm T \approx20$ K (green dashed line). We did not consider temperatures lower than 20 K because the deexcitation rate of \Hmolecular\ obtained by \cite{wiesenfeld_2014}, and used in Eq.~(\ref{eq:carbon_model_htotal}), is inaccurate at such low temperatures. To summarize for the orange, black, and yellow points we find that $n = 50-100$ \VolDens\ and $f(H_{2}) \leq 0.4$, while for the magenta point $n = 200-500$ and $f(H_{2}) \approx 0.4 - 0.85$ \VolDens.

\subsubsection{Overview of the gas phases}
\label{subsubsec:overview_gas_phase}

Here, we summarize the gas phase properties that we inferred for the target cloud from the analysis above. Referring to Fig.~\ref{fig:hi_polarization}, as we move across the main ridge of the cloud from the orange-star (RA = 09:48:00) to the magenta-star LOS (RA = 09:30:00), gas transitions from \HI\ to \Hmolecular; the transition seems to happen between the yellow-star and the magenta-star LOSs where the CO clump is. 

In Fig.~\ref{fig:pdr_model}, we show the adopted \HI\ and \Hmolecular\ column densities at different positions across the long axis of the target cloud. With the red solid line we show \NHatom\ versus RA. The color of the horizontal axis labels corresponds to the LOS marked with a colored-star LOS of the same color in Fig.~\ref{fig:hi_polarization}. With the black solid line, we show the assumed gas kinetic temperature, while with the green solid line the observed integrated CO (J=1-0) intensity ($\rm{I_{CO}}$) in K \kms, multiplied by $10^{20}$ for visualization purposes. The \Hmolecular\ column density is indirectly traced by \CII, and CO and is given by the following equation,
\begin{equation}
    \label{eq:NH2}
        \rm{N_{H_{2}}} = \frac{ \textit{f}(H_{2})}{2[1 - \textit{f}(H_{2})]} \rm{N_{H_{I}}} + X_{CO} I_{CO}.
\end{equation}
The first term corresponds to CO-dark \Hmolecular\ and is inferred by the $\textit{f}(H_{2})$ values that we obtained from the \CII\ observations, while the second term corresponds to CO-bright \Hmolecular. $\rm{X_{CO}}$ is the conversion factor between $\rm{H_{2}}$ and CO, and has a typical value in our Galaxy equal to $\rm{X_{CO}} = 2\times 10^{20}$ cm$^{-2}$ K \kms\ \citep{bolato_2013}, although variations can be significant in individual cases \citep[e.g.,][]{barriault_2011_co}. In the same figure, we also show the column density of CO-dark \Hmolecular\ with the blue dotted line, and of CO-bright \Hmolecular\ with the blue dash-dotted line.

There are three major uncertainties introduced in our estimated \NHmol\ values (Eq.~\ref{eq:NH2}). Firstly, the derived $f(\rm{H_{2}})$ values are degenerate with $n$ and $\rm T$. Secondly, $\rm{X_{CO}}$ can vary significantly throughout our Galaxy \citep[e.g.,][]{barriault_2011_co}, hence the mean $\rm{X_{CO}}$ that we adopted may not be representative for the target cloud. Thirdly, there is \Hmolecular\ gas traced only by \CI\ \citep{pineda_2017}; the contribution of this layer was neglected in the estimated column densities since we do not have \CI\ observations for this cloud. We cannot overcome these problems with the existing dataset, and for this reason we note that the values shown in Fig.~\ref{fig:pdr_model} offer a qualitative characterization of the gas phase properties of this cloud.

\cite{kalberla_2020} created a full-sky column density map of the CO-dark \Hmolecular\ gas. They estimated the total column densities with data from the HI4PI survey \citep{hi4pi} and the extinction map of \cite{Schlegel_1998}. Our adopted picture shown in Fig.~\ref{fig:pdr_model} is consistent with their map; the CO-dark \Hmolecular\ column densities increase towards the green-star and blue-star LOSs. Since the data from this map are similar to our properties for the \NHmol, it gives us confidence that Fig.~\ref{fig:pdr_model} accurately represents the atomic-to-molecular gas transition of this cloud. However, a direct comparison between our inferred values and their map may not be meaningful due to following main reasons: 1) There is a significant difference in the angular resolution of the \cite{kalberla_2020} map ($11\arcmin$) and our data ($0.55 \arcmin$ and $1.18\arcmin$ for the SOFIA and the ISO data respectively), 2) Their map is based on the assumption that the E(B-V)/$\rm{N_{H}}$ ratio is constant throughout our Galaxy, and 3) $f(\rm{H_{2}})$, which is inserted in our \NHmol\ computation, is degenerate with $n$ and T.

\subsection{Magnetic field strength}
\label{sec:bpos_strength_computation}

\begin{table*}
\caption{Magnetic field strength and Alfvénic Mach number estimation.}             
\label{table:bfield_parameters}      
\centering                          
\begin{tabular}{c c c c c c c c c c c}        
\hline\hline                 
Region & $\delta \chi_{int}$ & $\sigma_{obs, i}$\tablefootmark{a} & $ \langle n \rangle $\tablefootmark{b}  &  $\rm{B}_{\rm{POS}}$ \tablefootmark{c} & $\rm{B}_{\rm{LOS}}$ \tablefootmark{c, d} & $\rm{B}_{\rm{tot}}$\tablefootmark{c} & $M_{A}$\\    
\hline                        
\\
Atomic  & 16.8 $\degr$ & 4.26$^{*}$ & 75$^{+25}_{-25}$  & 25$^{+4}_{-4}$  & $10^{+3}_{-3}$ & $27^{+5}_{-5}$ & 1.2 \vspace{0.2cm} 
\\ 
\multirow{2}{*}{Molecular} & \multirow{2}{*}{14.4$\degr$} & 3.00$^{*}$ & \multirow{2}{*}{290$^{+210}_{-90}$} & \multirow{2}{*}{$19^{+12}_{-6}$}  &  
\multirow{2}{*}{$8^{+4}_{-2}$} & \multirow{2}{*}{$21^{+12}_{-6}$} & \multirow{2}{*}{1.1} \\ 
& & 1.00$^{**}$ & 
\\
\hline 
\end{tabular}
\tablefoot{\tablefoottext{a}{Measured in \kms.}
           \tablefoottext{b}{Measured in \VolDens.}
           \tablefoottext{c}{Measured in $\mu$G.}
           \tablefoottext{d}{Data from \cite{myers_1995}.}\\
           In the $\sigma_{obs, i}$ column, a star ($^{*}$) is used for \HI\ velocities while double star ($^{**}$) for CO (J=1-0) velocities.}
\end{table*}

We employed the method of \cite{skalidis_2020} (ST) to estimate the strength of the plane-of-the-sky (POS) magnetic field component, \Bpos. In contrast to the widely applied method of \cite{davis_1951} and \cite{chandra_fermi} (DCF) which assumes that the dispersion of polarization angles ($\delta \chi$) is induced by incompressible waves, this method assumes that compressible fluctuations are the dominant; ST was found to give better magnetic field strength estimates than DCF when tested against MHD numerical simulations which do not include self-gravity, even for super-Alfvénic models with $M_{A} = 2.0$ \citep{skalidis_2021}. The POS magnetic field strength can be estimated as,
\begin{equation}
    \label{eq:st}
    \rm{B_{POS}} = \sqrt{2 \pi \rho} \frac{\sigma_{v, turb}}{\sqrt{\delta \chi}},
\end{equation}
where $\rho$ is the gas volume density and $\rm \sigma_{v, turb}$ the gas turbulent velocity. This equation is applicable when $\delta \chi \ll 1$ rads.

\subsubsection{Polarization angle dispersion estimation}

    \begin{figure}[h]
        \centering
        \includegraphics[width=\hsize]{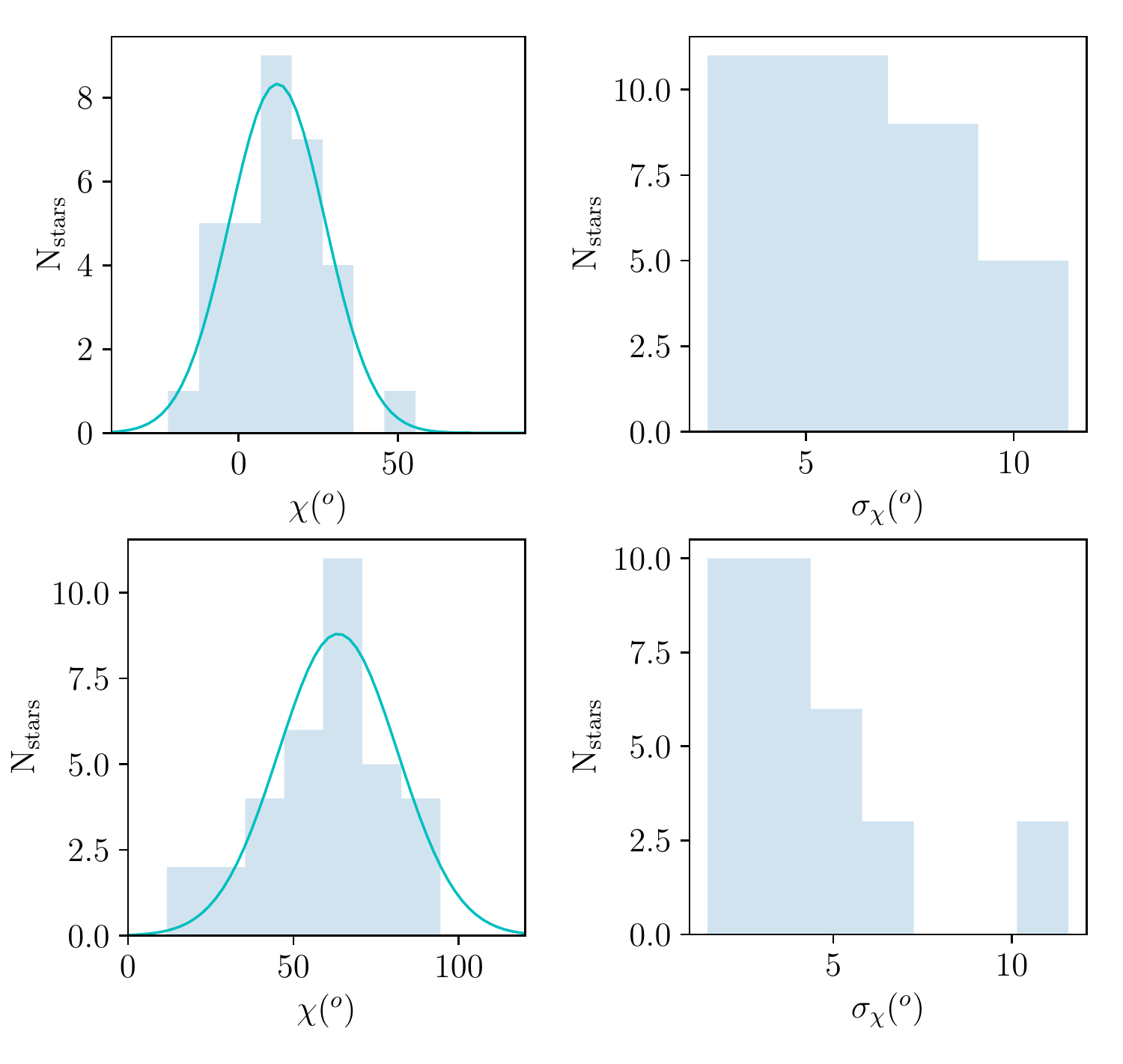}
        \caption{Upper panels: The left panel shows the dispersion of polarization angles for the atomic region. The cyan solid profile corresponds to the fitted Gaussian. The mean ($\mu$), standard deviation ($\sigma$), and amplitude ($\alpha$) of the fit are ($\mu$, $\sigma$, $\alpha$) = (63$\degr$, 18$\degr$, 9). The right panel shows the polarization angle error distribution. Bottom: Same as for the upper panels, but for the molecular region. The free parameters of the fitted Gaussian are ($\mu$, $\sigma$, $\alpha$) = (12$\degr$, 15$\degr$, 8).}
        \label{fig:EVPA_distrib}
    \end{figure}

It is evident from Fig.~\ref{fig:hi_polarization} that the mean magnetic field orientation is not uniform within the cloud, which complicates the computation of $\delta \chi$. At the orange-star LOS, the magnetic field follows the \NHI\ structure of the cloud, while between the black- and yellow-star LOSs a transition happens; the mean field orientation tends to be perpendicular to the \NHI\ structure of the cloud; this is discussed in more detail in Sect.~\ref{subsec:rht_polarization}. This change in the mean field orientation induces extra spread in the distribution of polarization angles, which is not related to MHD waves. 

In order to accurately constrain the MHD wave-induced dispersion (intrinsic dispersion, $\delta \chi_{intr}$), we removed measurements located between the black-star and yellow-star LOS in Fig.~\ref{fig:hi_polarization}; this is the transition zone where the mean magnetic field orientation changes. Then, we classified two distinct sub-regions within the cloud. The first region corresponds to measurements located between RA=09:53:00 and RA=09:39:22 (black-star LOS in Fig.~\ref{fig:hi_polarization}). The mean magnetic field orientation of these measurements is parallel to the \NHI\ structure of the cloud, and gas is mostly atomic there. Hereafter, we refer to this region as "atomic". The second classified region includes all measurements between the yellow-star and blue-star LOSs (Fig.~\ref{fig:hi_polarization}), from RA=09:34:39 to RA=09:22:41. The mean magnetic field orientation of these measurements tends to be perpendicular to the \NHI\ structure of the cloud, and gas is mostly molecular there. We will refer to this region as "molecular".

In both regions, we applied three statistical criteria in order to minimize the various sources of uncertainty that bias $\delta \chi$ towards larger values. Firstly, we considered only measurements with S/N $\geq 2.5$. Secondly, we considered only polarization measurements of stars located behind the target cloud, but not very far away to be affected by the IVC cloud; the target cloud is located at 300pc (Appendix~\ref{sec:cloud_distance}), while the IVC cloud at 1kpc \citep{tritsis_2019}. In this step, we included stars with distances $300 \leq d \leq 1000$ pc. This guarantees that the polarization measurements are not affected by the IVC cloud, which despite its weaker \HI\ emission (Fig.~\ref{fig:spectra}) may significantly contribute in the total polarization, as shown by \cite{panopoulou_2019_tom}. Thirdly, we discarded stars with polarization angles offset by more than $60\degr$ from the mean orientation. This ensures that our analysis is not affected by potentially intrinsically polarized stars. These measurements do not probe the morphology of the ISM magnetic field and they are outliers in the polarization angle or degree of polarization distributions. In total, we found only two outliers in the polarization angle distribution of the atomic region. The measurements used for the \Bpos\ estimation in the atomic region are shown as blue segments in Fig.~\ref{fig:hi_polarization}, while measurements included \Bpos\ estimation of the molecular region are shown in yellow. Measurements with S/N $\geq 2.5$ that did not meet the aforementioned selection criteria and were not included in the \Bpos\ estimation are shown in red.

The distribution of polarization angles ($\chi$) is shown in  Fig.~\ref{fig:EVPA_distrib}. The upper left panel corresponds to the distribution of $\chi$ for the atomic region, while the upper right panel shows their corresponding observational uncertainties ($\sigma_{\chi}$). The two lower panels show the same quantities for the molecular region. We performed an optimal binning to these distributions following Knuth's rule Bayesian approach as implemented in \texttt{astropy}. We fitted Gaussian profiles in the $\chi$ distributions as shown with the cyan solid lines. In the atomic region, the spread of the fit is equal to $18\degr$, which is very close to the raw spread ($20\degr$) of the distribution. In the molecular region, both the spread of the fit and the raw spread of the distribution are approximately equal to $15\degr$. Observational uncertainties tend to bias the spread of the $\chi$ distribution to larger values. For this reason, we corrected for the observational uncertainties following  \cite{crutcher_2004} and \cite{panopoulou_2016} as,
\begin{equation}
    \label{eq:std_intr}
    \delta\chi_{int}^{2} = \delta\chi_{obs}^{2} - \hat{\sigma_{\chi}}^{2},
\end{equation}
where $\hat{\delta\chi}$ is the mean observational uncertainty. The mean uncertainty in the atomic region is $\hat{\sigma_{\chi}}=6.3\degr$, while in the molecular region it is $\hat{\sigma_{\chi}}=4.0\degr$. For the atomic and molecular regions the derived intrinsic spreads corrected for the observational uncertainties are $\delta\chi_{int} = 16.8\degr$ and $14.4\degr$ respectively, as shown in Table~\ref{table:bfield_parameters}. 

\subsubsection{Turbulent gas velocity and density estimation}

Gas turbulent velocities, $\sigma_{u, turb}$, were computed from the \HI\ and CO emission lines respectively. \HI\ spectra probe the atomic gas layers kinematics, while CO probes the molecular layers. We computed the average spectra in both regions and fitted their profiles with Gaussians. We subtracted the thermal broadening via a quadrature subtraction,
\begin{equation}
    \label{eq:thermal_subtraction}
    \rm \sigma_{v, turb}^{2} = \sigma^{2}_{v, obs} - \frac{2kT}{m_{i}},
\end{equation}
where $\rm{\sigma_{v, obs}}$ is the standard deviation of the Gaussian fitting and $i$ refers to the mass of the emitting species, which is either \HI\ or \CO. Following our estimates in Sect.~\ref{sec:gas_properties}, the mean gas temperature in the atomic region is $T=50$ K (orange and black star in Fig.~\ref{fig:carbon_models}), while in the molecular region it is $T \approx 300$ K (magenta, green, blue and cyan stars in Fig.~\ref{fig:carbon_models}) for the CO-dark layer, while for the CO-bright layer we assumed that $T = 15$ K. Overall, our calculations are not very sensitive to temperature variations. In Table~\ref{table:bfield_parameters} we show the turbulent velocity spreads in \kms\ for both regions. 

The emission of chemical tracers originates from local regions (layers) within a cloud, hence it traces the local and not the average cloud kinematics. On the other hand, dust polarization is averaged along every gas layer in the cloud, and as a result it represents the average magnetic field fluctuations of the cloud. Thus, there is an inconsistency between the two observables (polarization and velocity broadening) since they do not trace the same regions of the cloud. In the atomic region the aforementioned problem is not so prominent, since there is a dominant \HI\ gas layer (Sect.~\ref{sec:gas_properties}). In that case the spread derived from the \HI\ emission line fitting should accurately represent the average gas kinematics. In the molecular region, however, there is both atomic and molecular, gas with the latter being the most abundant. We computed the weighted rms turbulent velocity in the molecular region as,
\begin{align}
    \label{eq:weighted_mean_sigma_turb}
    \sigma_{v, turb} = \Bigg\{ f_{\rm{HI}}~[\sigma_{turb, H_{I}}]^{2}
    + f_{\rm{H_{2}}}^{dark}~[\sigma_{turb, H_{2}}^{dark}]^{2} 
    + f_{\rm{H_{2}}}^{bright}~[\sigma_{turb, H_{2}}^{bright}]^{2} \Bigg\}^{1/2},
\end{align}
where the first term corresponds to the kinematics contribution from the \HI\ layer, the second to a CO-dark \Hmolecular\ layer, and the third term to a CO-bright \Hmolecular\ layer; $f_{\rm{HI}}$, $f_{\rm{H_{2}}}^{dark}$, and $f_{\rm{H_{2}}}^{bright}$ are the fractional abundances of the \HI, CO-dark \Hmolecular, and CO-bright \Hmolecular\ layers respectively, which are defined as,
\begin{align}
   & f_{\rm{HI}} = \frac{N_{H_{\rm{I}}}}{N_{H_{\rm{I}}} + 2N_{H_{2}}^{dark} + 2N_{H_{2}}^{bright}}, \\
   & f_{\rm{H_{2}}}^{dark} = \frac{2N_{H_{2}}^{dark}}{N_{H_{\rm{I}}} + 2N_{H_{2}}^{dark} + 2N_{H_{2}}^{bright}}, \\
   & f_{\rm{H_{2}}}^{bright} = \frac{2N_{H_{2}}^{bright}}{N_{H_{\rm{I}}} + 2N_{H_{2}}^{dark} + 2N_{H_{2}}^{bright}} . 
\end{align}
The motivation behind Eq.~(\ref{eq:weighted_mean_sigma_turb}) is that the molecular region is characterized by three different gas layers. Eq.~(\ref{eq:weighted_mean_sigma_turb}) can be verified by computing the mean spread of three Gaussians centered at the same velocity with different spreads, and amplitudes which are proportional to the relative abundance of each species.

In Fig.~\ref{fig:pdr_model} we show our inferred column densities of each layer. Based on these estimates we computed the mean abundance of each layer, which are $f_{\rm{HI}} \approx 0.07$, $f_{\rm{H_{2}}}^{dark} \approx 0.66$, and $f_{\rm{H_{2}}}^{bright} \approx 0.27$; $\sigma_{turb, H_{I}}$ is the \HI\ broadening, while $\sigma_{turb, H_{2}}^{bright}$ the CO (J=1-0) broadening. In Eq.~(\ref{eq:weighted_mean_sigma_turb}) the only parameter which is not observationally constrained is $\sigma_{turb, H_{2}}^{dark}$. For this reason, we assumed similarly to past works \citep[e.g.,][]{panopoulou_2016} that the CO broadening is an accurate proxy for the averaged \Hmolecular\ kinematics, which means that both the CO-dark and CO-bright \Hmolecular\ layers share similar kinematics properties, hence $\sigma_{turb, H_{2}}^{bright}\approx \sigma_{turb, H_{2}}^{dark}$. We derived that the mean turbulent broadening is $1.15$ \kms\ in the molecular region.

We averaged the density over all layers in the molecular region as,
\begin{equation}
    \rho =  m_{p}  \times \left (
             \mu_{H_{\rm{I}}}  f_{\rm{HI}}               n_{H_{I}}
           + \mu_{H_{2}}       f_{\rm{H_{2}}}^{dark}     n_{\rm{H_{2}}}^{dark}
           + \mu_{H_{2}}       f_{\rm{H_{2}}}^{bright}   n_{\rm{H_{2}}}^{bright} \right ), 
\end{equation}
where $m_{p}$ is the proton mass, $\mu_{H_{\rm{I}}}=1.36$ is the mean atomic mass, and $\mu_{H_{2}}=2.33$ is the mean molecular weight, $n_{H_{I}}$ the mean volume density of the \HI\ layer, $n_{\rm{H_{2}}}^{dark}$ the mean volume density of the CO-dark \Hmolecular\ layer, and  $n_{\rm{H_{2}}}^{bright}$ the mean volume density of the CO-bright \Hmolecular\ layer. From the analysis in Sect.~\ref{sec:gas_properties}, we find that the average density and molecular fractional abundance are $f(H_{2}) \approx 0.86$, and $n \approx 300$ \VolDens respectively; these two quantities, however, refer only to the \HI\ and CO-dark \Hmolecular\ layers, which are probed by the \CII\ emission line data. Using these values, we computed $n_{H_{I}}$ and $n_{\rm{H_{2}}}^{dark}$ as, $n_{H_{I}} = n \times [1-f(H_{2})] = 42$ \VolDens, and $n_{\rm{H_{2}}}^{dark} = n \times f(H_{2})  = 258$ \VolDens. 

In order to estimate $n_{\rm{H_{2}}}^{bright}$ we need to make some assumptions regarding the 3D shape of the CO clump. For dynamically important magnetic fields, like ours (Sect.~\ref{subsec:Btotal}), oblate shapes are favored; this has been verified observationally by \cite{tassis_2009}. The size of the major axis of the CO clump on the sky is $\sim 0.4\degr$ which corresponds to 2.0 pc, while its minor axis is $\sim 0.1\degr$ and corresponds to 0.5 pc. Given an oblate 3D shape, it is reasonable to assume that the depth of the \Hmolecular-CO layer is closer to the size of its major axis than its minor axis. For this reason, we assumed that the LOS dimension of the core is 1.5 pc. For the molecular region, our estimated mean CO-bright \NHmol\ column density is $8.6\times10^{20}$ \ColDens, which yields $n_{\rm{H_{2}}}^{bright} \approx 400$ \VolDens. Then the estimated weighted mean number density in the molecular region is $\langle n \rangle \approx 290$ \VolDens. In the atomic region there is negligible \Hmolecular, hence $f(H_{2}) \approx 0$, $f_{\rm{H_{2}}}^{dark} \approx 0$, and $f_{\rm{H_{2}}}^{bright} \approx 0$. Density uncertainties are propagated to the final \Bpos\ estimates as shown in Table~\ref{table:bfield_parameters}.

\subsubsection{POS and total magnetic field strength}
\label{subsec:Btotal}

There are three major uncertainties affecting the \Bpos\ estimation, mainly caused by variations in  temperature, molecular fractional abundance, and density within the two regions (atomic and molecular); with gas density being the dominant source of uncertainty. In the atomic region, the number density inferred from our \CII\ data varies from 50 to 100 \VolDens, while in the molecular from 200 to 500 \VolDens, Sect.~\ref{sec:gas_properties}. We estimated the limits on \Bpos\ based on the full range of density values found for each region. However, for the molecular region, $f(H_{2})$ is also an important source of uncertainty, since it affects the estimated turbulent velocity in Eq.~(\ref{eq:weighted_mean_sigma_turb}). Thus, for the molecular region we varied both $n$ and $f(H_{2})$ in the range that we established from our \CII\ models in Sect.~\ref{subsubsec:gas_phase_uncertainties}. The lower limit on \Bpos\ is found for $f(H_{2})=1.0$ and $n_{\rm{H_{2}}}^{dark}=n_{\rm{H_{2}}}^{bright}=200$  \VolDens\ and the upper limit is found for $f(H_{2})=0.4$, $n_{\rm{H_{2}}}^{dark}=n_{\rm{H_{2}}}^{bright}=500$  \VolDens, and $n_{H_{I}} = 100$ \VolDens. In Table~\ref{table:bfield_parameters} we show the estimated \Bpos\ with corresponding limits for both regions; the POS magnetic field strength is consistent within uncertainties between the two regions.

For the current cloud, there are archival \HI\ Zeeman data \citep{myers_1995} tracing the LOS magnetic field strength (\Blos). We combined the \Blos\ measurements from this dataset with our \Bpos\ estimated values and derived the total magnetic field strength of the cloud, $\rm{B_{tot}}$. We computed the mean \Blos\ of both the atomic and molecular regions; the minimum and maximum \Blos\ values of each region were used as lower and upper limits respectively. For both regions, it is true that \Blos\ $<$ \Bpos, which indicates that the magnetic field of the cloud is mostly in the POS. We computed the $34\%$ and $83\%$ percentiles of their \Blos\ distribution in every region and used these upper (and lower) limits with the corresponding limits from our \Bpos\ estimates to derive the upper (and lower) limits of  $\rm{B_{tot}}$. We find that the $\rm{B_{tot}}$ estimated values between the two regions are consistent within our uncertainties. We note, however, that in the molecular region \Blos\ may not be representative of the total LOS magnetic field component. The existing Zeeman data trace only the \HI\ layer and not the bulk of the gas, which is mostly molecular there (Sect.~\ref{sec:gas_properties}). Thus, in the molecular region $B_{\rm{tot}}$ should be treated as a lower limit of the total magnetic field strength.

\subsubsection{Estimating the Alfvén Mach number}

In addition, we estimated the Alfvénic Mach number \Ma\ of the cloud. According to ST the projected Alfvén Mach number is $M_{A}^{1\rm{D}} \approx \sqrt{2\delta \chi}$. We find that $M_{A}^{1\rm{D}} \approx 0.76$ in the atomic region, while $M_{A}^{1\rm{D}} \approx 0.70$ in the molecular region. The total ($\rm{3D}$) Alfvén Mach number will be $M_{A} = C M_{A}^{1\rm{D}}$, where $C$ is a constant. For isotropic turbulence $C=\sqrt{3}$. However, recent ideal-MHD numerical simulations suggest that magnetic fluctuations become fully isotropic when $M_{A} \geq 10$ \citep{beattie_2020}. According to their results (Fig.~6), parallel magnetic fluctuations are $\sim 0.5 - 0.6$ times smaller than perpendicular when $0.5 \leq M_{A} \leq 2$. If we also consider that the dispersion of polarization angles probe perpendicular magnetic field fluctuations \citep[e.g.,][]{skalidis_2021}, we derive $C \approx \sqrt{2.6}$; this is a more reasonable value than simply assuming that turbulence is isotropic and $C=\sqrt{3}$. 

Then we find that $M_{A}\approx 1.2$, and $M_{A} \approx 1.1$ for the atomic and molecular region respectively. For both regions $M_{A} \sim 1$, which indicates that turbulence is trans-Alfvénic in this cloud, which means that the magnetic field is dynamically important. It is not straightforward to impose uncertainties on this estimate, hence \Ma\ could be slightly below or above one. This uncertainty, however, cannot affect our conclusion about the relative importance of the magnetic field in the cloud dynamics; in order to consider the magnetic field as dynamically unimportant (or equally that turbulence is purely hydro), $M_{A}$ should be larger than two \citep{beattie_2020}. Even if we set $C=\sqrt{3}$, which would be an upper limit, the estimated \Ma\ is less than 1.4 for both regions. Our results are consistent with \cite{planck_xxxv_2016}, who found that turbulence in the diffuse ISM is sub-, trans- Alfvénic.

\subsection{Is the magnetic field morphology correlated with the \HI\ velocity gradients?}
\label{subsec:Vc_gradients}

    \begin{figure*}
        \includegraphics[width=\hsize]{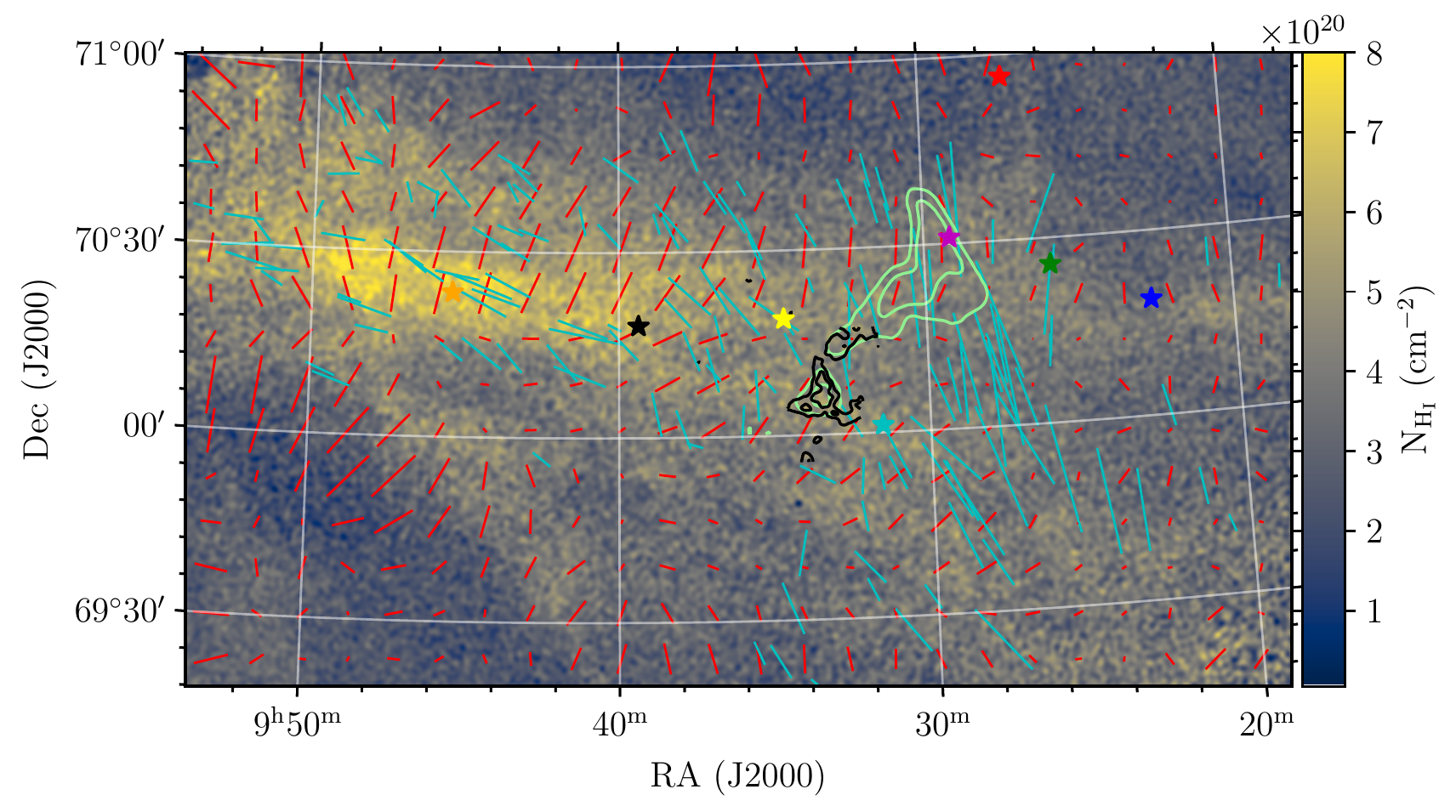}
        \caption{Velocity gradients overplotted on the \NHI\ column density map of our target cloud. Cyan segments show our polarization data, while red segments show the velocity gradients. The CO contours and the colored stars are the same as in Fig.~\ref{fig:hi_polarization}.}
        \label{fig:velocity_gradients_polarization}
    \end{figure*}

Our estimates indicate that turbulence is trans-Alfvénic in the target cloud (Sect.~\ref{sec:bpos_strength_computation}). This means that the magnetic field should play an important role in the dynamics of the cloud. Theoretical scenarios suggest that, for dynamically important fields, the morphology of the field lines is correlated with the gas velocity gradients \citep[e.g.,][]{gonzalez_lazarian_2017, girichidis_2021}. There are two favored topological states towards which the magnetic field tends to: the orientation of the field is either perpendicular or parallel to the orientation of the gas velocity gradient. According to \cite{gonzalez_lazarian_2017} the orthogonality between the magnetic field and the velocity gradient field is due to the strong Alfvénic motions that develop within a cloud. The same configuration can be achieved in clouds formed in the boundaries of bubbles when gas is expanding perpendicular to the background ISM magnetic field \citep[e.g.,][]{girichidis_2021}. When gravity takes over, velocity gradients tend to become parallel to the magnetic field \citep[e.g.,][]{lazarian_2018, hu_2020_vgt_gravity, girichidis_2021}. We tested these scenarios by exploring the correlation between the magnetic field morphology and the \HI\ velocity gradients in the target cloud.

The \HI\ LOS velocity component, $\rm V_{c} (\alpha, \delta)$, is the first moment of the brightness temperature map, $\rm T_{b}(\alpha, \delta)$, and we computed it as in \cite{miville_2003_fbm},
\begin{equation}
    \label{eq:Vc_definition}
    \rm V_{c} (\alpha, \delta) = \frac{\sum_{v} v T_{b}(\alpha, \delta, v)  \, \Delta v}{\sum_{v} T_{b}(\alpha, \delta, v) \, \Delta v},
\end{equation}
where the summation is performed within the velocity range relevant to the target cloud, which is [-22.1, 20.8] \kms\ (Fig.~\ref{fig:spectra}). We computed the $\rm V_{c}$ gradient field using second order central differences for every pixel. We smoothed the $\rm V_{c}$ map with a Gaussian kernel with standard deviation equal to $5.8\arcmin$ in order to reduce the noise in the map. We computed the velocity gradient orientations\footnote{Orientations are measured with respect to the horizontal axis of Fig.~\ref{fig:velocity_gradients_polarization} increasing counter-clockwise.} as,
\begin{equation}
    \label{eq:gradient_angles}
    \psi_{c}(\alpha, \delta) = \rm{arctan} \left( \frac{\partial V_{c}}{\partial \delta}/ \frac{\partial V_{c}}{\partial \alpha} \right),
\end{equation}
and the amplitudes as,
\begin{equation}
    \label{eq:gradient_magnitude}
    \rm |\vec{\nabla} V_{c}| =  \sqrt{ \left( \frac{\partial \rm V_{c} }{\partial \alpha} \right) ^{2} + \left( \frac{\partial \rm V_{c} }{\partial \delta} \right)^{2}}.
\end{equation}
Fluctuations in $\rm T_{b}$ can lead to $ \rm |\vec{\nabla} V_{c}| \neq 0$, even in the absence of a gradient; this can happen when there are multiple gas components with varying intensities. As we show in Appendix~\ref{sec:rho_vel_corr} this does not seem to be the case for the target cloud, where there is a dominant \HI\ component. Thus, the \HI\ gradients computed with the equations above accurately represent the projected \HI\ kinematics of the target cloud.

In Fig.~\ref{fig:velocity_gradients_polarization} we show the velocity gradient field with red segments overplotted on the \NHI\ map of the cloud; cyan segments correspond to our polarization measurements. In the atomic part of the cloud towards the orange-star LOS, velocity gradient orientations tend to be perpendicular to the magnetic field orientation. This behavior seems to hold in the majority of the cloud, as for example close to the cyan-star and yellow-star LOSs. But, towards the CO clump and close to the magenta-star LOS, velocity gradients tend to be parallel to the magnetic field. This alignment happens locally above the CO clump. Simulations suggest that velocity gradients are parallel to the local magnetic field orientation when gas is undergoing gravitational collapse \citep[e.g.,][]{lazarian_2018, hu_2020_vgt_gravity}. Recently, \cite{girichidis_2021} found that gravity takes over at $n \sim 400$ \VolDens, which is consistent with our inferred density for the CO-bright region of the target cloud derived by assuming an oblate triaxial 3D shape for the CO clump. Thus, it is probable that the observed \HI\ velocity gradient alignment with the local magnetic field morphology above the CO clump is due to gas which accretes onto the clump.

 \begin{figure}
        \centering
        \includegraphics[width=\hsize]{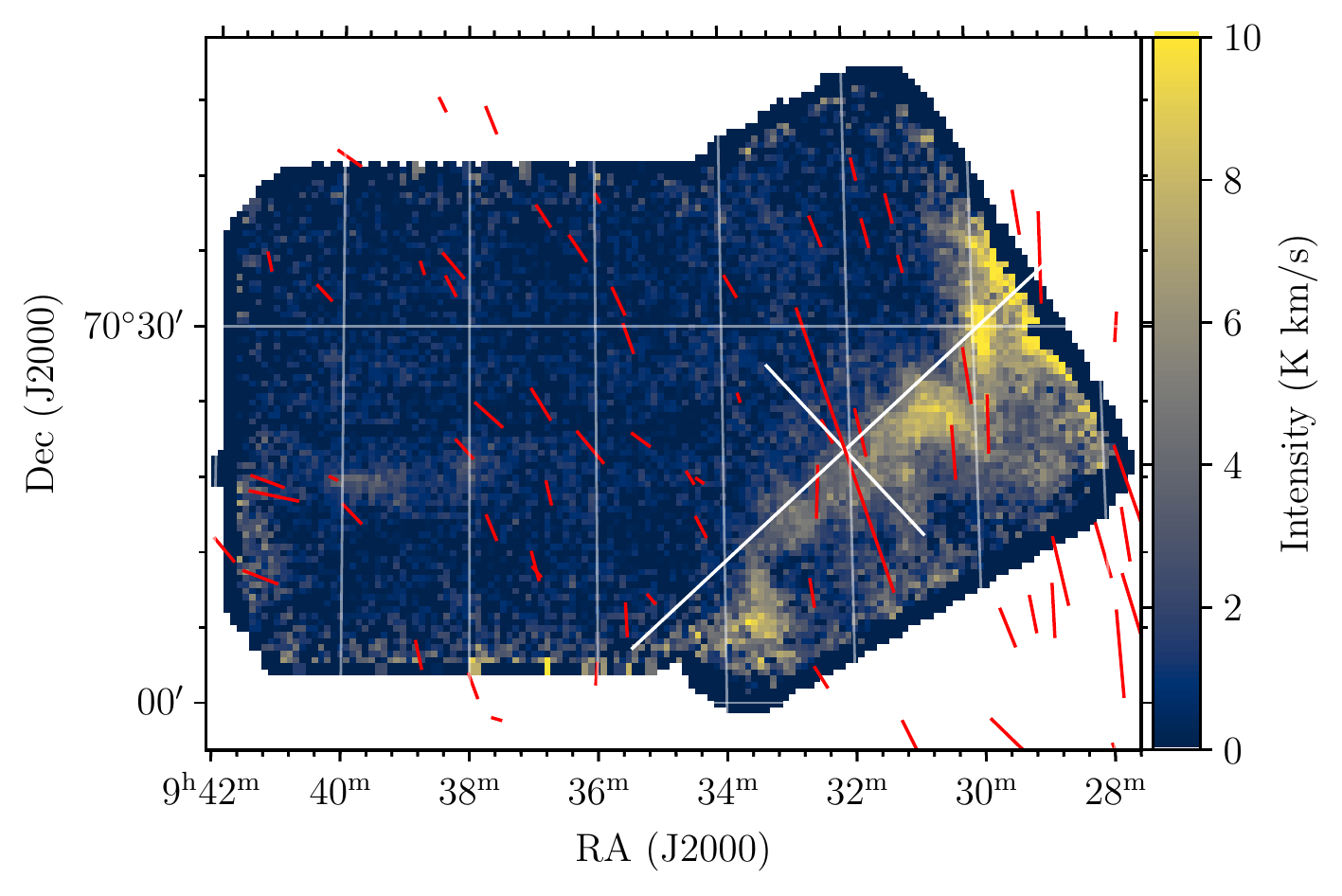}
        \caption{CO (J=1-0) integrated intensity. White lines mark the major and minor axis of the CO clump. Red segments show the polarization measurements as in Fig.~\ref{fig:hi_polarization}. The long red line at the center of the core corresponds to the mean magnetic field orientation.}
         \label{fig:CO_core_polarization}
    \end{figure}

  \begin{figure*}
        \centering
        \includegraphics[width=\textwidth]{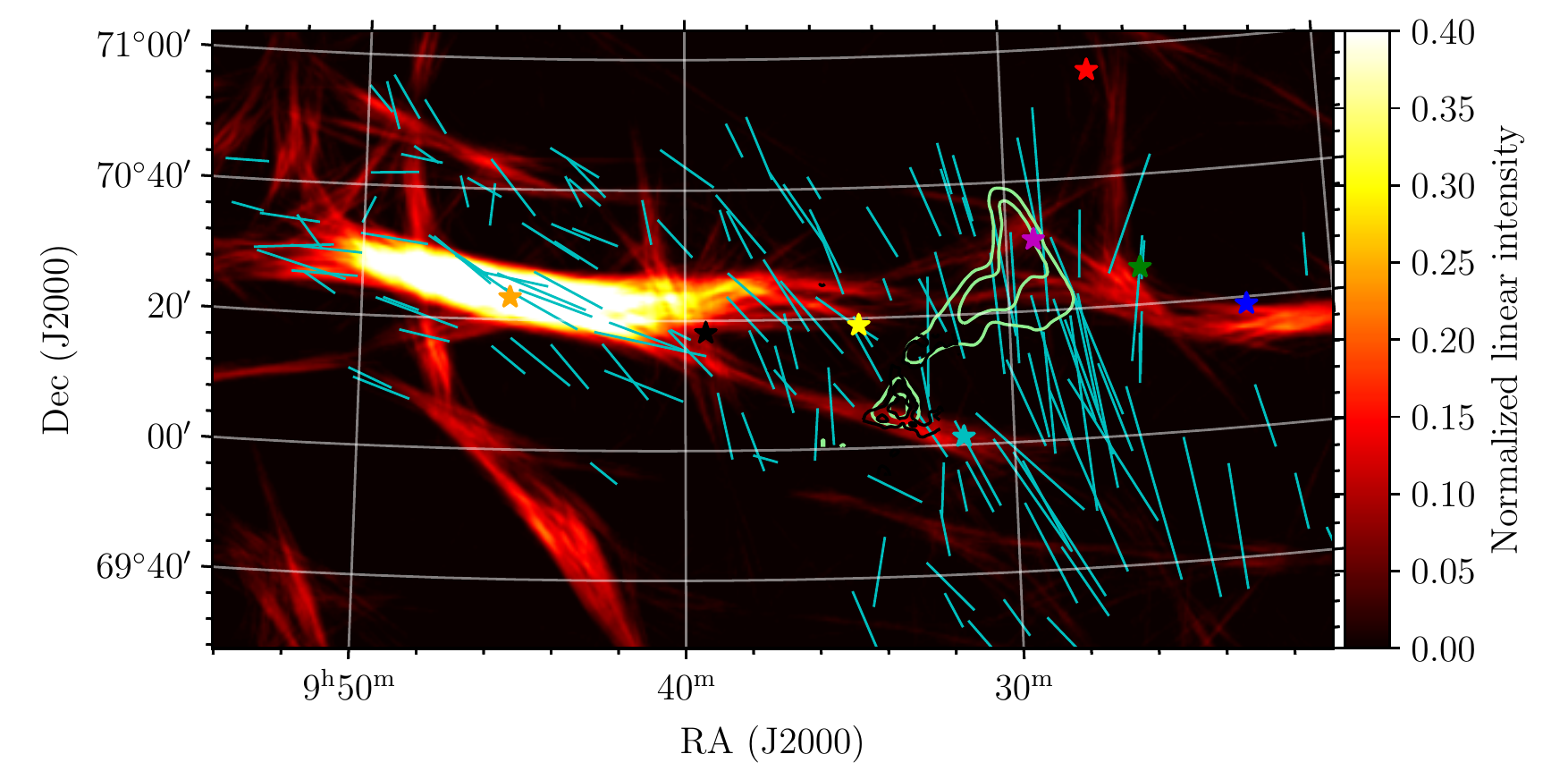}
        \caption{Our starlight polarization measurements (cyan segments) overplotted on the RHT output image. The colorbar shows the normalized intensity measured from the RHT. Stars and CO contours are the same as in Fig~\ref{fig:hi_polarization}.}
         \label{fig:rht_result}
    \end{figure*}

    \begin{figure}
        \centering
            \includegraphics[width=\hsize]{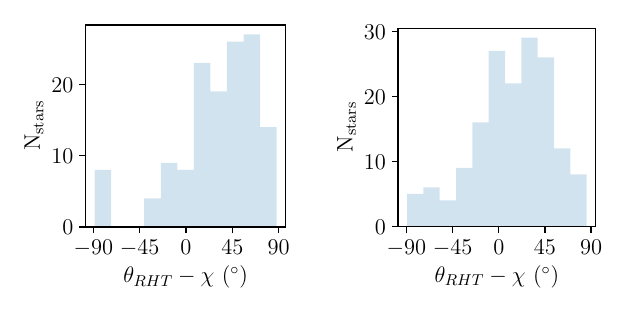}
        \caption{\textbf{Left panel:} Distribution of the difference between the orientation of the \NHI\ structure ($\theta_{RHT}$) and the local polarization orientation ($\chi$). The circular mean of this distribution is $43\degr$. \textbf{Right panel:} Same as in left panel, but for the difference between the orientation of the \NH\ structure and the local polarization orientation. The circular mean of the distribution is $20\degr$.}
        \label{fig:rht_angle_distrib}
    \end{figure}

\subsection{Does the magnetic field affect the CO clump shape?}
\label{subsec:core_shape_polarization_correlation}

The magnetic field is considered to affect the accumulation of gas \citep[e.g.,][]{mouschovias_1978, heitsch_2009}. When the field is dynamically important, it can support a cloud against its own self-gravity; magnetic forces are exerted perpendicular to the field lines, hence the magnetic field supports a cloud against its self-contraction perpendicular to the field lines. As a result, gas preferentially accumulates parallel to field lines, and the molecular clump is flattened with its small axis being parallel to the mean magnetic field orientation. For dynamically unimportant magnetic fields, no correlation between the mean field orientation and the clump shape is expected. For this reason, we explored if the shape of the CO clump is connected with the magnetic field as predicted by theoretical scenarios. 

Following \cite{tassis_2009} we computed the center and the shape of the CO clump using the first and second moments of the CO (J=1-0) integrated intensity map, $\rm{I_{CO}(x, y)}$, shown in Fig.~\ref{fig:CO_core_polarization}. The long and short white lines correspond to the CO clump major and minor axis respectively. We found that the aspect ratio between the two principal axis is $0.4$ which implies that the structure is slightly flattened; an aspect ratio equal to $1$ means that the clump is circular. Our polarization measurements are shown with the red segments in Fig.~\ref{fig:CO_core_polarization}. The long red segment at the center of the core corresponds to the mean polarization angle of the measurements close to the CO clump, i.e. of the region defined by the following boundaries: $09:28:00 \leq $ RA $\leq 09:36:00$, and $+70:00:00 \leq $ Dec $\leq 70:40:00$. 

The mean magnetic field orientation is closer to the minor principal axis of the clump than the major axis,  with a $24\degr$ offset. Although the alignment of the mean field orientation with the minor clump axis is not exact, our observations indicate that the magnetic field is likely responsible for the asymmetric shape of CO clump. This, in combination with the alignment that we found between the \HI\ gradients with the magnetic field above the clump, suggests that the CO clump could be self-gravitating. 

Our results are in line with \cite{tassis_2009}, who find that the mean magnetic field orientation is $\sim 20\degr$ offset from the minor axis of asymmetric molecular cores. These authors targeted cores with densities $10^{3} - 10^{4}$ \VolDens\ which is significantly larger than our inferred densities ($\sim 400$ \VolDens\ for the CO-bright clump). Our inferred alignment is slightly weaker than the average alignment found by \cite{tassis_2009}. A possible explanation is that the gas volume density of our cloud is one (or in some cases two) orders of magnitude less than the core densities studied by \cite{tassis_2009}. This implies that self-gravity may have a weaker role in the dynamics of our clump than in the dense cores of their sample.

Finally, we note that the CO survey of \cite{pound_1997}, towards the same cloud, showed that this clump is slightly extended further from the boundaries of our map, towards RA=09:28:00 and Dec=+70:30:00. However, the difference of the integrated intensity maps between the two datasets is not significant and does not affect the qualitative conclusions on the clump shape. In fact, using the data from \cite{pound_1997} we found that the offset between the minor axis of the clump and the mean magnetic field orientation is reduced by $6\degr$; we refer to Fig.~1 of \cite{miville_2002} for a visual comparison of our inferred clump shape with the data from \cite{pound_1997}.

    \begin{figure*}
        \includegraphics[width=\hsize]{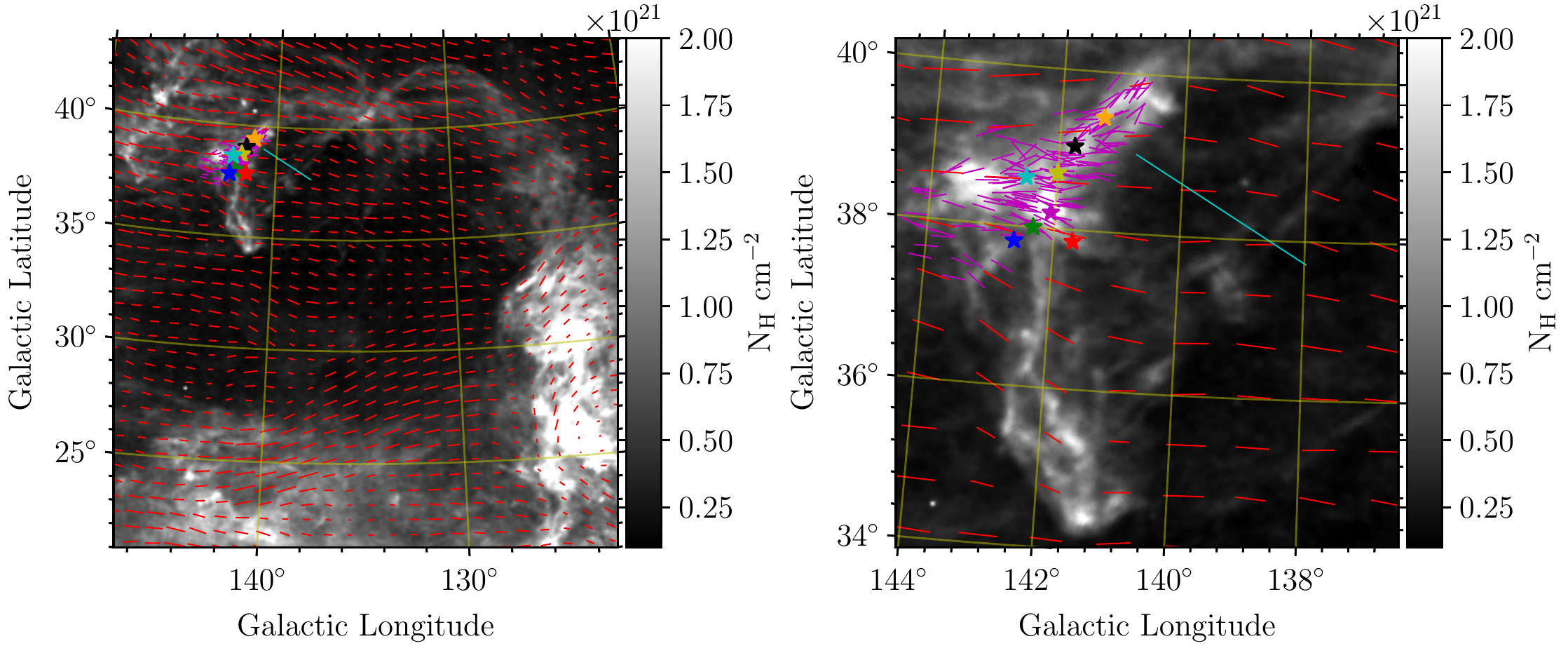}
        \caption{Magnetic field morphology overplotted on the \textit{Planck}-based \NH\ map of the NCPL. \textbf{Left panel:} Red segments show the dust emission polarization rotated by $90\degr$; the length of each segment is proportional to the polarized intensity. Magenta segments correspond to our polarization measurements; all segments have a fixed size for visualization purposes. There is an offset in the magnetic field orientation as traced by optical and sub-mm polarization due to the large beam difference between the two observables. Colored stars correspond to the target C+ LOSs, as shown in Fig.~\ref{fig:hi_polarization}.
        \textbf{Right panel:} Zoomed in region towards the target cloud. The cyan solid segment shows the $\rm{H_{\alpha}}$ canal found by \cite{McCullough_2001}.}
        \label{fig:av_ncpl}
    \end{figure*}

\section{Is the magnetic field morphology correlated with the \HI\ structure of the cloud?}
\label{subsec:rht_polarization}

Diffuse \HI\ elongated clouds were found to be statistically aligned with the Galactic magnetic field, as probed by dust polarization \citep{clark_2014, clark_2015, kalberla_2016}. Our polarization data (Fig.~\ref{fig:hi_polarization}) follow the main \HI\ structure in the left (atomic) part of the cloud, RA $>$ 09:36:00, while in the right (molecular) part, RA$\leq$ 09:36:00, they tend to be perpendicular to the cloud axis. 

We employed the Rolling Hough Transform (RHT) \citep{clark_2014} in order to quantify the relative alignment between our polarization orientations and the \HI\ structure. The input of this algorithm is the column density map of Fig.~\ref{fig:hi_polarization}. The algorithm uses a circular kernel of radius $D_{w}$ and smooths the image. A kernel with radius $D_{k}$ scans the image and measures the probability of each pixel being part of a linear structure along different orientations ($\theta$). Orientation angles, $\theta$, with probability larger than a given threshold, $Z$, are identified as linear structures. We set the free parameters of the RHT algorithm close to the authors recommended values, $D_{w}=95\arcmin$, $D_{k}=30\arcmin$ and $Z=70\%$. The output of the RHT is shown in Fig.~\ref{fig:rht_result}.

The colormap shows the linearity intensity normalized in the range $[0, 1]$. The algorithm has detected the main structure of the cloud at RA $\leq$ 09:40:00, which then splits into two branches of lower linearity at RA $\sim$ 09:40:00 close to the black-star LOS. Cyan segments correspond to our polarization data, contours to the CO integrated intensity, and colored stars to the LOSs where \CII\ are available, as in Fig.~\ref{fig:hi_polarization}. We quantified the relative difference between the polarization angles and the \HI\ cloud orientation. We computed the orientation angle of the \HI\ structure as suggested by \cite{clark_2014}, within circular regions of radius equal to $15\arcmin$ and centered at the position of each polarization measurement. The \HI\ orientation angles, or RHT angle, are denoted as $\theta_{RHT}$ and the polarization angles as $\chi$. We computed the difference between $\theta_{RHT}$ and $\chi$ restricted in the range [$-90\degr$, $90\degr$]. In Fig.~\ref{fig:rht_angle_distrib} we show the distribution of differences between the two quantities. The circular mean of the distribution is $43\degr$, which in striking contrast to \cite{clark_2014}, shows that the relative alignment of the \HI\ structure with our optical polarization orientation is weak; the loss of alignment happens close to the CO clump. Thus, the \HI\ structure of the cloud is not a good proxy of the magnetic field orientation in the molecular region of the target cloud. 

\subsection{What caused the magnetic field turn in the target cloud?}
\label{subsec:ncpl_magnetic_field}

In the molecular region of the target cloud, the magnetic field orientation turns and does not follow the \HI\ structure. This turn could be caused by the self-gravity of the CO clump. However, the onset of the magnetic field turn happens at RA = 09:40:00, close to the black-star LOS in Fig.~\ref{fig:rht_result}. This is $\sim 0.8\degr$ away from the center of the CO clump and corresponds to an actual distance of $\sim 4$ pc. This offset distance is significant, hence it seems unlikely that self-gravity of the dense CO-bright core could have significantly distorted the magnetic field lines of the cloud at such a large distance from it. Thus we examined other scenarios which could potentially explain the origin of this turn in the magnetic field morphology.

The target cloud is located close to the Ursa Major arc, which is a 30 degree long structure observed in ultraviolet wavelengths. As was suggested by the authors who discovered this arc \citep{braco_2020}, it is plausible that it has been formed by a radiative shock. If that is the case, then the magnetic field morphology of the target cloud may have been distorted by a large-scale explosion. Indeed, the cloud is at the edges of the so-called North Celestial Pole Loop (NCPL) \citep{heiles_1989}, which is considered to be an expanding cylinder or sphere \citep{meyerdierks_1990}. Thus, the NCPL explosion could have caused the observed turn of the cloud magnetic field morphology with respect to the \NHI\ axis. We explored if simple energetics arguments support this idea. For, if the magnetic field were sufficiently strong, it could prevent gas from expanding and the magnetic field lines would be relatively straight. In the opposite case, the expanding gas could drag the field lines, and hence create large-scale magnetic field distortions. 

The NCPL is expanding with a velocity $u_{\rm{exp}} \sim 20$ \kms\ \citep{meyerdierks_1990}. If we assume that the density of the expanding gas in the boundaries of the bubble is $\sim 100$ \VolDens\ \footnote{This is a lower limit for the volume density, since we have found that the volume density in the molecular sub-region of the current cloud can be up to $350$ \VolDens\ (Sect.~\ref{sec:gas_properties}).}, then the kinetic energy density of the expanding gas is $3.2\times 10^{-10}$ $\rm{gr}~\rm{cm}^{-1}~\rm{s}^{2}$. The magnetic field energy density, adopting from Sect.~\ref{sec:bpos_strength_computation} that $|\rm{B_{tot}}| \sim 30 \mu$G, is $1.6 \times 10^{-11}$ $\rm{gr}~\rm{cm}^{-1}~\rm{s}^{2}$; the ratio of the kinetic over the magnetic energy density is $\sim 10$. Thus, the NCPL expansion has enough energy to distort the magnetic field morphology of the cloud. Since the observed magnetic field turn of the cloud is probably related to the NCPL expansion, we explored if the magnetic field morphology of the NCPL is consistent with an expanding bubble or cylinder. 

In the left panel of Fig.~\ref{fig:av_ncpl} we show the \NH\ map of the NCPL. For the construction of the \NH\ map we used the extinction map from \cite{planck_2016_av}, $\rm{A_{V}}$, and the conversion factor $\rm{N_{H}/E(B-V)} = 6\times 10^{21}$ \ColDens\ mag$^{-1}$ \citep{kalberla_2020}, where $\rm{A_{V}/E(B-V)} = 3.1$. A characteristic hole is prominent at $l, b \sim (135\degr, 35\degr)$ possibly due to the expansion of the NCPL. Our target cloud is located at the edges of the loop at coordinates, $l, b =  142\degr, 39\degr$. In the right panel of Fig.~\ref{fig:av_ncpl}, we show a zoomed-in region towards the target cloud; magenta segments correspond to our polarization\footnote{We note that, in Fig.~\ref{fig:av_ncpl}, the reference frame and the orientation of the cloud is different from Fig.~\ref{fig:hi_polarization}.} measurements.

We probed the magnetic field morphology of the NCPL using the $353$ GHz dust emission polarization data from the Planck satellite \citep{planck_2020}. We smoothed the Stokes parameter maps at a resolution of $1\degr$ in order to increase the S/N (as described in \citealt{skalidis_2019}). In the left panel of Fig.~\ref{fig:av_ncpl}, we show, with the red segments, the Planck polarization rotated by $90\degr$, in order to probe the POS magnetic field orientation, with the length of each segment being proportional to the polarized intensity which is,
\begin{equation}
    P_{s} = \frac{\sqrt{Q_{s}^{2} + U_{s}^{2}}}{I_{s}},
\end{equation}
where $I_{s}$, $Q_{s}$ and $U_{s}$ are the Stokes parameters measured by Planck. The polarization orientation is,
\begin{equation}
    \chi_{s} = 0.5~\rm{arctan} \left (- \frac{U_{s}}{Q_{s}} \right ),
\end{equation}
measured in the Galactic reference frame, and following the IAU convention angles increase counterclockwise. 

    \begin{figure}
        \includegraphics[width=\hsize]{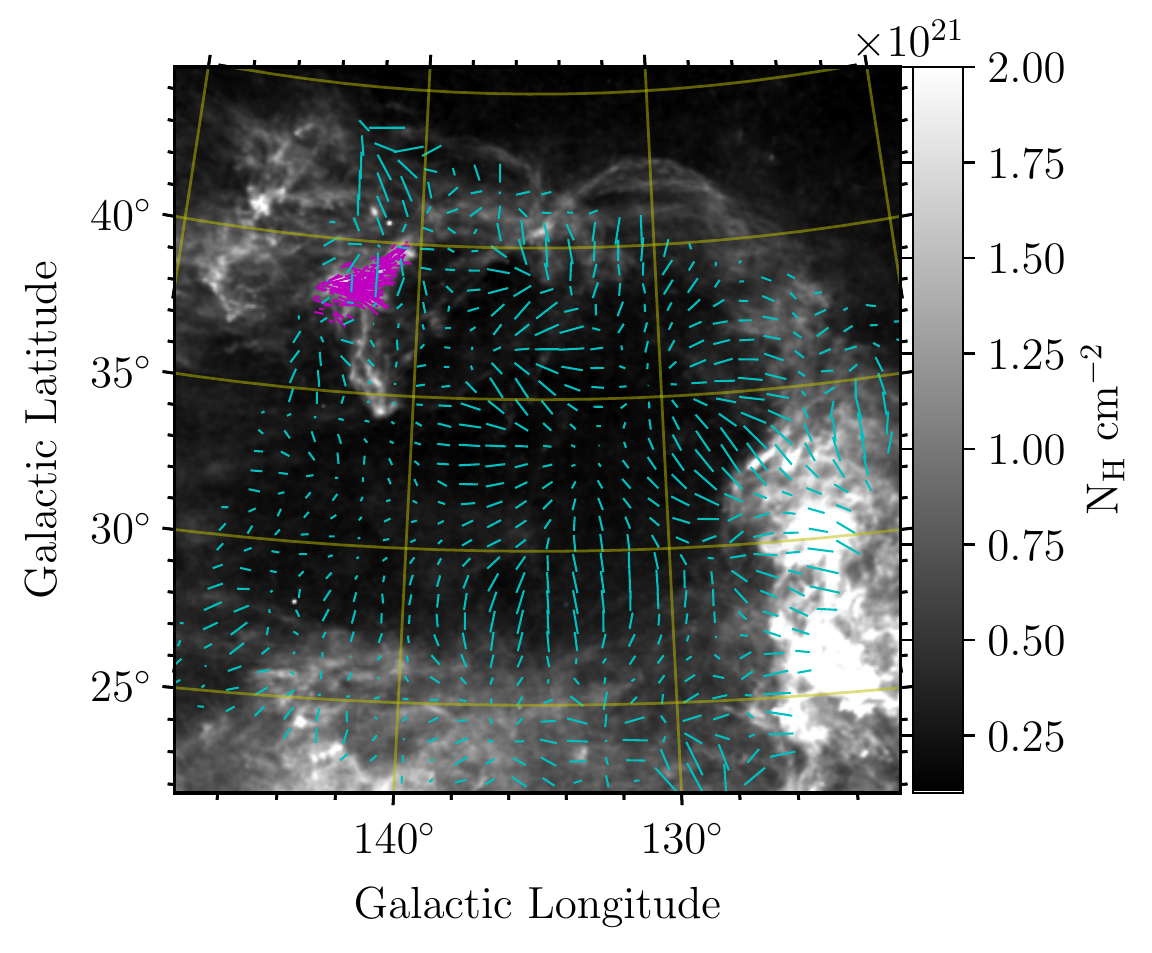} 
        \caption{Velocity gradients (cyan segments) overplotted on the \NH\ map of the NCPL. Magenta segments correspond to our polarization measurements.}
        \label{fig:av_grads_ncpl}
    \end{figure}

The magnetic field lines are squeezed in the right edge of the NCPL $l, b \sim 126\degr, 33\degr$, while in the left part $l, b \sim 142 \degr, 33\degr$ they are well ordered and parallel to each other; this indicates that NCPL expanded asymmetrically. In the left part, $140\degr \leq l \leq 145\degr$ and $35\degr \leq b \leq 40\degr$, gas seems to have escaped from the bubble without distorting the magnetic field lines, while in the right part, the swept-up gas could have collided with the background ISM and created regions with enhanced column density, such as those at $l, b \sim 126\degr, 30\degr$. The background ISM magnetic field is almost parallel to the Galactic Longitude axis at $136\degr \leq l \leq 145\degr$ and $38\degr \leq b \leq 40\degr$. Overall the magnetic field morphology is consistent with an expanding bubble or cylinder; simulated morphologies from \cite{Krumholz_2007} and \cite{ntormousi_2017} can be used for a visual comparison. The magnetic field morphology of our target cloud, magenta segments, exhibits a turn with respect to the background field of this area (right panel in Fig. \ref{fig:av_ncpl}): it forms an angle with the latter at $b > 39^\circ$ and becomes parallel to the ambient field at $b < 39^\circ$. Thus, it is likely that the NCPL expansion is responsible for the observed turn of the magnetic field orientation with respect to the \NHI\ axis in the molecular region of the target cloud, Fig.~\ref{fig:hi_polarization}.

    \begin{figure}
        \includegraphics[width=\hsize]{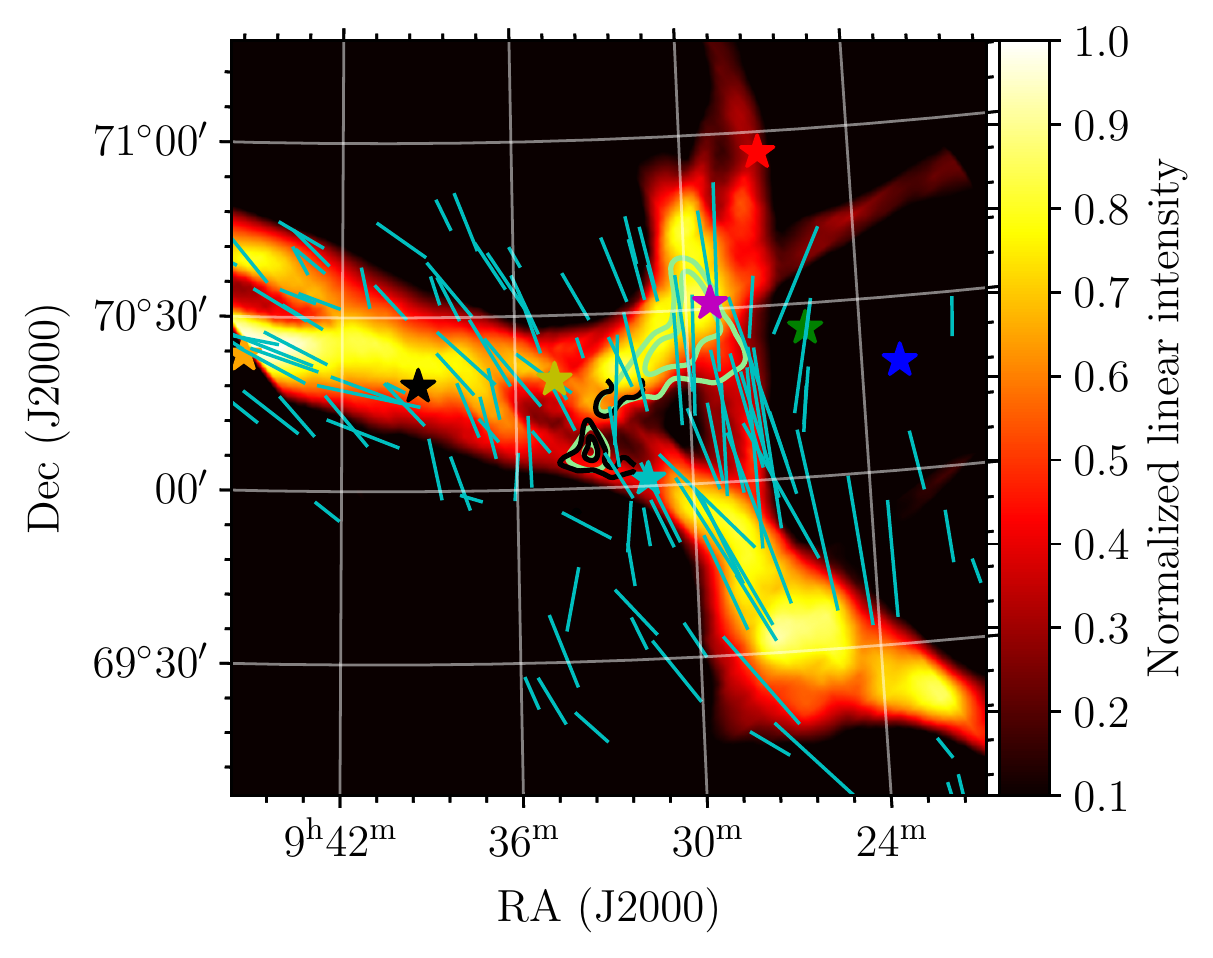} 
        \caption{RHT output for the \NH\ map of our target cloud. Symbols are as in Fig~\ref{fig:rht_result}.}
        \label{fig:rht_Avmap}
    \end{figure}

There is also a possibility that a smaller-scale expansion, other than the NCPL, affected the magnetic field morphology of the cloud. In the right panel of Fig.~\ref{fig:av_ncpl} we see that the target cloud is close to another cloud located at $l, b \sim 142\degr, 34.5\degr$. This cloud also has significant amount of CO \citep{pound_1997} and its magnetic field morphology (red segments) also turns with respect to the ISM background magnetic field. The \NH\ morphology of this cloud combined with our target cloud's structure create a bow-like structure, assuming that the center is at $\sim 138\degr, 37.5\degr$, which is typical of clouds located at the boundaries of expanding bubbles \citep{inoue_2008, inoue_2009}. One more piece of evidence which indicates that the two clouds could be at the boundaries of a bubble unrelated to the NCPL is that \cite{McCullough_2001} found a narrow and filamentary $\rm{H_{\alpha}}$ canal close to our target cloud, shown with the cyan segment in the right panel of Fig.~\ref{fig:av_ncpl}. From the various models they explored they concluded that this canal has been formed most probably due to a trail of a star or other compact object which moves through the ISM. This canal points towards our target cloud and could have been formed by an object kicked by an explosion; the same explosion which swept-up the gas and formed our target cloud at the boundaries of this bubble. The magnetic field lines (red segments) follow the bow-like shape of the two clouds and they seem to form a bottleneck at $l, b \sim 144\degr, 38\degr$ where they get squeezed and turn, in order to connect with the background ISM field. Thus, it is plausible that expanding gas, which is not related to the NCPL, dragged the magnetic field lines of our target cloud and created the observed offset between the \NHI\ axis of the cloud and its magnetic field orientation in the molecular region.

\subsection{Correlation of the magnetic field morphology with total column density}
    
    \begin{figure}
        \includegraphics[width=\hsize]{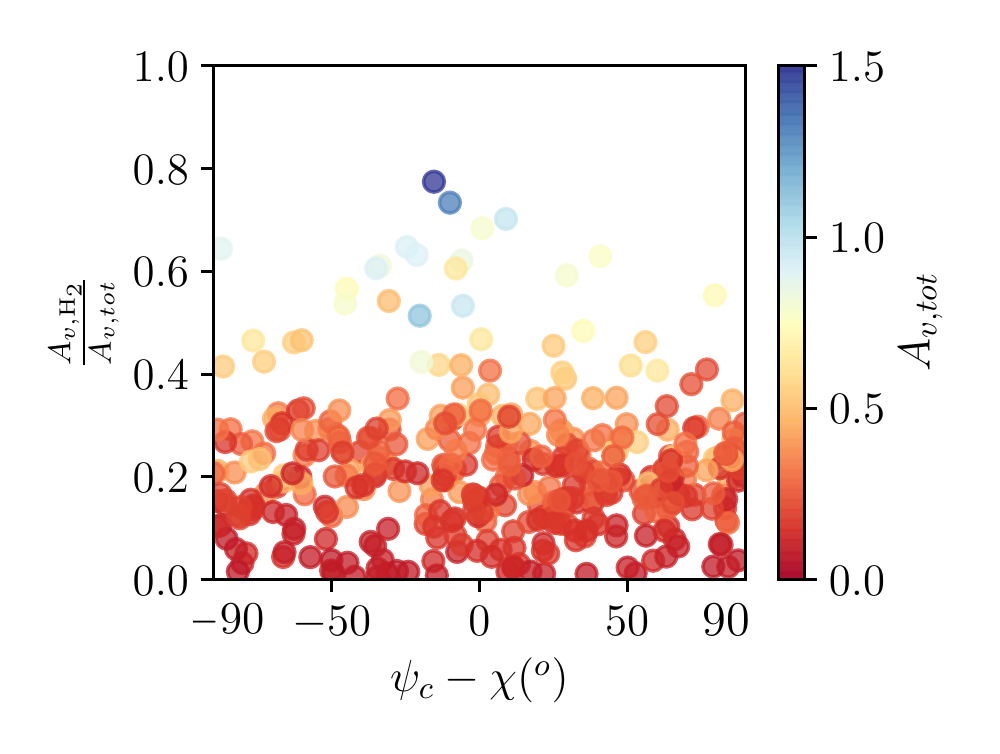} 
        \caption{Molecular fraction versus the relative difference between the  POS \HI\ velocity gradient angle ($\psi_{c}$) and the magnetic field orientation ($\chi$), inferred from the Planck data. Colorbar shows the total $A_{v}$.}
        \label{fig:NCPL_H2_angles}
    \end{figure}
    
We applied the RHT algorithm to the \NH\ map of our target cloud shown in Fig.~\ref{fig:av_ncpl} with the following parameters $D_{w}=41\arcmin$, $D_{k}=25\arcmin$ and $Z=70\%$. The output of the RHT analysis is shown in Fig.~\ref{fig:rht_Avmap} with our polarization segments overplotted in cyan. In the atomic region of the cloud, defined between the orange and black stars, the \NH-RHT morphology tends to be parallel to the magnetic field orientation. Since atomic gas is more abundant there, the \NHI\ map probes almost the total column density (\NH), hence both the RHT-\NHI\ (Fig.~\ref{fig:rht_result}), and RHT-\NH\ (Fig.~\ref{fig:rht_Avmap}) maps are consistent there. In the molecular region, there is a major difference between the RHT structures of \NHI\  and \NH\ when compared with the local magnetic field orientation.

In the molecular region of the cloud (between the yellow and black stars), the RHT-\NH\ structure splits into two branches; the first branch is towards the CO clump (green contours), where the RHT structure tends to be perpendicular to the magnetic field orientation, and the second branch is towards the cyan point, where the RHT structure tends to align with the magnetic field. On the other hand, in the molecular region, the RHT-\NHI\ map splits into two branches which both are perpendicular to the mean magnetic field orientation. The distribution of the difference between the RHT angles and our polarization data for the \NH\ map is shown in the right panel of Fig.~\ref{fig:rht_angle_distrib}; the mean of the distribution is $20\degr$, which implies that the total column density of the cloud tends to be parallel to the local magnetic field orientation. Although, the alignment between the POS magnetic field orientation and the \NHtot\ shape of the cloud is not perfect (20$\degr$ offset), it is stronger than the alignment between the magnetic field morphology and the \NHI\ shape (43$\degr$ offset).

\section{Does the magnetic field affect molecule formation in the NCPL?}
\label{subsection:ncpl_B_molecule_formation}

The magnetic field is considered to play an important role in the formation of \Hmolecular\ in expanding bubbles \citep[e.g.][]{silc_I, silc_II, girichidis_2021}. Gas which is expanding parallel to the magnetic field, can free stream and finally escape from the bubble. This leads to the formation of low column density regions (tunnels). On the other hand, gas which is expanding perpendicular to the magnetic field, compresses the field lines and induces a magnetic pressure gradient which is exerted against the gas expansion. The magnetic field prevents gas from escaping, and thus forming regions with enhanced column density which define the boundaries of the bubble. These regions are candidates for hosting molecular clumps, if self-shielding is sufficient to prevent the \Hmolecular\ photo-destruction. Thus, one would expect that \Hmolecular\ is preferentially formed near the bubble boundary regions, where the relative orientation between the magnetic field and the \HI\ velocity gradients is perpendicular. When gas accumulation becomes important, self-gravity takes over and then \HI\ velocity gradients align with the magnetic field \citep[e.g.,][]{hu_2020_vgt_gravity}. Although, it has been suggested that the magnetic field could determine the location of  molecular sites in expanding structures \citep[e.g.,][]{dawson_2011}, to our knowledge no systematic observational study has ever been done regarding this topic. For this reason, we used the correlation between the magnetic field and the \HI\ velocity gradients in order to explore if the relative orientation between the two is connected to the molecular gas in the wider NCPL region. 

We used data from the HI4PI all-sky survey \citep{hi4pi} in order to study the \HI\ kinematics of the NCPL; the higher resolution DHIGLS survey \citep{DHIGLS} does not cover the full region of the NCPL. The resolution of the HI4PI data is $10.8\arcmin$. We also employed the full-sky \NHmol\ map of \cite{kalberla_2020} in order to infer the NCPL \Hmolecular\ sites and the full-sky map of total hydrogen column density, \NHtot, provided by the same authors. These maps allowed us to probe the molecular fractional abundance towards the NCPL. 

The observed \HI\ spectra include contribution from multiple sources that may be spatially distinct, or from clouds at different ISM phases, such as the warm (WNM) and cold neutral medium (CNM). In order to study the kinematics of the NCPL we have to identify the \HI\ component which corresponds to the NCPL. For doing so, we decomposed the \HI\ spectra into different Gaussian components and identified the CNM component which matches the kinematics of the NCPL; WNM is expected to not have \Hmolecular\ and for this reason we neglected it. We employed the ROHSA algorithm \citep{rohsa}, which performs the \HI\ decomposition by taking into account the POS correlations of each component. The input of the algorithm is the \HI\ PPV cube and the output is a decomposition of the observed spectrum into $N$ distinct Gaussian components, which are spatially correlated. The algorithm is characterized by five free parameters, including the number of Gaussians $N$, which are defined by the user. We run the algorithm with the authors recommended values, $\lambda_{\alpha}=1000$, $\lambda_{\mu} = 1000$, $\lambda_{\sigma} = 1000$, $\lambda_{\sigma \prime} = 100$ and by setting $N = 5$; this is the minimum number of Gaussians required to produce accurate fitting of the NCPL spectra. 

In Fig.~\ref{fig:hi_decomposition} we show a typical \HI\ spectrum at a random LOS of the NCPL, decomposed into the five Gaussians. In Fig.~\ref{fig:ncpl_gaussian_components} we show the ROHSA algorithm output. We computed the POS velocity gradients of the Gaussian component related to the NCPL and transformed the \HI\ velocity gradient angles ($\psi_{c}$) to the Galactic reference frame (Appendix~\ref{sec:gradient_angles_transformation}) in order to compare them with the magnetic field orientation as inferred from the Planck data. In Fig.~\ref{fig:NCPL_H2_angles} we show, with the colored points, the molecular fractional abundance, $\rm A_{V, H_{2}}/A_{V, H}$, computed from the \cite{kalberla_2020} maps, as a function of the relative difference between the \HI\ POS velocity gradient angle with the dust polarization angle for multiple LOSs towards the NCPL. The color of each point corresponds to the total visual extinction, $\rm A_{V, \rm{tot}}$, as shown in the colorbar of the figure. All maps were smoothed to a common resolution equal to $1\degr$ \footnote{We applied the same analysis to data with better than $1\degr$ accuracy, but they were too noisy and we found no clear trend. We also tested these results using a larger smoothing radius and the results were similar to Fig.~\ref{fig:NCPL_H2_angles}.}.

Points with low $\rm A_{V, \rm{tot}}$, shown with the red color, are uniformly distributed and no correlation is observed. Points with large $\rm A_{V, \rm{tot}}$, shown in blue (light and dark), are preferentially clustered around zero. This could be because high $\rm A_{V, \rm{tot}}$ regions are dominated by self-gravity. This result is consistent with our observations, where we found that the \HI\ velocity gradients tend to be parallel to the magnetic field orientation above the CO clump in the target cloud (Fig.~\ref{fig:velocity_gradients_polarization}). There is also a point with relatively high extinction ($\rm A_{V, \rm{tot}} \sim 1$), but close to $-90\degr$. There, \Hmolecular\ formation could be at an early phase where gas has been accumulated due to the expansion of the gas, but self-gravity is not sufficient to align the velocity gradients with the magnetic field yet. One more piece of evidence, which supports that self-gravity may not be so important towards this point, is that the $\rm A_{V, \rm{tot}}$ of this point is smaller compared to the majority of points around $\psi_{c} - \chi = 0\degr$ in Fig.~\ref{fig:NCPL_H2_angles}. Overall, regions with relatively high extinction ($\geq 8$), and fractional abundance ($\geq 0.5$) tend to have gas flows which are either parallel or perpendicular to the local magnetic field orientation. These regions could probe the different phases of the \HI - \Hmolecular\ formation around the bubble, as presented in recent simulations \citep[e.g.,][]{girichidis_2021}, with the magnetic field playing an important role in this transition process; for, if the magnetic field were dynamically unimportant, one would expect the distribution of $\psi_{c} - \chi$ angles to be completely random.

It should be taken into account that projection effects are very important and can significantly affect these results \citep{girichidis_2021}. For if, the magnetic field were pointing towards the observer along a given LOS, $\chi$ would not be representative of the mean field orientation. It would rather represent a random angle of the fluctuating magnetic field component. In addition, if gas motions are mostly in the POS, then the observed \HI\ spectra, which trace only the LOS velocity component, and hence the velocity gradients, would not accurately probe the bulk motion of the gas. If any of these conditions apply, then the observed $\psi_{c} - \chi$ difference could be shifted towards any value independently of the molecular fractional abundance. Thus, these results should be interpreted with caution.


\section{Discussion}
\label{sec:discussion}

We have employed a suite of observations to probe the magnetic field, kinematics and chemical composition of a selected filamentary structure of the Ursa Major Cirrus. We found a change in the relative orientation between the structure's long axis and the magnetic field: the atomic sub-region of the structure shows alignment, whereas a shift in the orientation occurs at the interface between \HI\ and \Hmolecular\ media. Previous studies of the relative orientation have been statistical, binning observed/simulated data of multiple structures. Our approach complements these studies by providing a detailed look at the properties of one structure in the transition from atomic to molecular gas. By observing \HI, \CO\ and \CII\ we have established that the magnetic field changes orientation with respect to the gas morphology, as does the orientation of velocity gradients.

\subsection{Magnetic field and \HI\ alignment}
\label{subsec:discuss_bfield_hi}

\subsubsection{Consequences for modeling Galactic foregrounds for CMB studies}

\cite{clark_2014, clark_2015} and \cite{kalberla_2016} compared the orientation of the POS component of the magnetic field with the orientation of  \HI\ filaments and found that at low column densities the two are statistically aligned. Based on this property, \cite{clark_2019} constructed a 3D map of the magnetized ISM over the full sky. In their map, the \HI\ filament orientation was used to indirectly trace the POS magnetic field and the \HI\ line radial velocity was used to disentangle the contribution from different magnetic field configurations along each LOS. The resulting map has been used to improve modeling of Galactic foreground emission in 3D \citep{pelgrims_2021}, a subject of high importance for cosmological studies \citep{tassis_2015, planck2016_dust}.

In contrast to what has been assumed in the aforementioned works, our polarization data suggest that in certain cases \HI\ structure is not a good proxy for the POS magnetic field geometry. Even though the \NHI\ structure of the target cloud is filamentary, the major axis of this filament shifts orientation with respect to the magnetic field (showing alignment in the parts of the structure that are atomic but forming near-orthogonal angles when the gas becomes molecular in Fig.~\ref{fig:rht_angle_distrib}). The results of \cite{clark_2014} and \cite{kalberla_2016} are statistical in nature and the alignment does not necessarily hold in individual cases, such as ours and in \cite{skalidis_2018}. 

An important result that comes up from our survey is that the alignment between the \NHI\ structure and the magnetic field orientation becomes weak when molecular gas is present. Our polarization data follow the shape of the cloud, as traced by the total (both atomic and molecular) column density. This, in the context of the magnetic field modelling with \HI\ data \citep{clark_2019}, indicates that the magnetic field orientation in clouds with significant \Hmolecular\ may not be accurately traced by the orientation of \HI\ gas. If confirmed with observations at other transition clouds, the results of such modeling should be cautiously interpreted wherever the medium turns molecular. Such regions can be selected by maps like those from \cite{planck_2016_av} or \cite{kalberla_2020}. A conservative approach for mitigating biases in foreground modeling due to misalignment of filaments with magnetic fields would be to assign larger uncertainties for inferred magnetic field angles at regions that exhibit the presence of \Hmolecular\ above some thresholds. However, we note that the results of \cite{clark_2019} refer to much larger angular scales. Therefore, the potential misalignment caused by the presence of molecular gas (which is more spatially compact than the diffuse medium) may be statistically weaker when the averaging takes place.

\subsubsection{Comparison with past works}

The relative angle between the polarization orientation and the column density morphology of clouds has been a topic widely studied in the past few years. \cite{soler_2014} were the first to propose, using numerical simulations, that the relative orientation between the magnetic field and the major axis of filamentary structures changes from being parallel to perpendicular as column density increases. \cite{Chen_2016} proposed that the change happens when the Alfvén Mach number changes from sub- to super- Alfvénic. But subsequently \cite{soler_2017} and \cite{Seifried_2020} found that the transition in their simulations happens for $M_{A} \gg 1$. Numerical simulations of \cite{kroetgen_2020} showed that this transition can also happen in high-density regions within a cloud, when turbulence is sub-Alfvénic, even if self-gravity is not included. Among these works, the simulated conditions are different, but all of them suggest that there is a threshold, in \Ma\ or $n$, above which the relative orientation between the magnetic field and the column density (or volume density) of the cloud changes. 

\cite{planck_xxxv_2016} explored this transition using observations of polarized dust emission. They found that the transition happens, when $\rm{log_{10}(N)} \approx 21.7$; their sample consisted mostly of regions with $\rm{N_{H}} \sim 10^{21} - 10^{23}$ \ColDens. \cite{fissel_2019} studied the young molecular cloud, Vela C, and found that the transition happens when $n$ exceeds $10^{3}$ \VolDens, while \cite{alina_2019} concluded that there is an extra dependence on the relative orientation; it can vary according to the column density contrast between the filament and its background medium. More precisely, they found that the alignment of the magnetic field with the filamentary structures is prominent when the column density contrast is lower than $\rm{N_{H}} < 4 \times 10^{20}$ \ColDens, while when a cloud exceeds this limit the relative orientations tend to become random. All these works, however, mainly focus on high density (or column density) environments, whereas our cloud is a transition diffuse cloud. 

In our target cloud, the maximum inferred density is 500 \VolDens, while its maximum total column density is $\sim 2\times 10^{21}$ \ColDens. We found that the magnetic field orientation is relatively well correlated with the total column density (atomic and molecular) shape of the cloud (right panel in Fig.~\ref{fig:rht_angle_distrib}). This is consistent with all the aforementioned works showing that the relative orientation transition, from being parallel to perpendicular, happens at much higher density and column density values. 

\subsection{Alternative magnetic field strength estimate}
\label{subsec:tritsis_bpos}

In the same cloud, \cite{tritsis_2019} estimated the magnetic field strength using a novel technique developed by \cite{tritsis_2018}, which is based on the properties of the compressible MHD modes. This method assumes that the observed column density and velocity variations perpendicular to the mean magnetic field orientation is due to the propagation of fast magnetosonic waves. \cite{tritsis_2019} argued that the target cloud shares similar observational properties with striations, which are quasi-linear gas structures observed in molecular clouds \citep{goldsmith_2008, panopoulou_2016, tritsis_2018_sc} and likely formed by the magnetosonic waves \citep{tritsis_2016}. In addition, they assumed that the magnetic field is parallel to the \HI\ structure, as suggested by \cite{clark_2014}. Our atomic and molecular regions coincide with Region 1 and 2 in \cite{tritsis_2019}, and hence we can directly compare our \Bpos\ values with their estimates. 

\cite{tritsis_2019} assumed that $n=10$ \VolDens; this, however, is inconsistent with our \CII\ observations (Sect.~\ref{sec:gas_properties}), which suggest that there the gas density is 50 - 100 \VolDens. We rescaled their magnetic field strength using our density estimates. The magnetic field strength scales as $B_{0} \propto \sqrt{n}$ (Eq.~\ref{eq:st}), and hence the rescaled values are ,
\begin{equation}
    \label{eq:rescaled_b}
    B_{0}^{\rm{resc}} = \sqrt{\frac{n_{\rm{ours}}}{n_{\rm{TR19}}}} B_{0}^{\rm{TR19}},
\end{equation}
where $n_{\rm{ours}}$ is our inferred density estimates, $n_{\rm{TR19}}$ is the density assumed by \cite{tritsis_2019}, and $B_{0}^{\rm{TR19}}$ is their estimated magnetic field strength. In the atomic region, they found that $B_{0}^{\rm{TR19}} = 4^{+1}_{-1}~\mu$G. Using Eq.~(\ref{eq:rescaled_b}) we derive that the rescaled magnetic field strength values are $B_{0}^{\rm{resc}} = 11^{+5}_{-4}~\mu$G; we used our lower and upper $n_{\rm{our}}$ estimates with their lower and upper $B_{0}^{\rm{TR19}}$ limits in order to compute the rescaled magnetic field strength limits. Their estimated values are lower than what we inferred for the same region using our polarization measurements, which is $B_{0} = 25^{+4}_{-4}~\mu$G (Table~\ref{table:bfield_parameters}). 

In the molecular region, \cite{tritsis_2019} estimated that $B_{0}^{\rm{TR19}} = 14^{+7}_{-2}\mu$G. But, we cannot directly compare our estimated values with theirs for two main reasons: Firstly, they assumed that the magnetic field orientation is parallel to the \NHI\ structure, while our polarization data revealed that the orientation of the field is almost perpendicular to the cloud in Figs.~\ref{fig:hi_polarization} and~\ref{fig:rht_result}. Secondly, their estimates are solely based on \HI\ data, while there is significant \Hmolecular\ gas there.

\subsection{Magnetic field morphology and the \HI\ velocity gradients}
\label{sec:disc_velo_grad}

\cite{gonzalez_lazarian_2017} developed a novel technique for studying the ISM magnetized turbulence. It is called the "Velocity Gradient Technique" (VGT) and is based on the properties of strong Alfvénic turbulence. This theory predicts that turbulent eddies form perpendicular to the local magnetic field orientation, and hence developing gas velocity gradients which are perpendicular to the mean magnetic field orientation. The same behaviour can be obtained when large-scale flows of gas  expand perpendicular to the magnetic field lines \citep[e.g.,][]{girichidis_2021}. In either case, when self-gravity takes over the velocity gradients tend to become parallel to the magnetic field \citep[e.g.][]{yuen_ka_ho_2017, yue_hu_2019, girichidis_2021}. 

In our target cloud, we found that the POS magnetic field orientation in the atomic sub-region is perpendicular to the \HI\ velocity gradients, while they tend to align close to the CO clump. We have argued that a large-scale expansion of \HI\ gas has significantly affected the dynamics of our target cloud. Thus, it is more plausible that the magnetic field is perpendicular to the velocity gradients because of a large scale expansion which compresses the magnetic field lines rather than attributing this behavior to Alfvénic distortions as expected from the VGT. The alignment of the \HI\ velocity gradients with the magnetic field close to the CO clump is probably caused by self-gravity. This alignment and the fact that our inferred density for the CO clump is $\sim 400$ \VolDens\ match very well with the results of \cite{girichidis_2021} who suggested that self-gravity starts dominating at $n \sim 400$ \VolDens.

\subsection{The magnetic field strength versus density relation}
\label{subsec:B_rho_relation}

The relation between the magnetic field strength and density is important for determining how dense cores collapse. \cite{cruthcer_2010} found, using Zeeman data, that the maximum magnetic field strength scales as,
\begin{equation}
   \label{eq:B_rho}
    B \propto  
   \begin{cases} 
    \rm{constant}, & \mbox{$n \leq$ 300 cm$^{-3}$}\\ 
    n^{2/3}, & \mbox{$n > $ 300 cm$^{-3}$} \\ 
   \end{cases}.
\end{equation}
They concluded that the $2/3$ scaling is consistent with spherically collapsing cores where the magnetic field has but a minor role in the cloud dynamics. However, their data are dominated by noise and in order to reach this conclusion they relied on several prior assumptions applied within a Bayesian framework (especially the underestimation of the uncertainties of the density estimates used). The 2/3 scaling was questioned by \cite{tritsis_2015} who concluded that the same data are better described by $B \propto n^{1/2}$; this scaling is consistent with the anisotropic core collapse model as presented by \cite{mouschovias_1976a, mouschovias_1976b}. \cite{Hangjin_huabai} also showed that the results of \cite{cruthcer_2010} are debatable since the Zeeman data are too noisy, hence their analysis is dominated by statistical biases. \cite{pattle_2019} compiled archival polarization data and with the DCF method they found that their results are consistent with the scalings in Eq.~(\ref{eq:B_rho}). However, the scatter of $B$ in their sample is an order of magnitude or more in some cases. Recently, \cite{myers_2021} used polarization data towards 16 cores and they also found results consistent with Eq.~(\ref{eq:B_rho}). They concluded that their targeted cores are spherical, but contrary to \cite{cruthcer_2010}, that the magnetic field energy can be relatively large compared to the gravitational energy of the cores. Their analysis is based on the DCF method with the $Q=0.5$ calibration factor found in numerical simulation \citep{ostriker_2001, padoan_2001, heitsch_2001}. However, the assumptions and the regime of applicability of the DCF method was recently questioned by \cite{skalidis_2020} and \cite{skalidis_2021}. They found that DCF can be highly inaccurate in sub-, trans- Alfvénic turbulence even if the $Q$ factor is applied, or if the method is combined with more sophisticated techniques, like the dispersion function analysis of \cite{hildebrand_2009} and \cite{houde_2009}. Altogether, the current observational constraints of the $B - \rho$ relation seem to be vague, because the results are either based on noisy data (Zeeman data, \citealt{cruthcer_2010}) or high quality data (dust polarization) with uncertain methods (DCF, \citealt{pattle_2019} and \citealt{myers_2021}).

Our inferred total magnetic field strength values are a factor of $\sim 3$ higher than the upper limit inferred by \cite{cruthcer_2010} in Eq.~(\ref{eq:B_rho}). The maximum inferred density between the atomic and molecular region increases significantly, from $\sim 100$ to $\sim 400$ \VolDens, but the magnetic field strength remains constant within our uncertainties.\footnote{This is true even if we assumed a prolate 3D shape for the CO clump and its depth being comparable to its minor axis, which is $\sim 0.5$ pc. In that case the estimated density of the CO-bright region would be $\sim 1150$ \VolDens, which yields a magnetic field which is $\sim 25 \mu$G; this is also consistent with the magnetic field strength estimate for the atomic region.} Thus, there seems to be no correlation between the magnetic field strength and the gas density although the \HI\ to \Hmolecular\ transition takes place between the two regions. This indicates that gas kinematics are dominated by flows along the magnetic field lines which accumulate gas and increase the gas density without increasing the strength of the field in this transition cloud.

\section{Conclusion}
\label{sec:conclusions}

We studied the role of the magnetic field in the \HI-\Hmolecular\ transition in a diffuse ISM cloud towards the Ursa Major cirrus. We probed the magnetic field properties of the cloud using optical polarization and the gas properties of the cloud by performing CO spectroscopic observations tracing the J=0-1, J=1-2 lines and the 157.6 $\mu$m \CII\ emission line.

The target \HI\ cloud is filamentary (Fig.~\ref{fig:hi_polarization}). The cloud consists of two sub-regions: 1) the atomic, where gas is mostly atomic, and 2) the molecular, where molecular gas is most abundant. Towards the molecular sub-region, we detected a CO clump and estimated its mean molecular fractional abundance being $f_{H_{2}} \approx 0.8 - 0.95$. We also found that there is significant amount of hot ($T\approx$ 300 - 400 K) and dense ($n \approx$ 350 - 500 \VolDens) CO-dark \Hmolecular\ gas at the edges of the cloud; CO-dark \Hmolecular\ seems to be more  abundant than CO-bright \Hmolecular\ there.

Using our polarization data, we estimated the POS magnetic field strength with the method of \cite{skalidis_2020}. In the atomic sub-region, we estimated the POS magnetic field strength to be $25^{+4}_{-4}$ $\mu$G; this combined with archival \HI\ Zeeman data yields a total magnetic field strength equal to $27^{+5}_{-7}$ $\mu$G. In the molecular sub-region the estimated magnetic field strength is $19^{+12}_{-6}$ $\mu$G, with the total strength being equal to $21^{+12}_{-6}$ $\mu$G, which is consistent to the atomic sub-region estimate within our uncertainties. We found no correlation between the magnetic field strength and the gas density between the two regions. This indicates that flows along the magnetic field lines tend to accumulate gas without affecting the magnetic field strength . The estimated Alfvén Mach number of the cloud is $M_{A} \approx 1$, which indicates that turbulence is trans-Alfvénic.

We compared the POS magnetic field orientation with the \HI\ velocity gradients. We found that in the atomic sub-region the local magnetic field tends to be perpendicular to the \HI\ velocity gradients, while close to the CO clump they tend to become parallel. The alignment between the two quantities probably marks the gravitational infall of the \HI\ gas towards the clump. All these pieces of evidence are consistent with theoretical expectations. The shape of the CO clump is asymmetric with an aspect ratio between its minor and major principal axis equal to 0.4. The mean polarization orientation is closer ($24\degr$ offset) to the short axis of the clump. This also supports that self-gravity is important for the molecular part of the cloud and is consistent with the observation that \HI\ accumulates preferentially along the magnetic field lines.

We compared the POS magnetic field orientation with the \NHI\ morphology of the cloud using the RHT algorithm \citep{clark_2014}. We found that the magnetic field is parallel to the \HI\ structure of the cloud in the atomic region, while it tends to be perpendicular to the \HI\ cloud structure in the molecular sub-region; we provided evidence that this relative shift of the magnetic field orientation was probably caused by expanding gas. On the other hand, the magnetic field tends to form smaller offset angle with the total column density structure of the cloud (\NHtot) in both the atomic and molecular sub-regions. We conclude that the magnetic field morphology of the cloud is better correlated with the \NHtot\ rather than the \NHI\ structure of the cloud. 

Our target cloud lies at the edges of the NCPL, which is a large-scale expanding structure. Using archival data, we explored if there is a correlation between the relative difference of the magnetic field with the \HI\ velocity gradients and the molecular abundance, as suggested by simulations \citep[e.g.,][]{girichidis_2021}. We provided evidence that the molecular abundance is enhanced in regions where the magnetic field is parallel to the \HI\ velocity gradients, probably because these are regions where self-gravity dominates. However, these results should be interpreted with caution since projection effects at these scales are important and more work is required to further constrain this scenario. 

\bibliographystyle{aa}
\bibliography{bibliography}

\begin{thebibliography}{130}
\expandafter\ifx\csname natexlab\endcsname\relax\def\natexlab#1{#1}\fi

\bibitem[{{Alina} {et~al.}(2019){Alina}, {Ristorcelli}, {Montier},
  {Abdikamalov}, {Juvela}, {Ferri{\`e}re}, {Bernard}, \&
  {Micelotta}}]{alina_2019}
{Alina}, D., {Ristorcelli}, I., {Montier}, L., {et~al.} 2019, \mnras, 485, 2825

\bibitem[{{Andersson} {et~al.}(2015){Andersson}, {Lazarian}, \&
  {Vaillancourt}}]{andersson_review}
{Andersson}, B.~G., {Lazarian}, A., \& {Vaillancourt}, J.~E. 2015, \araa, 53,
  501

\bibitem[{{Bailer-Jones} {et~al.}(2021){Bailer-Jones}, {Rybizki}, {Fouesneau},
  {Demleitner}, \& {Andrae}}]{bailer_jones_2021}
{Bailer-Jones}, C.~A.~L., {Rybizki}, J., {Fouesneau}, M., {Demleitner}, M., \&
  {Andrae}, R. 2021, \aj, 161, 147

\bibitem[{{Barinovs} {et~al.}(2005){Barinovs}, {van Hemert}, {Krems}, \&
  {Dalgarno}}]{barinovs_2005}
{Barinovs}, {\u{G}}., {van Hemert}, M.~C., {Krems}, R., \& {Dalgarno}, A. 2005,
  \apj, 620, 537

\bibitem[{{Barriault} {et~al.}(2011){Barriault}, {Joncas}, \&
  {Plume}}]{barriault_2011_co}
{Barriault}, L., {Joncas}, G., \& {Plume}, R. 2011, \mnras, 416, 1250

\bibitem[{{Beattie} {et~al.}(2020){Beattie}, {Federrath}, \&
  {Seta}}]{beattie_2020}
{Beattie}, J.~R., {Federrath}, C., \& {Seta}, A. 2020, \mnras, 498, 1593

\bibitem[{{Bellomi} {et~al.}(2020){Bellomi}, {Godard}, {Hennebelle},
  {Valdivia}, {Pineau des For{\^e}ts}, {Lesaffre}, \&
  {P{\'e}rault}}]{bellomi_2020}
{Bellomi}, E., {Godard}, B., {Hennebelle}, P., {et~al.} 2020, \aap, 643, A36

\bibitem[{{Bialy} {et~al.}(2017){Bialy}, {Burkhart}, \&
  {Sternberg}}]{Bialy2017}
{Bialy}, S., {Burkhart}, B., \& {Sternberg}, A. 2017, \apj, 843, 92

\bibitem[{{Bieging} \& {Peters}(2011)}]{Bieging_2011}
{Bieging}, J.~H. \& {Peters}, W.~L. 2011, \apjs, 196, 18

\bibitem[{{Bieging} {et~al.}(2010){Bieging}, {Peters}, \&
  {Kang}}]{Bieging_2010}
{Bieging}, J.~H., {Peters}, W.~L., \& {Kang}, M. 2010, \apjs, 191, 232

\bibitem[{{Bieging} {et~al.}(2014){Bieging}, {Revelle}, \&
  {Peters}}]{Bieging_2014}
{Bieging}, J.~H., {Revelle}, M., \& {Peters}, W.~L. 2014, \apjs, 214, 7

\bibitem[{{Blagrave} {et~al.}(2017){Blagrave}, {Martin}, {Joncas}, {Kothes},
  {Stil}, {Miville-Desch{\^e}nes}, {Lockman}, \& {Taylor}}]{DHIGLS}
{Blagrave}, K., {Martin}, P.~G., {Joncas}, G., {et~al.} 2017, \apj, 834, 126

\bibitem[{{Blinov} {et~al.}(2021){Blinov}, {Kiehlmann}, {Pavlidou},
  {Panopoulou}, {Skalidis}, {Angelakis}, {Casadio}, {Einoder}, {Hovatta},
  {Kokolakis}, {Kougentakis}, {Kus}, {Kylafis}, {Kyritsis}, {Lalakos},
  {Liodakis}, {Maharana}, {Makrydopoulou}, {Mandarakas}, {Maragkakis},
  {Myserlis}, {Papadakis}, {Paterakis}, {Pearson}, {Ramaprakash}, {Readhead},
  {Reig}, {S{\l}owikowska}, {Tassis}, {Xexakis}, {{\.Z}ejmo}, \&
  {Zensus}}]{blinov_2021}
{Blinov}, D., {Kiehlmann}, S., {Pavlidou}, V., {et~al.} 2021, \mnras, 501, 3715

\bibitem[{Bolatto {et~al.}(2013)Bolatto, Wolfire, \& Leroy}]{bolato_2013}
Bolatto, A.~D., Wolfire, M., \& Leroy, A.~K. 2013, Annual Review of Astronomy
  and Astrophysics, 51, 207

\bibitem[{{Bracco} {et~al.}(2020){Bracco}, {Benjamin}, {Alves}, {Lehmann},
  {Boulanger}, {Montier}, {Mittelman}, {di Cicco}, \& {Walker}}]{braco_2020}
{Bracco}, A., {Benjamin}, R.~A., {Alves}, M.~I.~R., {et~al.} 2020, \aap, 636,
  L8

\bibitem[{{Chandrasekhar} \& {Fermi}(1953)}]{chandra_fermi}
{Chandrasekhar}, S. \& {Fermi}, E. 1953, \apj, 118, 113

\bibitem[{Chen {et~al.}(2016)Chen, King, \& Li}]{Chen_2016}
Chen, C.-Y., King, P.~K., \& Li, Z.-Y. 2016, The Astrophysical Journal, 829, 84

\bibitem[{{Chen} \& {Ostriker}(2014)}]{chen_ostriker_2014}
{Chen}, C.-Y. \& {Ostriker}, E.~C. 2014, \apj, 785, 69

\bibitem[{{Chen} \& {Ostriker}(2015)}]{chen_ostriker_2015}
{Chen}, C.-Y. \& {Ostriker}, E.~C. 2015, \apj, 810, 126

\bibitem[{{Clark} \& {Hensley}(2019)}]{clark_2019}
{Clark}, S.~E. \& {Hensley}, B.~S. 2019, \apj, 887, 136

\bibitem[{{Clark} {et~al.}(2015){Clark}, {Hill}, {Peek}, {Putman}, \&
  {Babler}}]{clark_2015}
{Clark}, S.~E., {Hill}, J.~C., {Peek}, J.~E.~G., {Putman}, M.~E., \& {Babler},
  B.~L. 2015, \prl, 115, 241302

\bibitem[{{Clark} {et~al.}(2014){Clark}, {Peek}, \& {Putman}}]{clark_2014}
{Clark}, S.~E., {Peek}, J.~E.~G., \& {Putman}, M.~E. 2014, \apj, 789, 82

\bibitem[{Colditz {et~al.}(2012)Colditz, Fumi, Geis, Hönle, Klein, Krabbe,
  Looney, Poglitsch, Raab, Savage, Rebell, \& Fischer}]{colditz}
Colditz, S., Fumi, F., Geis, N., {et~al.} 2012, in Ground-based and Airborne
  Instrumentation for Astronomy IV, ed. I.~S. McLean, S.~K. Ramsay, \&
  H.~Takami, Vol. 8446, International Society for Optics and Photonics (SPIE),
  392 -- 402

\bibitem[{{Crutcher} {et~al.}(2004){Crutcher}, {Nutter}, {Ward-Thompson}, \&
  {Kirk}}]{crutcher_2004}
{Crutcher}, R.~M., {Nutter}, D.~J., {Ward-Thompson}, D., \& {Kirk}, J.~M. 2004,
  \apj, 600, 279

\bibitem[{{Crutcher} {et~al.}(2010){Crutcher}, {Wandelt}, {Heiles},
  {Falgarone}, \& {Troland}}]{cruthcer_2010}
{Crutcher}, R.~M., {Wandelt}, B., {Heiles}, C., {Falgarone}, E., \& {Troland},
  T.~H. 2010, \apj, 725, 466

\bibitem[{Davis(1951)}]{davis_1951}
Davis, L. 1951, Phys. Rev., 81, 890

\bibitem[{{Dawson} {et~al.}(2011){Dawson}, {McClure-Griffiths}, {Kawamura},
  {Mizuno}, {Onishi}, {Mizuno}, \& {Fukui}}]{dawson_2011}
{Dawson}, J.~R., {McClure-Griffiths}, N.~M., {Kawamura}, A., {et~al.} 2011,
  \apj, 728, 127

\bibitem[{Dickey \& Lockman(1990)}]{dickey_lockman}
Dickey, J.~M. \& Lockman, F.~J. 1990, Annual Review of Astronomy and
  Astrophysics, 28, 215

\bibitem[{{Fadda} \& {Chambers}(2018)}]{sospex}
{Fadda}, D. \& {Chambers}, E.~T. 2018, in American Astronomical Society Meeting
  Abstracts, Vol. 231, American Astronomical Society Meeting Abstracts \#231,
  150.11

\bibitem[{{Fischer} {et~al.}(2018){Fischer}, {Beckmann}, {Bryant}, {Colditz},
  {Fumi}, {Geis}, {Hamidouche}, {Henning}, {H{\"o}nle}, {Iserlohe}, {Klein},
  {Krabbe}, {Looney}, {Poglitsch}, {Raab}, {Rebell}, {Rosenthal}, {Savage},
  {Schweitzer}, {Trinh}, \& {Vacca}}]{fisher_2018}
{Fischer}, C., {Beckmann}, S., {Bryant}, A., {et~al.} 2018, Journal of
  Astronomical Instrumentation, 7, 1840003

\bibitem[{{Fissel} {et~al.}(2019){Fissel}, {Ade}, {Angil{\`e}}, {Ashton},
  {Benton}, {Chen}, {Cunningham}, {Devlin}, {Dober}, {Friesen}, {Fukui},
  {Galitzki}, {Gandilo}, {Goodman}, {Green}, {Jones}, {Klein}, {King},
  {Korotkov}, {Li}, {Lowe}, {Martin}, {Matthews}, {Moncelsi}, {Nakamura},
  {Netterfield}, {Newmark}, {Novak}, {Pascale}, {Poidevin}, {Santos}, {Savini},
  {Scott}, {Shariff}, {Soler}, {Thomas}, {Tucker}, {Tucker}, {Ward-Thompson},
  \& {Zucker}}]{fissel_2019}
{Fissel}, L.~M., {Ade}, P. A.~R., {Angil{\`e}}, F.~E., {et~al.} 2019, \apj,
  878, 110

\bibitem[{{Gaia Collaboration} {et~al.}(2021){Gaia Collaboration}, {Brown},
  {Vallenari}, {Prusti}, {de Bruijne}, {Babusiaux}, {Biermann}, {Creevey},
  {Evans}, {Eyer}, {Hutton}, {Jansen}, {Jordi}, {Klioner}, {Lammers},
  {Lindegren}, {Luri}, {Mignard}, {Panem}, {Pourbaix}, {Randich}, {Sartoretti},
  {Soubiran}, {Walton}, {Arenou}, {Bailer-Jones}, {Bastian}, {Cropper},
  {Drimmel}, {Katz}, {Lattanzi}, {van Leeuwen}, {Bakker}, {Cacciari},
  {Casta{\~n}eda}, {De Angeli}, {Ducourant}, {Fabricius}, {Fouesneau},
  {Fr{\'e}mat}, {Guerra}, {Guerrier}, {Guiraud}, {Jean-Antoine Piccolo},
  {Masana}, {Messineo}, {Mowlavi}, {Nicolas}, {Nienartowicz}, {Pailler},
  {Panuzzo}, {Riclet}, {Roux}, {Seabroke}, {Sordo}, {Tanga}, {Th{\'e}venin},
  {Gracia-Abril}, {Portell}, {Teyssier}, {Altmann}, {Andrae}, {Bellas-Velidis},
  {Benson}, {Berthier}, {Blomme}, {Brugaletta}, {Burgess}, {Busso}, {Carry},
  {Cellino}, {Cheek}, {Clementini}, {Damerdji}, {Davidson}, {Delchambre},
  {Dell'Oro}, {Fern{\'a}ndez-Hern{\'a}ndez}, {Galluccio}, {Garc{\'\i}a-Lario},
  {Garcia-Reinaldos}, {Gonz{\'a}lez-N{\'u}{\~n}ez}, {Gosset}, {Haigron},
  {Halbwachs}, {Hambly}, {Harrison}, {Hatzidimitriou}, {Heiter},
  {Hern{\'a}ndez}, {Hestroffer}, {Hodgkin}, {Holl}, {Jan{\ss}en}, {Jevardat de
  Fombelle}, {Jordan}, {Krone-Martins}, {Lanzafame}, {L{\"o}ffler}, {Lorca},
  {Manteiga}, {Marchal}, {Marrese}, {Moitinho}, {Mora}, {Muinonen}, {Osborne},
  {Pancino}, {Pauwels}, {Petit}, {Recio-Blanco}, {Richards}, {Riello},
  {Rimoldini}, {Robin}, {Roegiers}, {Rybizki}, {Sarro}, {Siopis}, {Smith},
  {Sozzetti}, {Ulla}, {Utrilla}, {van Leeuwen}, {van Reeven}, {Abbas}, {Abreu
  Aramburu}, {Accart}, {Aerts}, {Aguado}, {Ajaj}, {Altavilla}, {{\'A}lvarez},
  {{\'A}lvarez Cid-Fuentes}, {Alves}, {Anderson}, {Anglada Varela}, {Antoja},
  {Audard}, {Baines}, {Baker}, {Balaguer-N{\'u}{\~n}ez}, {Balbinot}, {Balog},
  {Barache}, {Barbato}, {Barros}, {Barstow}, {Bartolom{\'e}}, {Bassilana},
  {Bauchet}, {Baudesson-Stella}, {Becciani}, {Bellazzini}, {Bernet}, {Bertone},
  {Bianchi}, {Blanco-Cuaresma}, {Boch}, {Bombrun}, {Bossini}, {Bouquillon},
  {Bragaglia}, {Bramante}, {Breedt}, {Bressan}, {Brouillet}, {Bucciarelli},
  {Burlacu}, {Busonero}, {Butkevich}, {Buzzi}, {Caffau}, {Cancelliere},
  {C{\'a}novas}, {Cantat-Gaudin}, {Carballo}, {Carlucci}, {Carnerero},
  {Carrasco}, {Casamiquela}, {Castellani}, {Castro-Ginard}, {Castro Sampol},
  {Chaoul}, {Charlot}, {Chemin}, {Chiavassa}, {Cioni}, {Comoretto}, {Cooper},
  {Cornez}, {Cowell}, {Crifo}, {Crosta}, {Crowley}, {Dafonte}, {Dapergolas},
  {David}, {David}, {de Laverny}, {De Luise}, {De March}, {De Ridder}, {de
  Souza}, {de Teodoro}, {de Torres}, {del Peloso}, {del Pozo}, {Delbo},
  {Delgado}, {Delgado}, {Delisle}, {Di Matteo}, {Diakite}, {Diener},
  {Distefano}, {Dolding}, {Eappachen}, {Edvardsson}, {Enke}, {Esquej}, {Fabre},
  {Fabrizio}, {Faigler}, {Fedorets}, {Fernique}, {Fienga}, {Figueras},
  {Fouron}, {Fragkoudi}, {Fraile}, {Franke}, {Gai}, {Garabato},
  {Garcia-Gutierrez}, {Garc{\'\i}a-Torres}, {Garofalo}, {Gavras}, {Gerlach},
  {Geyer}, {Giacobbe}, {Gilmore}, {Girona}, {Giuffrida}, {Gomel}, {Gomez},
  {Gonzalez-Santamaria}, {Gonz{\'a}lez-Vidal}, {Granvik},
  {Guti{\'e}rrez-S{\'a}nchez}, {Guy}, {Hauser}, {Haywood}, {Helmi}, {Hidalgo},
  {Hilger}, {H{\l}adczuk}, {Hobbs}, {Holland}, {Huckle}, {Jasniewicz},
  {Jonker}, {Juaristi Campillo}, {Julbe}, {Karbevska}, {Kervella}, {Khanna},
  {Kochoska}, {Kontizas}, {Kordopatis}, {Korn}, {Kostrzewa-Rutkowska},
  {Kruszy{\'n}ska}, {Lambert}, {Lanza}, {Lasne}, {Le Campion}, {Le Fustec},
  {Lebreton}, {Lebzelter}, {Leccia}, {Leclerc}, {Lecoeur-Taibi}, {Liao},
  {Licata}, {Lindstr{\o}m}, {Lister}, {Livanou}, {Lobel}, {Madrero Pardo},
  {Managau}, {Mann}, {Marchant}, {Marconi}, {Marcos Santos}, {Marinoni},
  {Marocco}, {Marshall}, {Martin Polo}, {Mart{\'\i}n-Fleitas}, {Masip},
  {Massari}, {Mastrobuono-Battisti}, {Mazeh}, {McMillan}, {Messina},
  {Michalik}, {Millar}, {Mints}, {Molina}, {Molinaro}, {Moln{\'a}r},
  {Montegriffo}, {Mor}, {Morbidelli}, {Morel}, {Morris}, {Mulone}, {Munoz},
  {Muraveva}, {Murphy}, {Musella}, {Noval}, {Ord{\'e}novic}, {Orr{\`u}},
  {Osinde}, {Pagani}, {Pagano}, {Palaversa}, {Palicio}, {Panahi}, {Pawlak},
  {Pe{\~n}alosa Esteller}, {Penttil{\"a}}, {Piersimoni}, {Pineau}, {Plachy},
  {Plum}, {Poggio}, {Poretti}, {Poujoulet}, {Pr{\v{s}}a}, {Pulone}, {Racero},
  {Ragaini}, {Rainer}, {Raiteri}, {Rambaux}, {Ramos}, {Ramos-Lerate}, {Re
  Fiorentin}, {Regibo}, {Reyl{\'e}}, {Ripepi}, {Riva}, {Rixon}, {Robichon},
  {Robin}, {Roelens}, {Rohrbasser}, {Romero-G{\'o}mez}, {Rowell}, {Royer},
  {Rybicki}, {Sadowski}, {Sagrist{\`a} Sell{\'e}s}, {Sahlmann}, {Salgado},
  {Salguero}, {Samaras}, {Sanchez Gimenez}, {Sanna}, {Santove{\~n}a},
  {Sarasso}, {Schultheis}, {Sciacca}, {Segol}, {Segovia}, {S{\'e}gransan},
  {Semeux}, {Shahaf}, {Siddiqui}, {Siebert}, {Siltala}, {Slezak}, {Smart},
  {Solano}, {Solitro}, {Souami}, {Souchay}, {Spagna}, {Spoto}, {Steele},
  {Steidelm{\"u}ller}, {Stephenson}, {S{\"u}veges}, {Szabados}, {Szegedi-Elek},
  {Taris}, {Tauran}, {Taylor}, {Teixeira}, {Thuillot}, {Tonello}, {Torra},
  {Torra}, {Turon}, {Unger}, {Vaillant}, {van Dillen}, {Vanel}, {Vecchiato},
  {Viala}, {Vicente}, {Voutsinas}, {Weiler}, {Wevers}, {Wyrzykowski}, {Yoldas},
  {Yvard}, {Zhao}, {Zorec}, {Zucker}, {Zurbach}, \& {Zwitter}}]{gaia_dr3}
{Gaia Collaboration}, {Brown}, A.~G.~A., {Vallenari}, A., {et~al.} 2021, \aap,
  649, A1

\bibitem[{{Gillmon} {et~al.}(2006){Gillmon}, {Shull}, {Tumlinson}, \&
  {Danforth}}]{Gillmon2006}
{Gillmon}, K., {Shull}, J.~M., {Tumlinson}, J., \& {Danforth}, C. 2006, \apj,
  636, 891

\bibitem[{{Girichidis}(2021)}]{girichidis_2021}
{Girichidis}, P. 2021, \mnras [\eprint[arXiv]{2106.12596}]

\bibitem[{{Girichidis} {et~al.}(2016){Girichidis}, {Walch}, {Naab}, {Gatto},
  {W{\"u}nsch}, {Glover}, {Klessen}, {Clark}, {Peters}, {Derigs}, \&
  {Baczynski}}]{silc_II}
{Girichidis}, P., {Walch}, S., {Naab}, T., {et~al.} 2016, \mnras, 456, 3432

\bibitem[{{Goldsmith}(2013)}]{goldsmith_2013}
{Goldsmith}, P.~F. 2013, \apj, 774, 134

\bibitem[{{Goldsmith} {et~al.}(2008){Goldsmith}, {Heyer}, {Narayanan}, {Snell},
  {Li}, \& {Brunt}}]{goldsmith_2008}
{Goldsmith}, P.~F., {Heyer}, M., {Narayanan}, G., {et~al.} 2008, \apj, 680, 428

\bibitem[{{Goldsmith} {et~al.}(2012){Goldsmith}, {Langer}, {Pineda}, \&
  {Velusamy}}]{goldsmith_2012}
{Goldsmith}, P.~F., {Langer}, W.~D., {Pineda}, J.~L., \& {Velusamy}, T. 2012,
  \apjs, 203, 13

\bibitem[{{Goldsmith} {et~al.}(2016){Goldsmith}, {Pineda}, {Langer}, {Liu},
  {Requena-Torres}, {Ricken}, \& {Riquelme}}]{goldsmith_2016}
{Goldsmith}, P.~F., {Pineda}, J.~L., {Langer}, W.~D., {et~al.} 2016, \apj, 824,
  141

\bibitem[{{Goldsmith} {et~al.}(2018){Goldsmith}, {Pineda}, {Neufeld},
  {Wolfire}, {Risacher}, \& {Simon}}]{goldsmith_2018}
{Goldsmith}, P.~F., {Pineda}, J.~L., {Neufeld}, D.~A., {et~al.} 2018, \apj,
  856, 96

\bibitem[{{Gonz{\'a}lez-Casanova} \& {Lazarian}(2017)}]{gonzalez_lazarian_2017}
{Gonz{\'a}lez-Casanova}, D.~F. \& {Lazarian}, A. 2017, \apj, 835, 41

\bibitem[{{Green} {et~al.}(2018){Green}, {Schlafly}, {Finkbeiner}, {Rix},
  {Martin}, {Burgett}, {Draper}, {Flewelling}, {Hodapp}, {Kaiser}, {Kudritzki},
  {Magnier}, {Metcalfe}, {Tonry}, {Wainscoat}, \& {Waters}}]{green_2018}
{Green}, G.~M., {Schlafly}, E.~F., {Finkbeiner}, D., {et~al.} 2018, \mnras,
  478, 651

\bibitem[{{Grenier} {et~al.}(2005){Grenier}, {Casandjian}, \&
  {Terrier}}]{Grenier_2005}
{Grenier}, I.~A., {Casandjian}, J.-M., \& {Terrier}, R. 2005, Science, 307,
  1292

\bibitem[{{Heiles}(1989)}]{heiles_1989}
{Heiles}, C. 1989, \apj, 336, 808

\bibitem[{{Heiles} \& {Troland}(2003)}]{heiles_troland1}
{Heiles}, C. \& {Troland}, T.~H. 2003, \apjs, 145, 329

\bibitem[{{Heitsch} {et~al.}(2009){Heitsch}, {Stone}, \&
  {Hartmann}}]{heitsch_2009}
{Heitsch}, F., {Stone}, J.~M., \& {Hartmann}, L.~W. 2009, \apj, 695, 248

\bibitem[{{Heitsch} {et~al.}(2001){Heitsch}, {Zweibel}, {Mac Low}, {Li}, \&
  {Norman}}]{heitsch_2001}
{Heitsch}, F., {Zweibel}, E.~G., {Mac Low}, M.-M., {Li}, P., \& {Norman}, M.~L.
  2001, \apj, 561, 800

\bibitem[{{HI4PI Collaboration} {et~al.}(2016){HI4PI Collaboration}, {Ben
  Bekhti}, {Fl{\"o}er}, {Keller}, {Kerp}, {Lenz}, {Winkel}, {Bailin},
  {Calabretta}, {Dedes}, {Ford}, {Gibson}, {Haud}, {Janowiecki}, {Kalberla},
  {Lockman}, {McClure-Griffiths}, {Murphy}, {Nakanishi}, {Pisano}, \&
  {Staveley-Smith}}]{hi4pi}
{HI4PI Collaboration}, {Ben Bekhti}, N., {Fl{\"o}er}, L., {et~al.} 2016, \aap,
  594, A116

\bibitem[{{Hildebrand} {et~al.}(2009){Hildebrand}, {Kirby}, {Dotson}, {Houde},
  \& {Vaillancourt}}]{hildebrand_2009}
{Hildebrand}, R.~H., {Kirby}, L., {Dotson}, J.~L., {Houde}, M., \&
  {Vaillancourt}, J.~E. 2009, \apj, 696, 567

\bibitem[{{Houde} {et~al.}(2009){Houde}, {Vaillancourt}, {Hildebrand},
  {Chitsazzadeh}, \& {Kirby}}]{houde_2009}
{Houde}, M., {Vaillancourt}, J.~E., {Hildebrand}, R.~H., {Chitsazzadeh}, S., \&
  {Kirby}, L. 2009, \apj, 706, 1504

\bibitem[{{Hu} {et~al.}(2020){Hu}, {Lazarian}, \& {Yuen}}]{hu_2020_vgt_gravity}
{Hu}, Y., {Lazarian}, A., \& {Yuen}, K.~H. 2020, \apj, 897, 123

\bibitem[{{Hu} {et~al.}(2019){Hu}, {Yuen}, {Lazarian}, {Ho}, {Benjamin},
  {Hill}, {Lockman}, {Goldsmith}, \& {Lazarian}}]{yue_hu_2019}
{Hu}, Y., {Yuen}, K.~H., {Lazarian}, V., {et~al.} 2019, Nature Astronomy, 3,
  776

\bibitem[{{Ingalls} {et~al.}(2002){Ingalls}, {Reach}, \&
  {Bania}}]{inglalls_2002}
{Ingalls}, J.~G., {Reach}, W.~T., \& {Bania}, T.~M. 2002, \apj, 579, 289

\bibitem[{{Inoue} \& {Inutsuka}(2008)}]{inoue_2008}
{Inoue}, T. \& {Inutsuka}, S.-i. 2008, \apj, 687, 303

\bibitem[{{Inoue} \& {Inutsuka}(2009)}]{inoue_2009}
{Inoue}, T. \& {Inutsuka}, S.-i. 2009, \apj, 704, 161

\bibitem[{{Jiang} {et~al.}(2020){Jiang}, {Li}, \& {Fan}}]{Hangjin_huabai}
{Jiang}, H., {Li}, H.-b., \& {Fan}, X. 2020, \apj, 890, 153

\bibitem[{{Kalberla} {et~al.}(2020){Kalberla}, {Kerp}, \&
  {Haud}}]{kalberla_2020}
{Kalberla}, P.~M.~W., {Kerp}, J., \& {Haud}, U. 2020, \aap, 639, A26

\bibitem[{{Kalberla} {et~al.}(2016){Kalberla}, {Kerp}, {Haud}, {Winkel}, {Ben
  Bekhti}, {Fl{\"o}er}, \& {Lenz}}]{kalberla_2016}
{Kalberla}, P.~M.~W., {Kerp}, J., {Haud}, U., {et~al.} 2016, \apj, 821, 117

\bibitem[{{King} {et~al.}(2014){King}, {Blinov}, {Ramaprakash}, {Myserlis},
  {Angelakis}, {Balokovi{\'c}}, {Feiler}, {Fuhrmann}, {Hovatta}, {Khodade},
  {Kougentakis}, {Kylafis}, {Kus}, {Modi}, {Paleologou}, {Panopoulou},
  {Papadakis}, {Papamastorakis}, {Paterakis}, {Pavlidou}, {Pazderska},
  {Pazderski}, {Pearson}, {Rajarshi}, {Readhead}, {Reig}, {Steiakaki},
  {Tassis}, \& {Zensus}}]{king_2014}
{King}, O.~G., {Blinov}, D., {Ramaprakash}, A.~N., {et~al.} 2014, \mnras, 442,
  1706

\bibitem[{Klein {et~al.}(2014)Klein, Beckmann, Bryant, Colditz, Fischer, Fumi,
  Geis, Hönle, Krabbe, Looney, Poglitsch, Raab, Rebell, \& Savage}]{klein}
Klein, R., Beckmann, S., Bryant, A., {et~al.} 2014, in Ground-based and
  Airborne Instrumentation for Astronomy V, ed. S.~K. Ramsay, I.~S. McLean, \&
  H.~Takami, Vol. 9147, International Society for Optics and Photonics (SPIE),
  1018 -- 1025

\bibitem[{{K{\"o}rtgen} \& {Soler}(2020)}]{kroetgen_2020}
{K{\"o}rtgen}, B. \& {Soler}, J.~D. 2020, \mnras, 499, 4785

\bibitem[{{Krumholz} {et~al.}(2007){Krumholz}, {Stone}, \&
  {Gardiner}}]{Krumholz_2007}
{Krumholz}, M.~R., {Stone}, J.~M., \& {Gardiner}, T.~A. 2007, \apj, 671, 518

\bibitem[{{Kutner} \& {Ulich}(1981)}]{kutner_1981}
{Kutner}, M.~L. \& {Ulich}, B.~L. 1981, \apj, 250, 341

\bibitem[{{Langer} {et~al.}(2010){Langer}, {Velusamy}, {Pineda}, {Goldsmith},
  {Li}, \& {Yorke}}]{langer_2010}
{Langer}, W.~D., {Velusamy}, T., {Pineda}, J.~L., {et~al.} 2010, \aap, 521, L17

\bibitem[{{Lazarian} \& {Yuen}(2018)}]{lazarian_2018}
{Lazarian}, A. \& {Yuen}, K.~H. 2018, \apj, 853, 96

\bibitem[{{Lord}(1992)}]{Lord_atran}
{Lord}, S.~D. 1992, {A new software tool for computing Earth's atmospheric
  transmission of near- and far-infrared radiation}, NASA Technical Memorandum
  103957

\bibitem[{{Marchal} {et~al.}(2019){Marchal}, {Miville-Desch{\^e}nes}, {Orieux},
  {Gac}, {Soussen}, {Lesot}, {d'Allonnes}, \& {Salom{\'e}}}]{rohsa}
{Marchal}, A., {Miville-Desch{\^e}nes}, M.-A., {Orieux}, F., {et~al.} 2019,
  \aap, 626, A101

\bibitem[{{McCullough} \& {Benjamin}(2001)}]{McCullough_2001}
{McCullough}, P.~R. \& {Benjamin}, R.~A. 2001, \aj, 122, 1500

\bibitem[{{Meyerdierks} {et~al.}(1991){Meyerdierks}, {Heithausen}, \&
  {Reif}}]{meyerdierks_1990}
{Meyerdierks}, H., {Heithausen}, A., \& {Reif}, K. 1991, \aap, 245, 247

\bibitem[{{Miville-Desch{\^e}nes} {et~al.}(2002){Miville-Desch{\^e}nes},
  {Boulanger}, {Joncas}, \& {Falgarone}}]{miville_2002}
{Miville-Desch{\^e}nes}, M.~A., {Boulanger}, F., {Joncas}, G., \& {Falgarone},
  E. 2002, \aap, 381, 209

\bibitem[{{Miville-Desch{\^e}nes}
  {et~al.}(2003{\natexlab{a}}){Miville-Desch{\^e}nes}, {Joncas}, {Falgarone},
  \& {Boulanger}}]{miville_2003}
{Miville-Desch{\^e}nes}, M.~A., {Joncas}, G., {Falgarone}, E., \& {Boulanger},
  F. 2003{\natexlab{a}}, \aap, 411, 109

\bibitem[{{Miville-Desch{\^e}nes}
  {et~al.}(2003{\natexlab{b}}){Miville-Desch{\^e}nes}, {Levrier}, \&
  {Falgarone}}]{miville_2003_fbm}
{Miville-Desch{\^e}nes}, M.~A., {Levrier}, F., \& {Falgarone}, E.
  2003{\natexlab{b}}, \apj, 593, 831

\bibitem[{{Monet} {et~al.}(2003){Monet}, {Levine}, {Canzian}, {Ables}, {Bird},
  {Dahn}, {Guetter}, {Harris}, {Henden}, {Leggett}, {Levison}, {Luginbuhl},
  {Martini}, {Monet}, {Munn}, {Pier}, {Rhodes}, {Riepe}, {Sell}, {Stone},
  {Vrba}, {Walker}, {Westerhout}, {Brucato}, {Reid}, {Schoening}, {Hartley},
  {Read}, \& {Tritton}}]{USNOB}
{Monet}, D.~G., {Levine}, S.~E., {Canzian}, B., {et~al.} 2003, \aj, 125, 984

\bibitem[{{Mouschovias}(1976{\natexlab{a}})}]{mouschovias_1976a}
{Mouschovias}, T.~C. 1976{\natexlab{a}}, \apj, 206, 753

\bibitem[{{Mouschovias}(1976{\natexlab{b}})}]{mouschovias_1976b}
{Mouschovias}, T.~C. 1976{\natexlab{b}}, \apj, 207, 141

\bibitem[{{Mouschovias}(1978)}]{mouschovias_1978}
{Mouschovias}, T.~C. 1978, in IAU Colloq. 52: Protostars and Planets, ed.
  T.~{Gehrels} \& M.~S. {Matthews}, 209

\bibitem[{{Myers} \& {Basu}(2021)}]{myers_2021}
{Myers}, P.~C. \& {Basu}, S. 2021, \apj, 917, 35

\bibitem[{{Myers} {et~al.}(1995){Myers}, {Goodman}, {Gusten}, \&
  {Heiles}}]{myers_1995}
{Myers}, P.~C., {Goodman}, A.~A., {Gusten}, R., \& {Heiles}, C. 1995, \apj,
  442, 177

\bibitem[{{Naghizadeh-Khouei} \& {Clarke}(1993)}]{Naghizadeh_clarke_1993}
{Naghizadeh-Khouei}, J. \& {Clarke}, D. 1993, \aap, 274, 968

\bibitem[{{Ntormousi} {et~al.}(2017){Ntormousi}, {Dawson}, {Hennebelle}, \&
  {Fierlinger}}]{ntormousi_2017}
{Ntormousi}, E., {Dawson}, J.~R., {Hennebelle}, P., \& {Fierlinger}, K. 2017,
  \aap, 599, A94

\bibitem[{{Ostriker} {et~al.}(2001){Ostriker}, {Stone}, \&
  {Gammie}}]{ostriker_2001}
{Ostriker}, E.~C., {Stone}, J.~M., \& {Gammie}, C.~F. 2001, \apj, 546, 980

\bibitem[{{Padoan} {et~al.}(2001){Padoan}, {Goodman}, {Draine}, {Juvela},
  {Nordlund}, \& {R{\"o}gnvaldsson}}]{padoan_2001}
{Padoan}, P., {Goodman}, A., {Draine}, B.~T., {et~al.} 2001, \apj, 559, 1005

\bibitem[{{Panopoulou} {et~al.}(2015){Panopoulou}, {Tassis}, {Blinov},
  {Pavlidou}, {King}, {Paleologou}, {Ramaprakash}, {Angelakis},
  {Balokovi{\'c}}, {Das}, {Feiler}, {Hovatta}, {Khodade}, {Kiehlmann}, {Kus},
  {Kylafis}, {Liodakis}, {Mahabal}, {Modi}, {Myserlis}, {Papadakis},
  {Papamastorakis}, {Pazderska}, {Pazderski}, {Pearson}, {Rajarshi},
  {Readhead}, {Reig}, \& {Zensus}}]{panopoulou_2015}
{Panopoulou}, G., {Tassis}, K., {Blinov}, D., {et~al.} 2015, \mnras, 452, 715

\bibitem[{{Panopoulou} {et~al.}(2019{\natexlab{a}}){Panopoulou}, {Hensley},
  {Skalidis}, {Blinov}, \& {Tassis}}]{panopoulou_2019_extreme}
{Panopoulou}, G.~V., {Hensley}, B.~S., {Skalidis}, R., {Blinov}, D., \&
  {Tassis}, K. 2019{\natexlab{a}}, \aap, 624, L8

\bibitem[{{Panopoulou} {et~al.}(2016){Panopoulou}, {Psaradaki}, \&
  {Tassis}}]{panopoulou_2016}
{Panopoulou}, G.~V., {Psaradaki}, I., \& {Tassis}, K. 2016, \mnras, 462, 1517

\bibitem[{{Panopoulou} {et~al.}(2019{\natexlab{b}}){Panopoulou}, {Tassis},
  {Skalidis}, {Blinov}, {Liodakis}, {Pavlidou}, {Potter}, {Ramaprakash},
  {Readhead}, \& {Wehus}}]{panopoulou_2019_tom}
{Panopoulou}, G.~V., {Tassis}, K., {Skalidis}, R., {et~al.} 2019{\natexlab{b}},
  \apj, 872, 56

\bibitem[{{Pattle} \& {Fissel}(2019)}]{pattle_2019}
{Pattle}, K. \& {Fissel}, L. 2019, Frontiers in Astronomy and Space Sciences,
  6, 15

\bibitem[{{Pelgrims} {et~al.}(2021){Pelgrims}, {Clark}, {Hensley},
  {Panopoulou}, {Pavlidou}, {Tassis}, {Eriksen}, \& {Wehus}}]{pelgrims_2021}
{Pelgrims}, V., {Clark}, S.~E., {Hensley}, B.~S., {et~al.} 2021, \aap, 647, A16

\bibitem[{{Pety}(2005)}]{2005sf2a.conf..721P}
{Pety}, J. 2005, in SF2A-2005: Semaine de l'Astrophysique Francaise, ed.
  F.~{Casoli}, T.~{Contini}, J.~M. {Hameury}, \& L.~{Pagani}, 721

\bibitem[{{Pineda} {et~al.}(2010){Pineda}, {Goldsmith}, {Chapman}, {Snell},
  {Li}, {Cambr{\'e}sy}, \& {Brunt}}]{pineda_2010}
{Pineda}, J.~L., {Goldsmith}, P.~F., {Chapman}, N., {et~al.} 2010, \apj, 721,
  686

\bibitem[{{Pineda} {et~al.}(2017){Pineda}, {Langer}, {Goldsmith}, {Horiuchi},
  {Kuiper}, {Muller}, {Hughes}, {Ott}, {Requena-Torres}, {Velusamy}, \&
  {Wong}}]{pineda_2017}
{Pineda}, J.~L., {Langer}, W.~D., {Goldsmith}, P.~F., {et~al.} 2017, \apj, 839,
  107

\bibitem[{{Pineda} {et~al.}(2013){Pineda}, {Langer}, {Velusamy}, \&
  {Goldsmith}}]{pineda_2013}
{Pineda}, J.~L., {Langer}, W.~D., {Velusamy}, T., \& {Goldsmith}, P.~F. 2013,
  \aap, 554, A103

\bibitem[{{Planck Collaboration} {et~al.}(2016{\natexlab{a}}){Planck
  Collaboration}, {Adam}, {Ade}, {Aghanim}, {Arnaud}, {Aumont}, {Baccigalupi},
  {Banday}, {Barreiro}, {Bartlett}, {Bartolo}, {Battaner}, {Benabed},
  {Benoit-L{\'e}vy}, {Bernard}, {Bersanelli}, {Bielewicz}, {Bonaldi},
  {Bonavera}, {Bond}, {Borrill}, {Bouchet}, {Boulanger}, {Bracco}, {Bucher},
  {Burigana}, {Butler}, {Calabrese}, {Cardoso}, {Catalano}, {Challinor},
  {Chamballu}, {Chary}, {Chiang}, {Christensen}, {Clements}, {Colombi},
  {Colombo}, {Combet}, {Couchot}, {Coulais}, {Crill}, {Curto}, {Cuttaia},
  {Danese}, {Davies}, {Davis}, {de Bernardis}, {de Zotti}, {Delabrouille},
  {Delouis}, {D{\'e}sert}, {Dickinson}, {Diego}, {Dolag}, {Dole}, {Donzelli},
  {Dor{\'e}}, {Douspis}, {Ducout}, {Dunkley}, {Dupac}, {Efstathiou}, {Elsner},
  {En{\ss}lin}, {Eriksen}, {Falgarone}, {Finelli}, {Forni}, {Frailis},
  {Fraisse}, {Franceschi}, {Frejsel}, {Galeotta}, {Galli}, {Ganga}, {Ghosh},
  {Giard}, {Giraud-H{\'e}raud}, {Gjerl{\o}w}, {Gonz{\'a}lez-Nuevo},
  {G{\'o}rski}, {Gratton}, {Gregorio}, {Gruppuso}, {Guillet}, {Hansen},
  {Hanson}, {Harrison}, {Helou}, {Henrot-Versill{\'e}},
  {Hern{\'a}ndez-Monteagudo}, {Herranz}, {Hivon}, {Hobson}, {Holmes},
  {Huffenberger}, {Hurier}, {Jaffe}, {Jaffe}, {Jewell}, {Jones}, {Juvela},
  {Keih{\"a}nen}, {Keskitalo}, {Kisner}, {Kneissl}, {Knoche}, {Knox},
  {Krachmalnicoff}, {Kunz}, {Kurki-Suonio}, {Lagache}, {Lamarre}, {Lasenby},
  {Lattanzi}, {Lawrence}, {Leahy}, {Leonardi}, {Lesgourgues}, {Levrier},
  {Liguori}, {Lilje}, {Linden-V{\o}rnle}, {L{\'o}pez-Caniego}, {Lubin},
  {Mac{\'\i}as-P{\'e}rez}, {Maffei}, {Maino}, {Mandolesi}, {Mangilli}, {Maris},
  {Martin}, {Mart{\'\i}nez-Gonz{\'a}lez}, {Masi}, {Matarrese}, {Mazzotta},
  {Meinhold}, {Melchiorri}, {Mendes}, {Mennella}, {Migliaccio}, {Mitra},
  {Miville-Desch{\^e}nes}, {Moneti}, {Montier}, {Morgante}, {Mortlock}, {Moss},
  {Munshi}, {Murphy}, {Naselsky}, {Nati}, {Natoli}, {Netterfield},
  {N{\o}rgaard-Nielsen}, {Noviello}, {Novikov}, {Novikov}, {Pagano}, {Pajot},
  {Paladini}, {Paoletti}, {Partridge}, {Pasian}, {Patanchon}, {Pearson},
  {Perdereau}, {Perotto}, {Perrotta}, {Pettorino}, {Piacentini}, {Piat},
  {Pierpaoli}, {Pietrobon}, {Plaszczynski}, {Pointecouteau}, {Polenta},
  {Ponthieu}, {Popa}, {Pratt}, {Prunet}, {Puget}, {Rachen}, {Reach}, {Rebolo},
  {Remazeilles}, {Renault}, {Renzi}, {Ricciardi}, {Ristorcelli}, {Rocha},
  {Rosset}, {Rossetti}, {Roudier}, {Rouill{\'e} d'Orfeuil},
  {Rubi{\~n}o-Mart{\'\i}n}, {Rusholme}, {Sandri}, {Santos}, {Savelainen},
  {Savini}, {Scott}, {Soler}, {Spencer}, {Stolyarov}, {Stompor}, {Sudiwala},
  {Sunyaev}, {Sutton}, {Suur-Uski}, {Sygnet}, {Tauber}, {Terenzi},
  {Toffolatti}, {Tomasi}, {Tristram}, {Tucci}, {Tuovinen}, {Valenziano},
  {Valiviita}, {Van Tent}, {Vibert}, {Vielva}, {Villa}, {Wade}, {Wandelt},
  {Watson}, {Wehus}, {White}, {White}, {Yvon}, {Zacchei}, \&
  {Zonca}}]{planck2016_dust}
{Planck Collaboration}, {Adam}, R., {Ade}, P.~A.~R., {et~al.}
  2016{\natexlab{a}}, \aap, 586, A133

\bibitem[{{Planck Collaboration} {et~al.}(2015{\natexlab{a}}){Planck
  Collaboration}, {Ade}, {Aghanim}, {Alina}, {Alves}, {Armitage-Caplan},
  {Arnaud}, {Arzoumanian}, {Ashdown}, {Atrio-Barand ela}, {Aumont},
  {Baccigalupi}, {Banday}, {Barreiro}, {Battaner}, {Benabed},
  {Benoit-L{\'e}vy}, {Bernard}, {Bersanelli}, {Bielewicz}, {Bock}, {Bond},
  {Borrill}, {Bouchet}, {Boulanger}, {Bracco}, {Burigana}, {Butler}, {Cardoso},
  {Catalano}, {Chamballu}, {Chary}, {Chiang}, {Christensen}, {Colombi},
  {Colombo}, {Combet}, {Couchot}, {Coulais}, {Crill}, {Curto}, {Cuttaia},
  {Danese}, {Davies}, {Davis}, {de Bernardis}, {de Gouveia Dal Pino}, {de
  Rosa}, {de Zotti}, {Delabrouille}, {D{\'e}sert}, {Dickinson}, {Diego},
  {Donzelli}, {Dor{\'e}}, {Douspis}, {Dunkley}, {Dupac}, {Efstathiou},
  {En{\ss}lin}, {Eriksen}, {Falgarone}, {Ferri{\`e}re}, {Finelli}, {Forni},
  {Frailis}, {Fraisse}, {Franceschi}, {Galeotta}, {Ganga}, {Ghosh}, {Giard},
  {Giraud-H{\'e}raud}, {Gonz{\'a}lez-Nuevo}, {G{\'o}rski}, {Gregorio},
  {Gruppuso}, {Guillet}, {Hansen}, {Harrison}, {Helou},
  {Hern{\'a}ndez-Monteagudo}, {Hildebrand t}, {Hivon}, {Hobson}, {Holmes},
  {Hornstrup}, {Huffenberger}, {Jaffe}, {Jaffe}, {Jones}, {Juvela},
  {Keih{\"a}nen}, {Keskitalo}, {Kisner}, {Kneissl}, {Knoche}, {Kunz},
  {Kurki-Suonio}, {Lagache}, {L{\"a}hteenm{\"a}ki}, {Lamarre}, {Lasenby},
  {Lawrence}, {Leahy}, {Leonardi}, {Levrier}, {Liguori}, {Lilje},
  {Linden-V{\o}rnle}, {L{\'o}pez-Caniego}, {Lubin}, {Mac{\'\i}as-P{\'e}rez},
  {Maffei}, {Magalh{\~a}es}, {Maino}, {Mandolesi}, {Maris}, {Marshall},
  {Martin}, {Mart{\'\i}nez-Gonz{\'a}lez}, {Masi}, {Matarrese}, {Mazzotta},
  {Melchiorri}, {Mendes}, {Mennella}, {Migliaccio}, {Miville-Desch{\^e}nes},
  {Moneti}, {Montier}, {Morgante}, {Mortlock}, {Munshi}, {Murphy}, {Naselsky},
  {Nati}, {Natoli}, {Netterfield}, {Noviello}, {Novikov}, {Novikov},
  {Oxborrow}, {Pagano}, {Pajot}, {Paladini}, {Paoletti}, {Pasian}, {Pearson},
  {Perdereau}, {Perotto}, {Perrotta}, {Piacentini}, {Piat}, {Pietrobon},
  {Plaszczynski}, {Poidevin}, {Pointecouteau}, {Polenta}, {Popa}, {Pratt},
  {Prunet}, {Puget}, {Rachen}, {Reach}, {Rebolo}, {Reinecke}, {Remazeilles},
  {Renault}, {Ricciardi}, {Riller}, {Ristorcelli}, {Rocha}, {Rosset},
  {Roudier}, {Rubi{\~n}o-Mart{\'\i}n}, {Rusholme}, {Sandri}, {Savini}, {Scott},
  {Spencer}, {Stolyarov}, {Stompor}, {Sudiwala}, {Sutton}, {Suur-Uski},
  {Sygnet}, {Tauber}, {Terenzi}, {Toffolatti}, {Tomasi}, {Tristram}, {Tucci},
  {Umana}, {Valenziano}, {Valiviita}, {Van Tent}, {Vielva}, {Villa}, {Wade},
  {Wandelt}, {Zacchei}, \& {Zonca}}]{planck_2015_XIX}
{Planck Collaboration}, {Ade}, P.~A.~R., {Aghanim}, N., {et~al.}
  2015{\natexlab{a}}, \aap, 576, A104

\bibitem[{{Planck Collaboration} {et~al.}(2016{\natexlab{b}}){Planck
  Collaboration}, {Ade}, {Aghanim}, {Alves}, {Aniano}, {Arnaud}, {Ashdown},
  {Aumont}, {Baccigalupi}, {Banday}, {Barreiro}, {Bartolo}, {Battaner},
  {Benabed}, {Benoit-L{\'e}vy}, {Bernard}, {Bersanelli}, {Bielewicz},
  {Bonaldi}, {Bonavera}, {Bond}, {Borrill}, {Bouchet}, {Boulanger}, {Burigana},
  {Butler}, {Calabrese}, {Cardoso}, {Catalano}, {Chamballu}, {Chiang},
  {Christensen}, {Clements}, {Colombi}, {Colombo}, {Couchot}, {Crill}, {Curto},
  {Cuttaia}, {Danese}, {Davies}, {Davis}, {de Bernardis}, {de Rosa}, {de
  Zotti}, {Delabrouille}, {Dickinson}, {Diego}, {Dole}, {Donzelli}, {Dor{\'e}},
  {Douspis}, {Draine}, {Ducout}, {Dupac}, {Efstathiou}, {Elsner}, {En{\ss}lin},
  {Eriksen}, {Falgarone}, {Finelli}, {Forni}, {Frailis}, {Fraisse},
  {Franceschi}, {Frejsel}, {Galeotta}, {Galli}, {Ganga}, {Ghosh}, {Giard},
  {Gjerl{\o}w}, {Gonz{\'a}lez-Nuevo}, {G{\'o}rski}, {Gregorio}, {Gruppuso},
  {Guillet}, {Hansen}, {Hanson}, {Harrison}, {Henrot-Versill{\'e}},
  {Hern{\'a}ndez-Monteagudo}, {Herranz}, {Hildebrandt}, {Hivon}, {Holmes},
  {Hovest}, {Huffenberger}, {Hurier}, {Jaffe}, {Jaffe}, {Jones},
  {Keih{\"a}nen}, {Keskitalo}, {Kisner}, {Kneissl}, {Knoche}, {Kunz},
  {Kurki-Suonio}, {Lagache}, {Lamarre}, {Lasenby}, {Lattanzi}, {Lawrence},
  {Leonardi}, {Levrier}, {Liguori}, {Lilje}, {Linden-V{\o}rnle},
  {L{\'o}pez-Caniego}, {Lubin}, {Mac{\'\i}as-P{\'e}rez}, {Maffei}, {Maino},
  {Mandolesi}, {Maris}, {Marshall}, {Martin}, {Mart{\'\i}nez-Gonz{\'a}lez},
  {Masi}, {Matarrese}, {Mazzotta}, {Melchiorri}, {Mendes}, {Mennella},
  {Migliaccio}, {Miville-Desch{\^e}nes}, {Moneti}, {Montier}, {Morgante},
  {Mortlock}, {Munshi}, {Murphy}, {Naselsky}, {Natoli}, {N{\o}rgaard-Nielsen},
  {Novikov}, {Novikov}, {Oxborrow}, {Pagano}, {Pajot}, {Paladini}, {Paoletti},
  {Pasian}, {Perdereau}, {Perotto}, {Perrotta}, {Pettorino}, {Piacentini},
  {Piat}, {Plaszczynski}, {Pointecouteau}, {Polenta}, {Ponthieu}, {Popa},
  {Pratt}, {Prunet}, {Puget}, {Rachen}, {Reach}, {Rebolo}, {Reinecke},
  {Remazeilles}, {Renault}, {Ristorcelli}, {Rocha}, {Roudier},
  {Rubi{\~n}o-Mart{\'\i}n}, {Rusholme}, {Sandri}, {Santos}, {Scott}, {Spencer},
  {Stolyarov}, {Sudiwala}, {Sunyaev}, {Sutton}, {Suur-Uski}, {Sygnet},
  {Tauber}, {Terenzi}, {Toffolatti}, {Tomasi}, {Tristram}, {Tucci}, {Umana},
  {Valenziano}, {Valiviita}, {Van Tent}, {Vielva}, {Villa}, {Wade}, {Wandelt},
  {Wehus}, {Ysard}, {Yvon}, {Zacchei}, \& {Zonca}}]{planck_2016_av}
{Planck Collaboration}, {Ade}, P.~A.~R., {Aghanim}, N., {et~al.}
  2016{\natexlab{b}}, \aap, 586, A132

\bibitem[{{Planck Collaboration} {et~al.}(2016{\natexlab{c}}){Planck
  Collaboration}, {Ade}, {Aghanim}, {Alves}, {Arnaud}, {Arzoumanian},
  {Ashdown}, {Aumont}, {Baccigalupi}, {Band ay}, {Barreiro}, {Bartolo},
  {Battaner}, {Benabed}, {Beno{\^\i}t}, {Benoit-L{\'e}vy}, {Bernard},
  {Bersanelli}, {Bielewicz}, {Bock}, {Bonavera}, {Bond}, {Borrill}, {Bouchet},
  {Boulanger}, {Bracco}, {Burigana}, {Calabrese}, {Cardoso}, {Catalano},
  {Chiang}, {Christensen}, {Colombo}, {Combet}, {Couchot}, {Crill}, {Curto},
  {Cuttaia}, {Danese}, {Davies}, {Davis}, {de Bernardis}, {de Rosa}, {de
  Zotti}, {Delabrouille}, {Dickinson}, {Diego}, {Dole}, {Donzelli}, {Dor{\'e}},
  {Douspis}, {Ducout}, {Dupac}, {Efstathiou}, {Elsner}, {En{\ss}lin},
  {Eriksen}, {Falceta-Gon{\c{c}}alves}, {Falgarone}, {Ferri{\`e}re}, {Finelli},
  {Forni}, {Frailis}, {Fraisse}, {Franceschi}, {Frejsel}, {Galeotta}, {Galli},
  {Ganga}, {Ghosh}, {Giard}, {Gjerl{\o}w}, {Gonz{\'a}lez-Nuevo}, {G{\'o}rski},
  {Gregorio}, {Gruppuso}, {Gudmundsson}, {Guillet}, {Harrison}, {Helou},
  {Hennebelle}, {Henrot-Versill{\'e}}, {Hern{\'a}ndez-Monteagudo}, {Herranz},
  {Hildebrand t}, {Hivon}, {Holmes}, {Hornstrup}, {Huffenberger}, {Hurier},
  {Jaffe}, {Jaffe}, {Jones}, {Juvela}, {Keih{\"a}nen}, {Keskitalo}, {Kisner},
  {Knoche}, {Kunz}, {Kurki-Suonio}, {Lagache}, {Lamarre}, {Lasenby},
  {Lattanzi}, {Lawrence}, {Leonardi}, {Levrier}, {Liguori}, {Lilje},
  {Linden-V{\o}rnle}, {L{\'o}pez-Caniego}, {Lubin}, {Mac{\'\i}as-P{\'e}rez},
  {Maino}, {Mandolesi}, {Mangilli}, {Maris}, {Martin},
  {Mart{\'\i}nez-Gonz{\'a}lez}, {Masi}, {Matarrese}, {Melchiorri}, {Mendes},
  {Mennella}, {Migliaccio}, {Miville-Desch{\^e}nes}, {Moneti}, {Montier},
  {Morgante}, {Mortlock}, {Munshi}, {Murphy}, {Naselsky}, {Nati},
  {Netterfield}, {Noviello}, {Novikov}, {Novikov}, {Oppermann}, {Oxborrow},
  {Pagano}, {Pajot}, {Paladini}, {Paoletti}, {Pasian}, {Perotto}, {Pettorino},
  {Piacentini}, {Piat}, {Pierpaoli}, {Pietrobon}, {Plaszczynski},
  {Pointecouteau}, {Polenta}, {Ponthieu}, {Pratt}, {Prunet}, {Puget}, {Rachen},
  {Reinecke}, {Remazeilles}, {Renault}, {Renzi}, {Ristorcelli}, {Rocha},
  {Rossetti}, {Roudier}, {Rubi{\~n}o-Mart{\'\i}n}, {Rusholme}, {Sandri},
  {Santos}, {Savelainen}, {Savini}, {Scott}, {Soler}, {Stolyarov}, {Sudiwala},
  {Sutton}, {Suur-Uski}, {Sygnet}, {Tauber}, {Terenzi}, {Toffolatti}, {Tomasi},
  {Tristram}, {Tucci}, {Umana}, {Valenziano}, {Valiviita}, {Van Tent},
  {Vielva}, {Villa}, {Wade}, {Wandelt}, {Wehus}, {Ysard}, {Yvon}, \&
  {Zonca}}]{planck_xxxv_2016}
{Planck Collaboration}, {Ade}, P.~A.~R., {Aghanim}, N., {et~al.}
  2016{\natexlab{c}}, \aap, 586, A138

\bibitem[{{Planck Collaboration} {et~al.}(2020){Planck Collaboration},
  {Aghanim}, {Akrami}, {Alves}, {Ashdown}, {Aumont}, {Baccigalupi},
  {Ballardini}, {Banday}, {Barreiro}, {Bartolo}, {Basak}, {Benabed}, {Bernard},
  {Bersanelli}, {Bielewicz}, {Bock}, {Bond}, {Borrill}, {Bouchet}, {Boulanger},
  {Bracco}, {Bucher}, {Burigana}, {Calabrese}, {Cardoso}, {Carron}, {Chary},
  {Chiang}, {Colombo}, {Combet}, {Crill}, {Cuttaia}, {de Bernardis}, {de
  Zotti}, {Delabrouille}, {Delouis}, {Di Valentino}, {Dickinson}, {Diego},
  {Dor{\'e}}, {Douspis}, {Ducout}, {Dupac}, {Efstathiou}, {Elsner},
  {En{\ss}lin}, {Eriksen}, {Falgarone}, {Fantaye}, {Fernandez-Cobos},
  {Ferri{\`e}re}, {Finelli}, {Forastieri}, {Frailis}, {Fraisse}, {Franceschi},
  {Frolov}, {Galeotta}, {Galli}, {Ganga}, {G{\'e}nova-Santos}, {Gerbino},
  {Ghosh}, {Gonz{\'a}lez-Nuevo}, {G{\'o}rski}, {Gratton}, {Green}, {Gruppuso},
  {Gudmundsson}, {Guillet}, {Handley}, {Hansen}, {Helou}, {Herranz}, {Hivon},
  {Huang}, {Jaffe}, {Jones}, {Keih{\"a}nen}, {Keskitalo}, {Kiiveri}, {Kim},
  {Krachmalnicoff}, {Kunz}, {Kurki-Suonio}, {Lagache}, {Lamarre}, {Lasenby},
  {Lattanzi}, {Lawrence}, {Le Jeune}, {Levrier}, {Liguori}, {Lilje},
  {Lindholm}, {L{\'o}pez-Caniego}, {Lubin}, {Ma}, {Mac{\'\i}as-P{\'e}rez},
  {Maggio}, {Maino}, {Mandolesi}, {Mangilli}, {Marcos-Caballero}, {Maris},
  {Martin}, {Mart{\'\i}nez-Gonz{\'a}lez}, {Matarrese}, {Mauri}, {McEwen},
  {Melchiorri}, {Mennella}, {Migliaccio}, {Miville-Desch{\^e}nes}, {Molinari},
  {Moneti}, {Montier}, {Morgante}, {Moss}, {Natoli}, {Pagano}, {Paoletti},
  {Patanchon}, {Perrotta}, {Pettorino}, {Piacentini}, {Polastri}, {Polenta},
  {Puget}, {Rachen}, {Reinecke}, {Remazeilles}, {Renzi}, {Ristorcelli},
  {Rocha}, {Rosset}, {Roudier}, {Rubi{\~n}o-Mart{\'\i}n}, {Ruiz-Granados},
  {Salvati}, {Sandri}, {Savelainen}, {Scott}, {Sirignano}, {Sunyaev},
  {Suur-Uski}, {Tauber}, {Tavagnacco}, {Tenti}, {Toffolatti}, {Tomasi},
  {Trombetti}, {Valiviita}, {Vansyngel}, {Van Tent}, {Vielva}, {Villa},
  {Vittorio}, {Wandelt}, {Wehus}, {Zacchei}, \& {Zonca}}]{planck_2020}
{Planck Collaboration}, {Aghanim}, N., {Akrami}, Y., {et~al.} 2020, \aap, 641,
  A12

\bibitem[{{Planck Collaboration} {et~al.}(2015{\natexlab{b}}){Planck
  Collaboration}, {Fermi Collaboration}, {Ade}, {Aghanim}, {Aniano}, {Arnaud},
  {Ashdown}, {Aumont}, {Baccigalupi}, {Banday}, {Barreiro}, {Bartolo},
  {Battaner}, {Benabed}, {Benoit-L{\'e}vy}, {Bernard}, {Bersanelli},
  {Bielewicz}, {Bonaldi}, {Bonavera}, {Bond}, {Borrill}, {Bouchet},
  {Boulanger}, {Burigana}, {Butler}, {Calabrese}, {Cardoso}, {Casand jian},
  {Catalano}, {Chamballu}, {Chiang}, {Christensen}, {Colombo}, {Combet},
  {Couchot}, {Crill}, {Curto}, {Cuttaia}, {Danese}, {Davies}, {Davis}, {de
  Bernardis}, {de Rosa}, {de Zotti}, {Delabrouille}, {D{\'e}sert}, {Dickinson},
  {Diego}, {Digel}, {Dole}, {Donzelli}, {Dor{\'e}}, {Douspis}, {Ducout},
  {Dupac}, {Efstathiou}, {Elsner}, {En{\ss}lin}, {Eriksen}, {Falgarone},
  {Finelli}, {Forni}, {Frailis}, {Fraisse}, {Franceschi}, {Frejsel}, {Fukui},
  {Galeotta}, {Galli}, {Ganga}, {Ghosh}, {Giard}, {Gjerl{\o}w},
  {Gonz{\'a}lez-Nuevo}, {G{\'o}rski}, {Gregorio}, {Grenier}, {Gruppuso},
  {Hansen}, {Hanson}, {Harrison}, {Henrot-Versill{\'e}},
  {Hern{\'a}ndez-Monteagudo}, {Herranz}, {Hildebrand t}, {Hivon}, {Hobson},
  {Holmes}, {Hovest}, {Huffenberger}, {Hurier}, {Jaffe}, {Jaffe}, {Jones},
  {Juvela}, {Keih{\"a}nen}, {Keskitalo}, {Kisner}, {Kneissl}, {Knoche}, {Kunz},
  {Kurki-Suonio}, {Lagache}, {Lamarre}, {Lasenby}, {Lattanzi}, {Lawrence},
  {Leonardi}, {Levrier}, {Liguori}, {Lilje}, {Linden-V{\o}rnle},
  {L{\'o}pez-Caniego}, {Lubin}, {Mac{\'\i}as-P{\'e}rez}, {Maffei}, {Maino},
  {Mand olesi}, {Maris}, {Marshall}, {Martin}, {Mart{\'\i}nez-Gonz{\'a}lez},
  {Masi}, {Matarrese}, {Mazzotta}, {Melchiorri}, {Mendes}, {Mennella},
  {Migliaccio}, {Miville-Desch{\^e}nes}, {Moneti}, {Montier}, {Morgante},
  {Mortlock}, {Munshi}, {Murphy}, {Naselsky}, {Natoli}, {N{\o}rgaard-Nielsen},
  {Novikov}, {Novikov}, {Oxborrow}, {Pagano}, {Pajot}, {Paladini}, {Paoletti},
  {Pasian}, {Perdereau}, {Perotto}, {Perrotta}, {Pettorino}, {Piacentini},
  {Piat}, {Plaszczynski}, {Pointecouteau}, {Polenta}, {Popa}, {Pratt},
  {Prunet}, {Puget}, {Rachen}, {Reach}, {Rebolo}, {Reinecke}, {Remazeilles},
  {Renault}, {Ristorcelli}, {Rocha}, {Roudier}, {Rusholme}, {Sandri}, {Santos},
  {Scott}, {Spencer}, {Stolyarov}, {Strong}, {Sudiwala}, {Sunyaev}, {Sutton},
  {Suur-Uski}, {Sygnet}, {Tauber}, {Terenzi}, {Tibaldo}, {Toffolatti},
  {Tomasi}, {Tristram}, {Tucci}, {Umana}, {Valenziano}, {Valiviita}, {Van
  Tent}, {Vielva}, {Villa}, {Wade}, {Wandelt}, {Wehus}, {Yvon}, {Zacchei}, \&
  {Zonca}}]{planck_2015_chameleon}
{Planck Collaboration}, {Fermi Collaboration}, {Ade}, P.~A.~R., {et~al.}
  2015{\natexlab{b}}, \aap, 582, A31

\bibitem[{{Plaszczynski} {et~al.}(2014){Plaszczynski}, {Montier}, {Levrier}, \&
  {Tristram}}]{Plaszczynski_2014}
{Plaszczynski}, S., {Montier}, L., {Levrier}, F., \& {Tristram}, M. 2014,
  \mnras, 439, 4048

\bibitem[{{Pound} \& {Goodman}(1997)}]{pound_1997}
{Pound}, M.~W. \& {Goodman}, A.~A. 1997, \apj, 482, 334

\bibitem[{{Ramaprakash} {et~al.}(2019){Ramaprakash}, {Rajarshi}, {Das},
  {Khodade}, {Modi}, {Panopoulou}, {Maharana}, {Blinov}, {Angelakis},
  {Casadio}, {Fuhrmann}, {Hovatta}, {Kiehlmann}, {King}, {Kylafis},
  {Kougentakis}, {Kus}, {Mahabal}, {Marecki}, {Myserlis}, {Paterakis},
  {Paleologou}, {Liodakis}, {Papadakis}, {Papamastorakis}, {Pavlidou},
  {Pazderski}, {Pearson}, {Readhead}, {Reig}, {S{\l}owikowska}, {Tassis}, \&
  {Zensus}}]{robopol_paper_2019}
{Ramaprakash}, A.~N., {Rajarshi}, C.~V., {Das}, H.~K., {et~al.} 2019, \mnras,
  485, 2355

\bibitem[{{Schlegel} {et~al.}(1998){Schlegel}, {Finkbeiner}, \&
  {Davis}}]{Schlegel_1998}
{Schlegel}, D.~J., {Finkbeiner}, D.~P., \& {Davis}, M. 1998, \apj, 500, 525

\bibitem[{{Schmidt} {et~al.}(1992){Schmidt}, {Elston}, \&
  {Lupie}}]{schmidt_1992}
{Schmidt}, G.~D., {Elston}, R., \& {Lupie}, O.~L. 1992, \aj, 104, 1563

\bibitem[{{Seifried} {et~al.}(2020){Seifried}, {Haid}, {Walch}, {Borchert}, \&
  {Bisbas}}]{Seifried_2020}
{Seifried}, D., {Haid}, S., {Walch}, S., {Borchert}, E.~M.~A., \& {Bisbas},
  T.~G. 2020, \mnras, 492, 1465

\bibitem[{{Shan} {et~al.}(2012){Shan}, {Yang}, {Shi}, {Yao}, {Zuo}, {Lin},
  {Chen}, {Zhang}, {Duan}, {Cao}, {Li}, {Li}, {Liu}, \&
  {Zhong}}]{2012ITTST...2..593S}
{Shan}, W., {Yang}, J., {Shi}, S., {et~al.} 2012, IEEE Transactions on
  Terahertz Science and Technology, 2, 593

\bibitem[{{Shull} {et~al.}(2021){Shull}, {Danforth}, \&
  {Anderson}}]{shull_2021}
{Shull}, J.~M., {Danforth}, C.~W., \& {Anderson}, K.~L. 2021, \apj, 911, 55

\bibitem[{{Skalidis} {et~al.}(2018){Skalidis}, {Panopoulou}, {Tassis},
  {Pavlidou}, {Blinov}, {Komis}, \& {Liodakis}}]{skalidis_2018}
{Skalidis}, R., {Panopoulou}, G.~V., {Tassis}, K., {et~al.} 2018, \aap, 616,
  A52

\bibitem[{{Skalidis} \& {Pelgrims}(2019)}]{skalidis_2019}
{Skalidis}, R. \& {Pelgrims}, V. 2019, \aap, 631, L11

\bibitem[{{Skalidis} {et~al.}(2021){Skalidis}, {Sternberg}, {Beattie},
  {Pavlidou}, \& {Tassis}}]{skalidis_2021}
{Skalidis}, R., {Sternberg}, J., {Beattie}, J.~R., {Pavlidou}, V., \& {Tassis},
  K. 2021, arXiv e-prints, arXiv:2109.10925

\bibitem[{{Skalidis} \& {Tassis}(2021)}]{skalidis_2020}
{Skalidis}, R. \& {Tassis}, K. 2021, \aap, 647, A186

\bibitem[{{Sofia} {et~al.}(2004){Sofia}, {Lauroesch}, {Meyer}, \&
  {Cartledge}}]{sofia_2001}
{Sofia}, U.~J., {Lauroesch}, J.~T., {Meyer}, D.~M., \& {Cartledge}, S. I.~B.
  2004, \apj, 605, 272

\bibitem[{{Soler} \& {Hennebelle}(2017)}]{soler_2017}
{Soler}, J.~D. \& {Hennebelle}, P. 2017, \aap, 607, A2

\bibitem[{{Soler} {et~al.}(2013){Soler}, {Hennebelle}, {Martin},
  {Miville-Desch{\^e}nes}, {Netterfield}, \& {Fissel}}]{soler_2014}
{Soler}, J.~D., {Hennebelle}, P., {Martin}, P.~G., {et~al.} 2013, \apj, 774,
  128

\bibitem[{{Sternberg} {et~al.}(2014){Sternberg}, {Le Petit}, {Roueff}, \& {Le
  Bourlot}}]{Sternberg2014}
{Sternberg}, A., {Le Petit}, F., {Roueff}, E., \& {Le Bourlot}, J. 2014, \apj,
  790, 10

\bibitem[{{Sun} {et~al.}(2018){Sun}, {Lu}, {Yang}, {Su}, {Zhang}, {Zhou}, \&
  {Lin}}]{2018AcASn..59....3S}
{Sun}, J.~X., {Lu}, D.~R., {Yang}, J., {et~al.} 2018, Acta Astronomica Sinica,
  59, 3

\bibitem[{{Tassis} {et~al.}(2009){Tassis}, {Dowell}, {Hildebrand}, {Kirby}, \&
  {Vaillancourt}}]{tassis_2009}
{Tassis}, K., {Dowell}, C.~D., {Hildebrand}, R.~H., {Kirby}, L., \&
  {Vaillancourt}, J.~E. 2009, \mnras, 399, 1681

\bibitem[{{Tassis} \& {Pavlidou}(2015)}]{tassis_2015}
{Tassis}, K. \& {Pavlidou}, V. 2015, \mnras, 451, L90

\bibitem[{{Tritsis} {et~al.}(2019){Tritsis}, {Federrath}, \&
  {Pavlidou}}]{tritsis_2019}
{Tritsis}, A., {Federrath}, C., \& {Pavlidou}, V. 2019, \apj, 873, 38

\bibitem[{{Tritsis} {et~al.}(2018){Tritsis}, {Federrath}, {Schneider}, \&
  {Tassis}}]{tritsis_2018}
{Tritsis}, A., {Federrath}, C., {Schneider}, N., \& {Tassis}, K. 2018, \mnras,
  481, 5275

\bibitem[{{Tritsis} {et~al.}(2015){Tritsis}, {Panopoulou}, {Mouschovias},
  {Tassis}, \& {Pavlidou}}]{tritsis_2015}
{Tritsis}, A., {Panopoulou}, G.~V., {Mouschovias}, T.~C., {Tassis}, K., \&
  {Pavlidou}, V. 2015, \mnras, 451, 4384

\bibitem[{{Tritsis} \& {Tassis}(2016)}]{tritsis_2016}
{Tritsis}, A. \& {Tassis}, K. 2016, \mnras, 462, 3602

\bibitem[{{Tritsis} \& {Tassis}(2018)}]{tritsis_2018_sc}
{Tritsis}, A. \& {Tassis}, K. 2018, Science, 360, 635

\bibitem[{{Turnshek} {et~al.}(1990){Turnshek}, {Bohlin}, {Williamson}, {Lupie},
  {Koornneef}, \& {Morgan}}]{turnshek_1990}
{Turnshek}, D.~A., {Bohlin}, R.~C., {Williamson}, R.~L., I., {et~al.} 1990,
  \aj, 99, 1243

\bibitem[{{Ulich} \& {Haas}(1976)}]{1976ApJS...30..247U}
{Ulich}, B.~L. \& {Haas}, R.~W. 1976, \apjs, 30, 247

\bibitem[{{Vaillancourt}(2006)}]{Vaillancourt_2006}
{Vaillancourt}, J.~E. 2006, \pasp, 118, 1340

\bibitem[{{Velusamy} {et~al.}(2010){Velusamy}, {Langer}, {Pineda}, {Goldsmith},
  {Li}, \& {Yorke}}]{velusamy_2010}
{Velusamy}, T., {Langer}, W.~D., {Pineda}, J.~L., {et~al.} 2010, \aap, 521, L18

\bibitem[{{Walch} {et~al.}(2015){Walch}, {Girichidis}, {Naab}, {Gatto},
  {Glover}, {W{\"u}nsch}, {Klessen}, {Clark}, {Peters}, {Derigs}, \&
  {Baczynski}}]{silc_I}
{Walch}, S., {Girichidis}, P., {Naab}, T., {et~al.} 2015, \mnras, 454, 238

\bibitem[{{Wiesenfeld} \& {Goldsmith}(2014)}]{wiesenfeld_2014}
{Wiesenfeld}, L. \& {Goldsmith}, P.~F. 2014, \apj, 780, 183

\bibitem[{{Yuen} {et~al.}(2021){Yuen}, {Ho}, \&
  {Lazarian}}]{yuen2021_VelRhoCorr}
{Yuen}, K.~H., {Ho}, K.~W., \& {Lazarian}, A. 2021, \apj, 910, 161

\bibitem[{{Yuen} \& {Lazarian}(2017)}]{yuen_ka_ho_2017}
{Yuen}, K.~H. \& {Lazarian}, A. 2017, \apjl, 837, L24

\end{thebibliography}

\begin{acknowledgements}
We thank the anonymous referee for their careful and constructive comments. We also thank P. F. Goldsmith, S. E. Clark, M. Shull, D. Seifried, A. Tritsis, M. Kopsacheili, and V. Pelgrims for fruitful comments on the manuscript. We are grateful to T. Diaz-Santos, D. Fadda, and R. Munch for their help with the SOFIA observations. We are grateful to M.-A. Miville-Deschénes for informing us about the DHIGLS data, and J. Ingalls for sharing their ISO data. We appreciate the assistance of the PMO-13.7 m staff during the observations. Particularly, we would like to thank Jixian Sun for scheduling our supplementary observations with the PMO-13.7 m telescope on very short notice. We greatly thank Qing-zeng Yan for sharing his CO (J=1-0) data prior to his publication. This project has received funding from the European Research Council (ERC) under the European Unions Horizon 2020 research and innovation programme under grant agreement No. 771282. Y.G. was supported by the National Key R\&D Program of China under grant 2017YFA0402702. This project has received funding from the European Union's Horizon 2020 research and innovation programme under the Marie Sklodowska-Curie RISE action, grant agreement No 691164 (ASTROSTAT).
\end{acknowledgements}

\begin{appendix}

    \begin{figure}
        \includegraphics[width=\hsize]{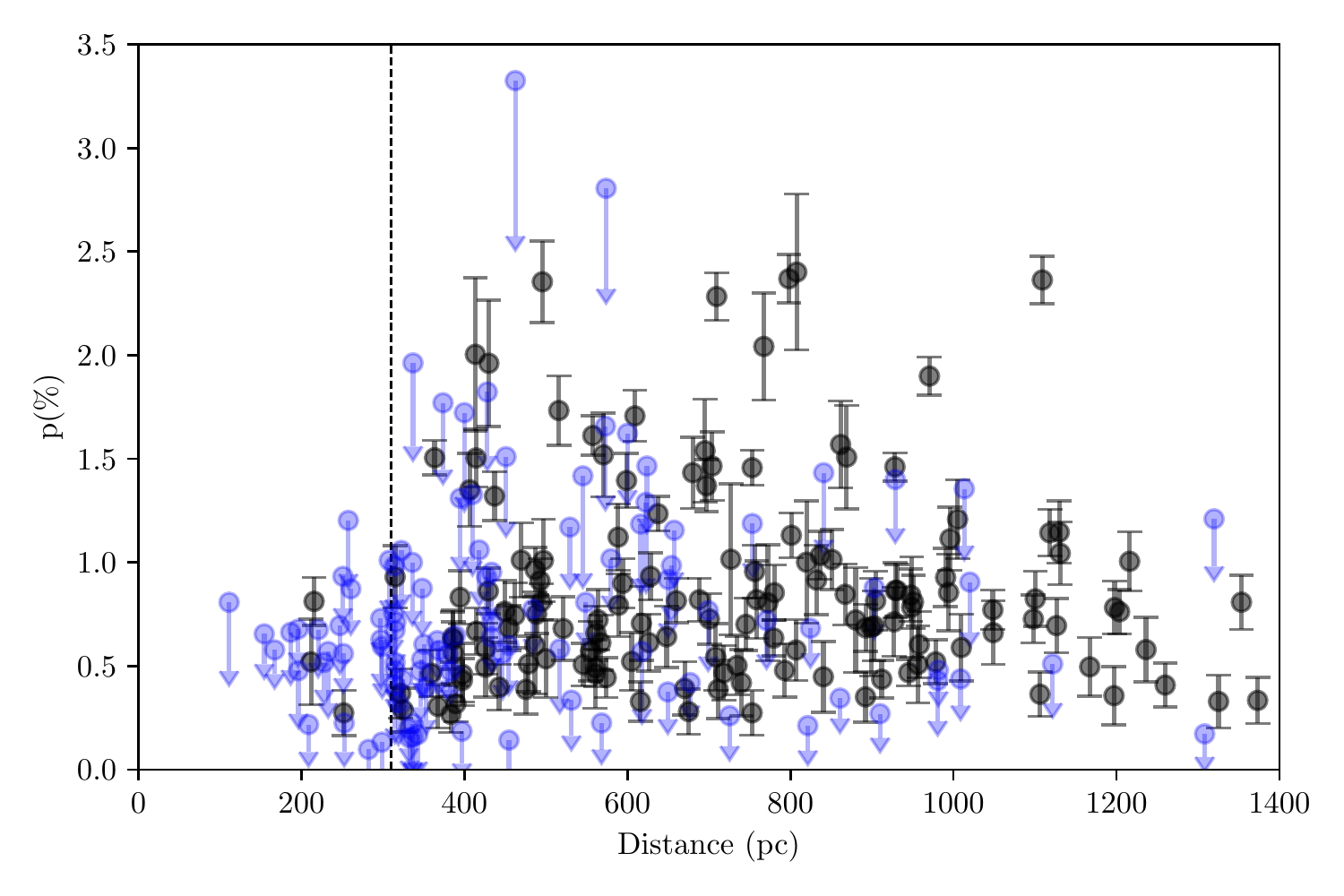}
        \caption{Degree of polarization versus star distance. Black dots indicate measurements with S/N$\geq 2.5$, blue dots measurements with S/N$<2.5$. The plot has been truncated at $1400$ pc for visualization purposes. The vertical line is at $310$ pc.}
        \label{fig:p_d}
    \end{figure}

\section{Cloud distance}
\label{sec:cloud_distance}

Optical dust polarization is a powerful tool for estimating the ISM cloud distances \citep[e.g.][]{panopoulou_2019_extreme, panopoulou_2019_tom}. Stars located in front of an ISM cloud will be un-polarized, because there is negligible dust material to polarize the starlight. On the other hand, starlight emitted from sources behind the cloud, will get linearly polarized as they pass through the cloud due to dichroic extinction \citep{andersson_review}. 

In Fig.~\ref{fig:p_d} we show the degree of polarization (biased) versus the star distances. Black dots indicate measurements with S/N $\geq 2.5$, while blue dots correspond to upper limits. Some measurements located further than 1400 pc are not shown for visualization purposes. There is a step-like increase in the degree of polarization at $\sim 300 - 400$ pc and this indicates the cloud distance. There are some measurements with low S/N at further distances, but this could be caused by large uncertainties due to low exposures or because the magnetic field is inclined with respect to the LOS axis.    

\section{Velocity gradient angles transformation to the Galactic reference frame}
\label{sec:gradient_angles_transformation}

    \begin{figure}
         \centering
         \includegraphics[width=\hsize]{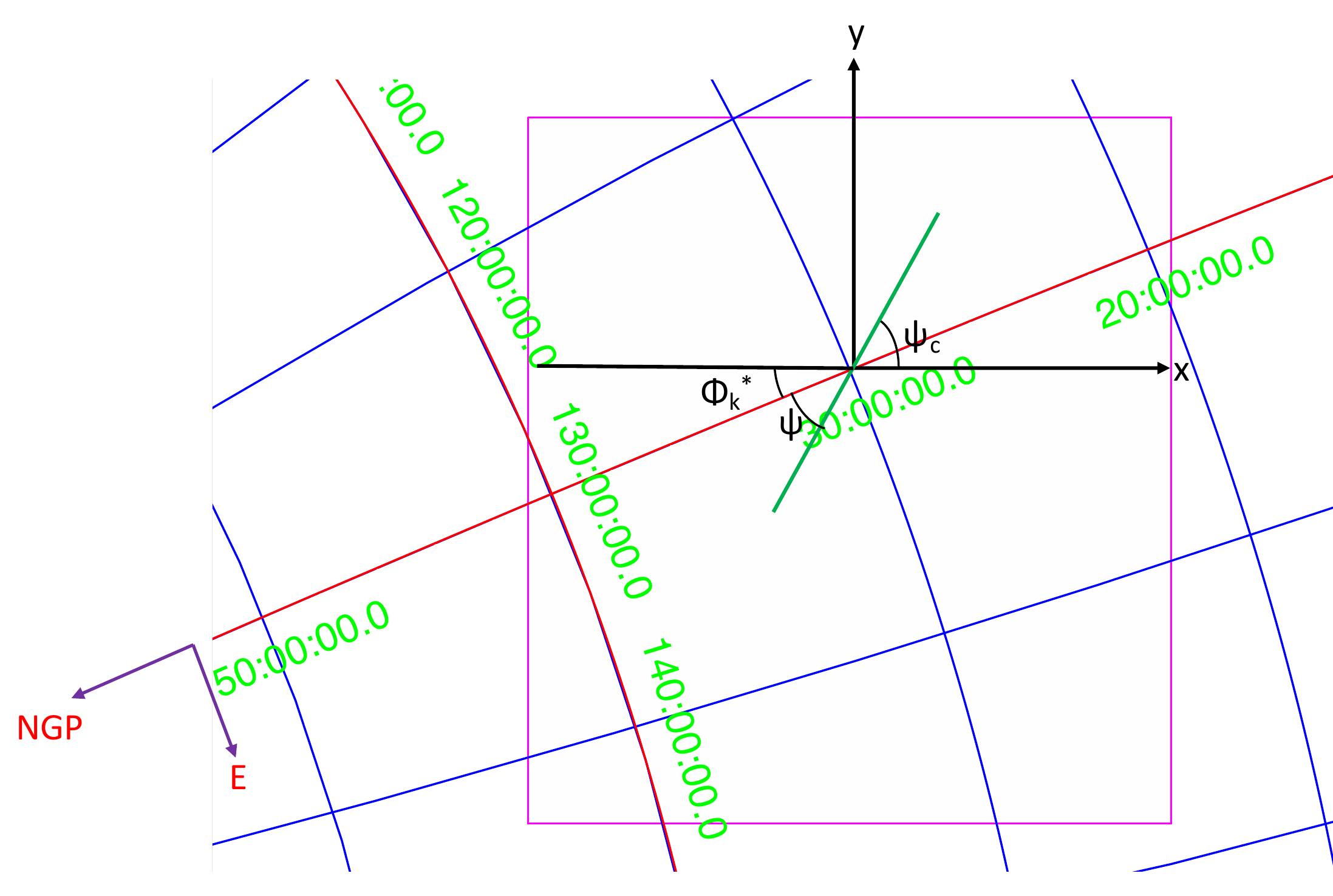}
         \caption{Magenta box corresponds to the HI4PI field towards the NCPL. $\psi_{c}$ is the velocity gradient angle measured with respect to the Cartesian $x-y$ plane of the target field, $\phi_{k}^{*}$ is the local angle between the $x$ axis and the coordinate grid, and $\psi$ is the velocity gradient angle measured with respect to the NGP.}
        \label{fig:velocity_gradient_transformation}
    \end{figure}

We used Eq.~(\ref{eq:gradient_angles}) to measure the \HI\ velocity gradients towards the NCPL. This equation, however, measures the velocity gradients with respect to the horizontal axis of the target field and lie within the $[-90\degr, +90\degr]$ range. On the other hand, the magnetic field orientation, as measured from Planck polarization data, are with respect to the North Galactic Pole (NGP), increasing towards East. We transformed the \HI\ velocity gradient angles ($\psi_{c}$) to the NGP reference frame in order to compute their difference with the magnetic field orientation. 

Fig.~\ref{fig:velocity_gradient_transformation} shows the geometry of this transformation. The magenta box corresponds to the HI4PI field towards the NCPL; $\psi_{c}$ is the velocity gradient angle measured with respect to the Cartesian $x-y$ plane of the image. In order to compute $\psi$, which is the velocity gradient angle measured in the Galactic reference frame, we computed $\phi_{k}^{*}$; this is the local angle between the $x$ axis of the image and the coordinate grid. For the $\phi_{k}^{*}$ computation we used Eq.~(A9) from the Appendix of \cite{planck_2015_XIX} which properly takes into account the local curvature of the grid. Then, we computed $\psi$ as
\begin{equation}
   \label{eq:MA_scaling}
   \psi =  
   \begin{cases} 
    \psi_{c} - \phi_{k}^{*}, & \mbox{$0 \leq \psi \leq 90\degr$} \\ 
    180 - \psi_{c} - \phi_{k}^{*}, & \mbox{$-90 \leq \psi < 0\degr$}. 
   \end{cases}
\end{equation}
The derived $\psi$ angle lies in the [$0\degr$, $180\degr$] range.

\section{\HI\ Gaussian decomposition}
\label{sec:rohsa}

    \begin{figure}
         \centering
         \includegraphics[width=\hsize]{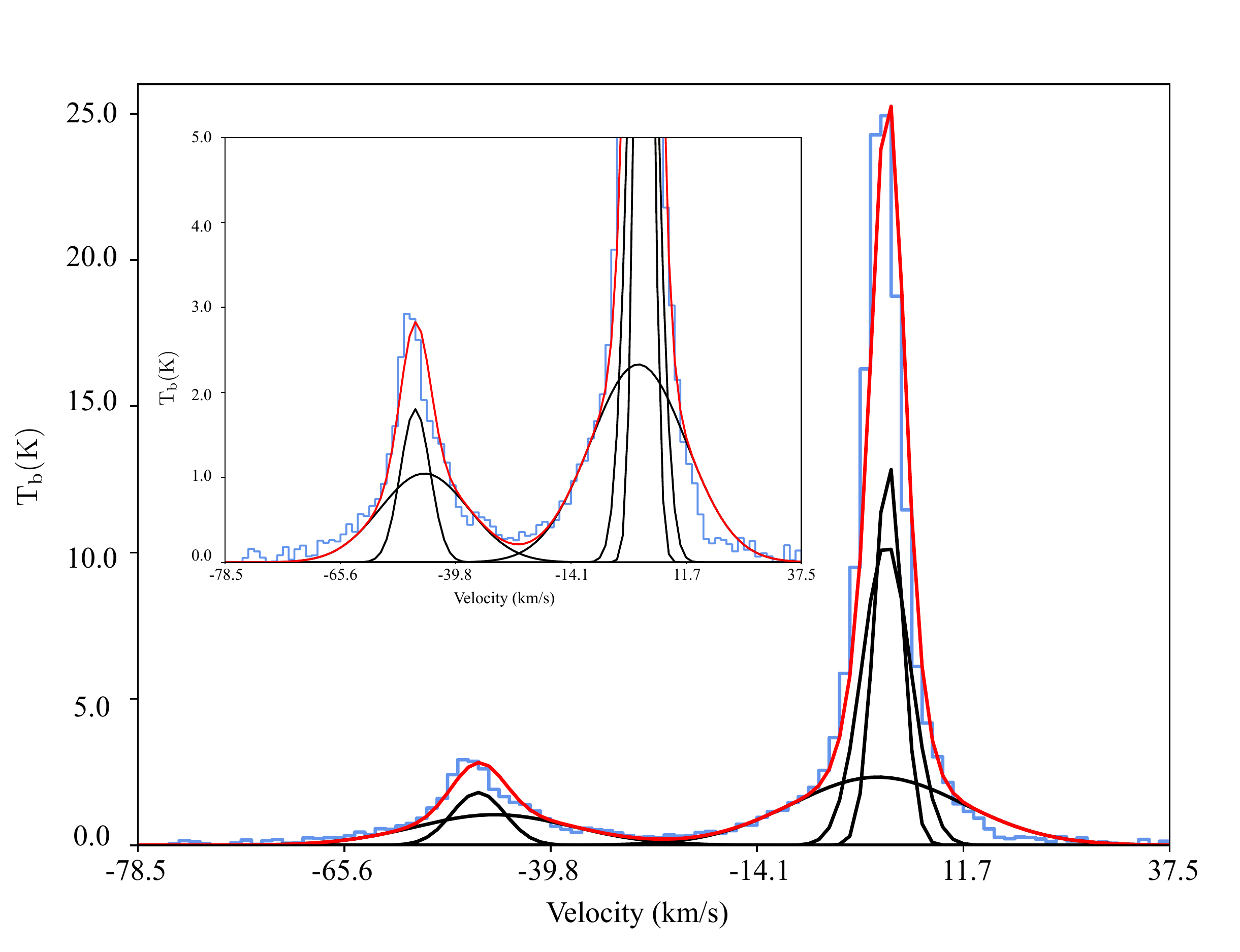}
         \caption{A characteristic \HI\ spectrum fitted by ROHSA. The vertical axis is intensity measured in antenna temperature units (K) and the horizontal axis is the gas velocity in \kms. Blue histogram denotes the observed \HI\ emission spectrum and the red line the corresponding fit produced by ROHSA. The black lines show the individual Gaussian components of the spectrum decomposition.}
         \label{fig:hi_decomposition}
    \end{figure}
    
A characteristic fitted spectrum with ROHSA towards a random LOS along the NCPL is shown with the blue histogram in Fig.~\ref{fig:hi_decomposition}. The individual Gaussian components fitted by ROHSA are shown with the black solid lines, while the red line is the total intensity of the fitted spectrum. In Fig.~\ref{fig:ncpl_gaussian_components} we show the five decomposed Gaussians used for the \HI\ fitting of the NCPL spectra. The first, second and third column corresponds to their zeroth (intensity), first (velocity centroid) and second (velocity spread) moment maps respectively.
  
\section{Contribution of intensity fluctuations in velocity gradients}
\label{sec:rho_vel_corr}
 
\HI\ spectra may show multiple peaks which vary in intensity at small velocity intervals. For example, in Fig.~\ref{fig:hi_decomposition}, the major peak at velocities close to 5 \kms\ is decomposed into three Gaussian components. Consider that each component corresponds to a distinct gas layer within the cloud,  moving at different, but constant velocity. If the intensity of each component changes, because the gas column density changes, then the velocity gradients computed with Eq.~(\ref{eq:gradient_angles}) will be non-zero; however, this gradient would arise from changes in $\rm T_{b}$, rather than velocity fluctuations.

In order to determine the relative contribution of the $\rm T_{b}$ fluctuations in the velocity gradients computation, we explored whether the intensities of the decomposed Gaussians are spatially correlated. If the $\rm T_{b}$ fluctuations between the different components are strongly correlated, then we can fit the spectrum with a single Gaussian, which is the sum of the individual components (as shown in Fig.~\ref{fig:hi_decomposition}). In this case the $\rm T_{b}$ fluctuations are minor, and $\rm V_{c}$ gradients accurately represent the averaged gas kinematics. On the other hand, if Gaussian intensities are not correlated (or anti-correlated), then fluctuations of $\rm T_{b}$ between the different components can be significant; in this case one has to decompose the observed signal into distinct components.

\begin{figure*}
      \centering
      \includegraphics[width=\hsize]{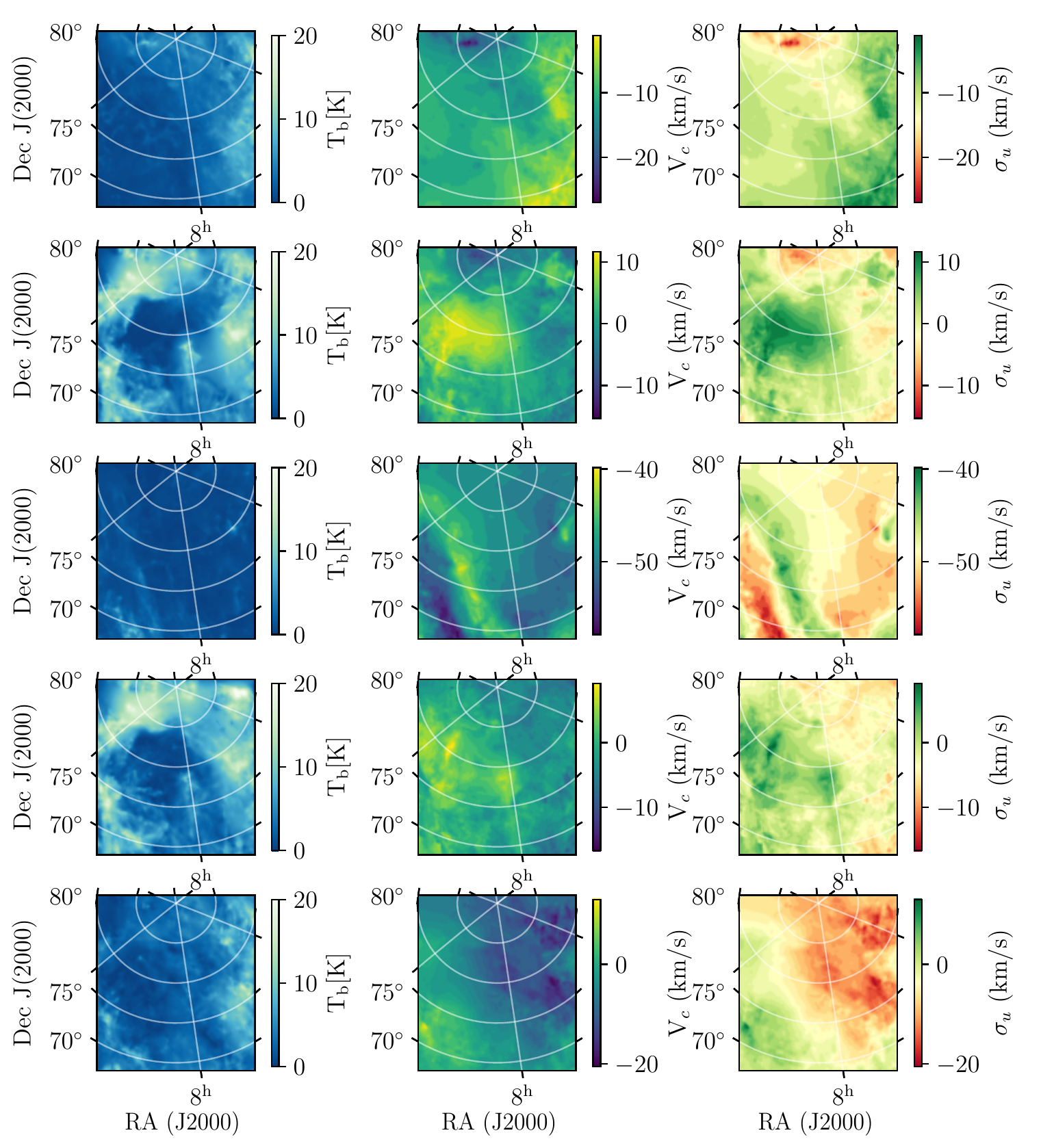}
      \caption{Each row represents the different Gaussian components used for the fitting of the \HI\ spectrum of the NCPL. \textbf{Left column:} Zeroth velocity moment map in antenna temperature units, \textbf{middle column:}, first moment (centroid velocity) map measured in \kms and \textbf{right column:} second moment (velocity dispersion) map measured in \kms.}
      \label{fig:ncpl_gaussian_components}
  \end{figure*}

We employed the \HI\ Gaussian decomposition that we performed with ROHSA in Sect.~\ref{subsection:ncpl_B_molecule_formation}. We computed the normalized correlation of the intensity maps ($\rm T_{b}$) between the three decomposed Gaussians at velocities $\sim 5$ \kms (Fig.~\ref{fig:hi_decomposition}) towards our target cloud; the $\rm T_{b}$ maps of the individuals components  are shown in the first column of the second, fourth, and fifth row in Fig.~\ref{fig:ncpl_gaussian_components}. We computed the correlation between two intensity maps, $\rm T_{b, A}$ and $\rm T_{b, B}$, as in \cite{yuen2021_VelRhoCorr}, 
\begin{equation}
    \rm{NCC(T_{A}, T_{B})} = \frac{\langle (T_{b, A} - \langle T_{b, A} \rangle) (T_{b.B} - \langle T_{b.B} \rangle) \rangle}{\sigma_{A} \sigma_{B}},
\end{equation}
where $\sigma^{2}_{j} = \langle (j - \langle j \rangle)^{2}$, and $j = \rm T_{b, A}, T_{b, B}$. If the two decomposed intensity maps were perfectly correlated, then $\rm{NCC(T_{b, A}, T_{b, B})} = 1$; if they were anti-correlated $\rm{NCC(T_{b, A}, T_{b, B})} = -1$, while if they were uncorrelated, then $\rm{NCC(T_{b, A}, T_{b, B})} = 0$. We consider that two maps are strongly correlated when $\rm{NCC(T_{A}, T_{B})} > 0.7$. Our results are the following: the  $\rm T_{b}$ maps correlation between the second and fourth Gaussian is $\rm{NCC(T_{b, A}, T_{b, B})} = 0.87$, between the second and fifth Gaussian is $\rm{NCC(T_{b, A}, T_{b, B})} = 0.75$, while between the fourth and fifth Gaussian is $\rm{NCC(T_{b, A}, T_{b, B})} = 0.78$. The $\rm T_{b}$ maps of the individual components are strongly correlated with each other. Thus, the observed \HI\ spectrum towards the target cloud can be fitted with a single Gaussian component. This suggests that $\rm T_{b}$ fluctuations have but a minor contribution in the observed \HI\ spectra. Thus, the \HI\ velocity gradients, computed in Sect.~\ref{subsec:Vc_gradients}, are an accurate proxy of the average kinematics of the cloud.

\end{appendix}
\end{document}